\documentclass{article}

    \PassOptionsToPackage{numbers, compress}{natbib}
 \usepackage[preprint]{neurips_2026}

\usepackage{multirow}
\usepackage{booktabs}
\usepackage{graphicx}
\usepackage[utf8]{inputenc} 
\usepackage[T1]{fontenc}    
\usepackage{hyperref}       
\usepackage{url}            
\usepackage{booktabs}       
\usepackage{amsfonts}       
\usepackage{nicefrac}       
\usepackage{microtype}      
\usepackage{xcolor}         
\usepackage{tikz}
\usepackage{pgfplots}
\usepackage{pgfplotstable}
\usepgfplotslibrary{groupplots}
\usepgfplotslibrary{colormaps}
\usepgfplotslibrary{colorbrewer}
\usepackage{pgfplots}
\pgfplotsset{compat=1.18}

\usepackage{xspace}
\usepackage{enumitem} 
\usepackage{subcaption}
\usepackage{booktabs}
\usepackage{wrapfig}
\usepackage{pifont}

\usepackage{adjustbox}

\usetikzlibrary{arrows.meta,positioning,fit}

\usepackage{enumitem}
\newenvironment{smitemize}{
  \begin{itemize}[topsep=1pt, partopsep=2pt, itemsep=3pt, parsep=0pt, leftmargin=10pt, itemindent=1pt]
}{\end{itemize}}

\usepackage{xcolor}
\usepackage{amssymb}
\newcommand{\inc}{\textcolor{green!60!black}{$\uparrow$}}
\newcommand{\dec}{\textcolor{red}{$\downarrow$}}
\newcommand{\nochange}{\phantom{\inc}}
\usepackage{multirow}
\usepackage{makecell}
\usepackage{array}
\usepackage[color=green]{todonotes}

\usepackage[most]{tcolorbox}
\usepgfplotslibrary{fillbetween}
\usetikzlibrary{plotmarks}
\usepackage[table]{xcolor}
\usepackage{xcolor}
\usepackage{pifont}
\usepackage{booktabs}

\definecolor{hcColor}{HTML}{1F77B4}
\definecolor{acColor}{HTML}{FF7F0E}
\definecolor{inColor}{HTML}{2CA02C}
\definecolor{perColor}{HTML}{D62728}
\definecolor{apiColor}{HTML}{9467BD}
\definecolor{bcColor}{HTML}{8C564B}
\definecolor{csColor}{HTML}{7F7F7F}
\definecolor{pgColor}{HTML}{17BECF}
\definecolor{netColor}{HTML}{006D77}
\definecolor{fsColor}{HTML}{A23E48}
\definecolor{sysColor}{HTML}{6A4C93}
\definecolor{procColor}{HTML}{F77F00}
\definecolor{cmdColor}{HTML}{003049}
\definecolor{modColor}{HTML}{2A9D8F}
\definecolor{dbColor}{HTML}{9D0208}

\newcommand{\HC}{\textbf{\textcolor{hcColor}{HC}}}
\newcommand{\AC}{\textbf{\textcolor{acColor}{AC}}}
\newcommand{\IN}{\textbf{\textcolor{inColor}{IN}}}
\newcommand{\PER}{\textbf{\textcolor{perColor}{PER}}}
\newcommand{\API}{\textbf{\textcolor{apiColor}{API}}}

\newcommand{\CS}{\textbf{\textcolor{csColor}{CS}}}
\newcommand{\PG}{\textbf{\textcolor{pgColor}{PG}}}
\newcommand{\NET}{\textbf{\textcolor{netColor}{NET}}}
\newcommand{\FS}{\textbf{\textcolor{fsColor}{FS}}}
\newcommand{\SYS}{\textbf{\textcolor{sysColor}{SYS}}}
\newcommand{\PROC}{\textbf{\textcolor{procColor}{PROC}}}
\newcommand{\CMD}{\textbf{\textcolor{cmdColor}{CMD}}}
\newcommand{\MOD}{\textbf{\textcolor{modColor}{MOD}}}
\newcommand{\DB}{\textbf{\textcolor{dbColor}{DB}}}

\newcommand{\BfPara}[1]{\vspace{0.1em}{\noindent\bf#1.}\xspace\xspace}

\newcommand{\meanstd}[2]{#1{\tiny$\pm${#2}}}

\newcommand{\system}{{McNdroid}\xspace}

\title{\system: A Longitudinal Multimodal Benchmark for Robust Drift Detection in Android Malware}

\author{%
  Md Mahmuduzzaman Kamol\thanks{Corresponding Author.}~~$^{1}$, Jesus Lopez$^{1}$, Saeefa Rubaiyet Nowmi$^{1}$, Emilia Rivas$^{1}$,\\ 
  \textbf{Md Ahsanul Haque$^{1}$, Edward Raff$^{2}$, Aritran Piplai$^{1}$, Mohammad Saidur Rahman$^{1}$} \\
    $^1$Department of Computer Science, University of Texas at El Paso\\
    $^2$CrowdStrike\\
   \texttt{\{mkamol,jlopez126,srnowmi,erivas6,mhaque3\}@miners.utep.edu}
   \\ \texttt{edward.raff@crowdstrike.com}, \texttt{\{apiplai,msrahman3\}@utep.edu}\\
}

\begin{document}


\maketitle

\begin{abstract}
\label{sec:abstract}
Machine learning (ML) in real-world systems must contend with concept drift, adversarial actors, and a spectrum of potential features with varying costs and benefits. Malware naturally exhibits all of these complexities, but for the same reason, it is challenging to curate and organize data to study these factors. We present {\system}, to our knowledge the largest longitudinal multimodal Android malware benchmark for malware detection and drift analysis. {\system} spans 2013--2025, excluding 2015, and represents each application with three aligned modalities--static features from manifests and smali code, dynamic behavioral features from sandbox execution, and graph-based features from function-call graphs. Using temporally separated splits, we evaluate standard ML and deep-learning detectors across increasing train--test time gaps. Results show clear temporal degradation, while multimodal fusion outperforms the best single modality across long-term temporal gaps. Cross-modal agreement also declines over time, suggesting that drift affects both individual feature spaces and the consistency among modalities. We further analyze modality-specific drift, malware-family evolution, and temporal changes in model explanations. We publicly release {\system}, benchmark splits, and code to support reproducible research on temporal generalization and robust multimodal learning in security-critical, non-stationary settings.

\noindent\textbf{Data:} \url{https://huggingface.co/datasets/IQSeC-Lab/McNdroid}\\
\textbf{Code:} \url{https://github.com/IQSeC-Lab/McNdroid/}
\end{abstract}

\section{Introduction}
\label{sec:introduction}

Malware detection is a large-scale and continuously evolving machine-learning (ML) problem~\cite{ember,vsrndic2013detection,chen2020training,arp2014drebin,chen2018detecting}. ENISA identifies malware as a major cybersecurity threat~\cite{ENISAThreatLandscape}, and services such as VirusTotal process about 1.8 million samples per day~\cite{virustotalstats}. Traditional methods, including sandbox analysis, hand-crafted features, and signature matching, require substantial manual effort and often struggle with obfuscation and rapidly changing malware~\cite{nguyen2024pbp,rastogi2013droidchameleon}. Although ML detectors can reduce reliance on manual rules, they face a different challenge--both benign and malicious software evolve over time, causing models trained on historical data to degrade as deployment distributions shift, a problem known as model aging or concept drift~\cite{lei2019evedroid,aonzo2020obfuscapk,apigraph,cai2020assessing,tesseract}.

Evaluating this degradation requires benchmarks that capture more than one view of an application. Static features provide broad coverage but can be affected by obfuscation, repackaging, API substitution, and structural manipulation~\cite{arp2014drebin,apigraph,he2023efficient,mpdroid,zhao2021structural,haque2025lamda}. Dynamic traces capture runtime behavior~\cite{dynamic-malware,xu2024dva}, while program graphs capture application structure and dependencies~\cite{malnetNeurIPS,higraph,hamilton2017inductive-graph,mpdroid}; however, both are costly to extract at scale. {\em A useful benchmark should therefore provide aligned static, dynamic, and graph-based views over time, enabling evaluation of how each modality changes and fails under malware evolution}.

Existing Android malware benchmarks provide useful foundations, but do not fully support multimodal analysis of malware evolution. DREBIN~\cite{arp2014drebin} enabled large-scale Android malware detection, and TESSERACT~\cite{tesseract} and APIGraph~\cite{apigraph} introduced temporal evaluation, but these benchmarks remain limited in coverage, representation, or long-term family-level analysis. LAMDA~\cite{haque2025lamda}, the closest related benchmark, provides 12 years of Android malware data but primarily uses static DREBIN-style features. Such static-only temporal benchmarks cannot reveal whether degradation stems from static artifacts, runtime behavior, program structure, or malware-family shifts. Hidden spatio-temporal biases can also lead to overly optimistic conclusions~\cite{chow2025breaking}. These gaps motivate a temporally structured multimodal benchmark with aligned static, dynamic, and graph-based views~\cite{malnetNeurIPS,higraph,hamilton2017inductive-graph,mpdroid}. Table~\ref{tab:compact_apk_characterization_dataset} summarizes this gap by comparing existing APK benchmarks across characterization features, modality coverage, dataset scale, temporal span, artifact release, and dataset composition.

Existing multimodal benchmarks such as MULTIBENCH and MultiZoo support representation learning, fusion, robustness, and reproducible evaluation across several domains~\cite{liang2021multibench,liang2023multizoo,castro2019towards,zadeh2016mosi,zadeh2018multimodal,hasan2019ur,johnson2016mimic,lee2020multimodal,lee2020making,leiva2020enrico,kay2017kinetics,arevalo2017gated,vielzeuf2018centralnet}. However, they are not designed for naturally occurring temporal shift, where modalities, labels, and cross-modal relationships evolve together. This limits their use for evaluating multimodal generalization under realistic deployment shift.

Android malware evolution offers a natural setting for this gap. Apps can be represented through static, dynamic, and graph-based modalities, while distributions shift as ecosystems, malware families, APIs, and attacker strategies evolve~\cite{bavota2014impact,yang2018android,fazzini2019automated}. This makes malware evolution a useful testbed for studying multimodal models under realistic deployment shift, including fusion, modality robustness, graph learning, long-tailed labels, and continual adaptation.

We therefore introduce {\system}, a large-scale benchmark for temporal multimodal evaluation in Android malware detection. {\system} supports time-aware evaluation by training models on historical samples and testing them on future samples from shifted distributions~\cite{anoshift}. Unlike single-modality benchmarks, it provides aligned static, dynamic, and graph-based views for each application, enabling analysis of how different modalities change and degrade over time. By preserving chronology, {\system} supports realistic evaluation of model aging, adaptation, fusion, and generalization under deployment-like shift~\cite{chow2025breaking}. It also serves a broader testbed for multimodal learning under evolving and adversarial distributions.

In summary, our key contributions are:
\begin{smitemize}
    \item We introduce {\system}, a large-scale benchmark for temporal multimodal Android malware detection, with 858{,}859 samples and 1{,}354 malware families.

    \item We provide three modalities--static, dynamic, and graph-based-- for each application, capturing application artifacts, runtime behavior, and program structure.
    
    \item We provide a time-aware evaluation splits that train models on historical samples and test them on future samples, enabling realistic analysis of concept drift, modality-specific degradation, adaptation, and family-level evolution.

    \item We release pre-extracted feature representations and benchmark protocols to support reproducible research on multimodal fusion, graph learning, long-tailed classification, modality robustness, and generalization under non-stationary malware distributions.
\end{smitemize}

\BfPara{Contributions in the Appendices} We include additional analyses and supporting experiments in the appendices: dataset statistics (Appendix~\ref{app:dataset_stats}), comparison with prior multimodal datasets (Appendix~\ref{app:prior-multimodal-datasets}), a motivating GodFather malware case study (Appendix~\ref{app:godfather_case_study}), model descriptions (Appendix~\ref{app:model-descriptions}), multimodal fusion strategies (Appendix~\ref{app:fusion_strategies}), extended temporal analysis (Appendix~\ref{app:full_temporal_results}), feature-space visualizations (Appendix~\ref{app:feature-space-visualization}), feature-importance analysis (Appendix~\ref{app:feature_importance}), sensitivity to missing modalities (Appendix~\ref{app:missing_modal}), prediction uncertainty under label drift (Appendix~\ref{app:label_drift}), additional concept-drift adaptation results (Appendix~\ref{app:extra-cda}), unsupervised family structure analysis (Appendix~\ref{app:unsupervised-family-classification}), continual learning (Appendix~\ref{app:continual-learning}), malware-family temporal stability (Appendix~\ref{app:family-temporal-stability}), computational resources (Appendix~\ref{app:computation}), broader impacts (Appendix~\ref{app:broaderImpacts}), and dataset documentation (Appendix~\ref{app:datasetdocument}).
\section{Related Work}

\BfPara{Malware Datasets and Benchmarks}
Large-scale Android malware datasets are essential for training and evaluating ML-based detectors. DREBIN~\cite{arp2014drebin} remains one of the most widely used Android malware datasets based on static features. Other work has explored dynamic and graph-based representations, including MaMaDroid~\cite{mariconti2017mamadroid}, MalScan~\cite{malscan}, and MalNet~\cite{malnetNeurIPS}. 
However, malware benchmarks can rapidly become outdated due to evolving obfuscation and dropper techniques~\cite{tesseract}, and may suffer from sampling bias, data snooping, label noise, mixed-source inconsistencies, and unrealistic evaluation protocols~\cite{dosanddont}. Prior work such as TESSERACT~\cite{tesseract} and Arp et al.~\cite{dosanddont} provides guidance for reliable temporal evaluation, while MPDroid~\cite{mpdroid} supports multimodal representation but not {\em not longitudinal evaluation}.
These limitations motivate benchmarks that jointly support multimodal malware characterization and temporal evaluation, a gap summarized in Table~\ref{tab:compact_apk_characterization_dataset}. 



\begin{table*}[t]
\centering
\small
\caption{Compact comparison of APK characterization features, modality, dataset scale, time coverage, artifact release, and dataset composition. 
}
\label{tab:compact_apk_characterization_dataset}
\renewcommand{\arraystretch}{1.12}

\begin{tabular}{l|p{4.4cm}| c | r |c| c}
\toprule
\textbf{Approach} &
\textbf{APK Characterization} &
\textbf{Modality} &
\textbf{Samples} &
\textbf{Years} &
\textbf{Artifact} \\
\midrule

Drebin~\cite{arp2014drebin} &
\HC, \AC, \IN, \PER, \API, \CS &
S &
129,013 &
3 &
\ding{51} \\

MamaDroid~\cite{mariconti2017mamadroid} &
\API, \PG &
G &
43,940 &
7 &
\ding{51} \\

HinDroid~\cite{hou2017hindroid} &
\API, \PG &
G &
2,334 &
1 &
\ding{55} \\

Kim et al.~\cite{kim2018multimodal} &
\HC, \AC, \IN, \PER, \API, \CS &
S &
41,260 &
-- &
\ding{55} \\

MalScan~\cite{malscan} &
\API, \PG &
G &
30,715 &
8 &
\ding{51} \\

SDAC~\cite{sdac} &
\API, \PG &
G &
70,142 &
6 &
\ding{55} \\

HomDroid~\cite{homdroid} &
\API, \PG &
G &
8,198 &
-- &
\ding{55} \\

Xmal~\cite{xmal} &
\PER, \API &
S &
35,690 &
-- &
\ding{51} \\

RAMDA~\cite{ramda} &
\IN, \PER, \API &
S &
58,483 &
-- &
\ding{51} \\

MSDroid~\cite{msdroid} &
\PER, \API, \CS, \PG &
G &
81,790 &
6 &
\ding{51} \\

MalNet~\cite{malnetCIKM,malnetNeurIPS} &
\API, \CS &
S/I &
1,262,024 &
-- &
\ding{51} \\

HiGraph~\cite{higraph} &
\PER, \API, \CS, \PG &
G &
595,211 &
11 &
\ding{51} \\

LAMDA~\cite{haque2025lamda} &
\HC, \API, \CS &
S &
1,008,381 &
12 &
\ding{51} \\

MPDroid~\cite{mpdroid} &
\HC, \API, \CS &
S/D &
10,000 &
-- &
\ding{55} \\ \midrule

\textbf{\cellcolor{green!10}\system} &
\textbf{\cellcolor{green!10}\HC, \AC, \IN, \PER, \API, \CS, \PG, \NET, \FS, \SYS, \PROC, \CMD, \MOD, \DB} &
\textbf{\cellcolor{green!10}S/D/G} &
\textbf{\cellcolor{green!10}858,859} &
\textbf{\cellcolor{green!10}12} &
\textbf{\cellcolor{green!10}\ding{51}} \\

\bottomrule
\end{tabular}
\vspace{1mm}
\begin{minipage}{0.98\textwidth}
\footnotesize
\textbf{Legend:}
\HC{} = Hardware / Resource Component;
\AC{} = Application Component;
\IN{} = Intent;
\PER{} = Permission;
\API{} = API Call;
\CS{} = Code String;
\PG{} = Program Graph Feature;
\NET{} = Network Artifact;
\FS{} = File-System Artifact;
\SYS{} = System Property;
\PROC{} = Process;
\CMD{} = Command Execution;
\MOD{} = Loaded Module;
\DB{} = Database Access.
\textbf{Modality:}
S = Static;
D = Dynamic;
G = Graph-based;
I = Image-based.
\end{minipage}
\end{table*}

\BfPara{Static and Graph-Based Analysis}
Static analysis extracts APK features such as permissions, API calls, and services~\cite{arp2014drebin}. Graph-based approaches use function-call graphs (FCGs), control-flow graphs (CFGs), and related representations to capture application structure and behavior~\cite{malscan,malnetCIKM,malnetNeurIPS,mariconti2017mamadroid,li2025revisiting}. API-sequence modeling further improves robustness to simple obfuscation~\cite{mariconti2017mamadroid}. LAMDA~\cite{haque2025lamda} is the closest longitudinal benchmark, but it primarily supports static analysis. As a result, it cannot evaluate how runtime behavior and program structure evolve alongside static features. \system addresses this limitation by aligning static, dynamic, and graph-based views over time.

\BfPara{Dynamic and Behavioral Analysis}
Dynamic analysis executes applications in controlled environments to observe runtime behavior, including system calls, API calls, dynamic loading, and network traffic~\cite{lashkari2018toward,mahdavifar2020dynamic}. TaintDroid~\cite{enck2014taintdroid}, DroidScope~\cite{yan2012droidscope}, CopperDroid~\cite{tam2015copperdroid}, and DVa~\cite{xu2024dva} show that runtime traces can reveal privacy leaks, high-level malicious behaviors, accessibility abuse, dynamic payloads, and persistence. VirusTotal reports also support static, dynamic, and behavioral studies~\cite{virustotal,lashkari2018toward,mahdavifar2020dynamic}. However, dynamic analysis remains costly at scale and can suffer from low code coverage and evasion~\cite{kondracki2022droid}. \system provides pre-extracted dynamic features aligned with static and graph-based views, enabling temporal behavioral analysis without rerunning large-scale sandboxing.


\BfPara{Multimodal and Fusion Analysis}
Static, dynamic, and graph-based analyses capture complementary views of Android malware, motivating multimodal fusion. Prior work provides strong foundations for individual modalities, including static features in DREBIN~\cite{arp2014drebin}, API-sequence behavior in MaMaDroid~\cite{mariconti2017mamadroid}, graph representations in MalNet~\cite{malnetNeurIPS}, and longitudinal evaluation in TESSERACT~\cite{tesseract}. Multimodal detectors combine permissions, APIs, opcodes, strings, components, native code, and URLs, sometimes with feature-level explanations~\cite{kim2018multimodal,zhumultimodal}; MPDroid~\cite{mpdroid} further integrates static function-call and dynamic API-call graphs through multimodal pre-training, alignment, and fusion. However, these efforts primarily target representation learning and downstream detection, rather than large-scale longitudinal benchmarking. \system addresses this gap with aligned static, dynamic, and graph-based views, temporal splits, and protocols for studying modality-specific degradation, fusion robustness, and family-level evolution over time.

\section{\system Creation}
\label{sec:creation}

\begin{figure}[!t]
\centering
\begin{subfigure}{0.44\textwidth}
    \centering
    \includegraphics[width=\linewidth]{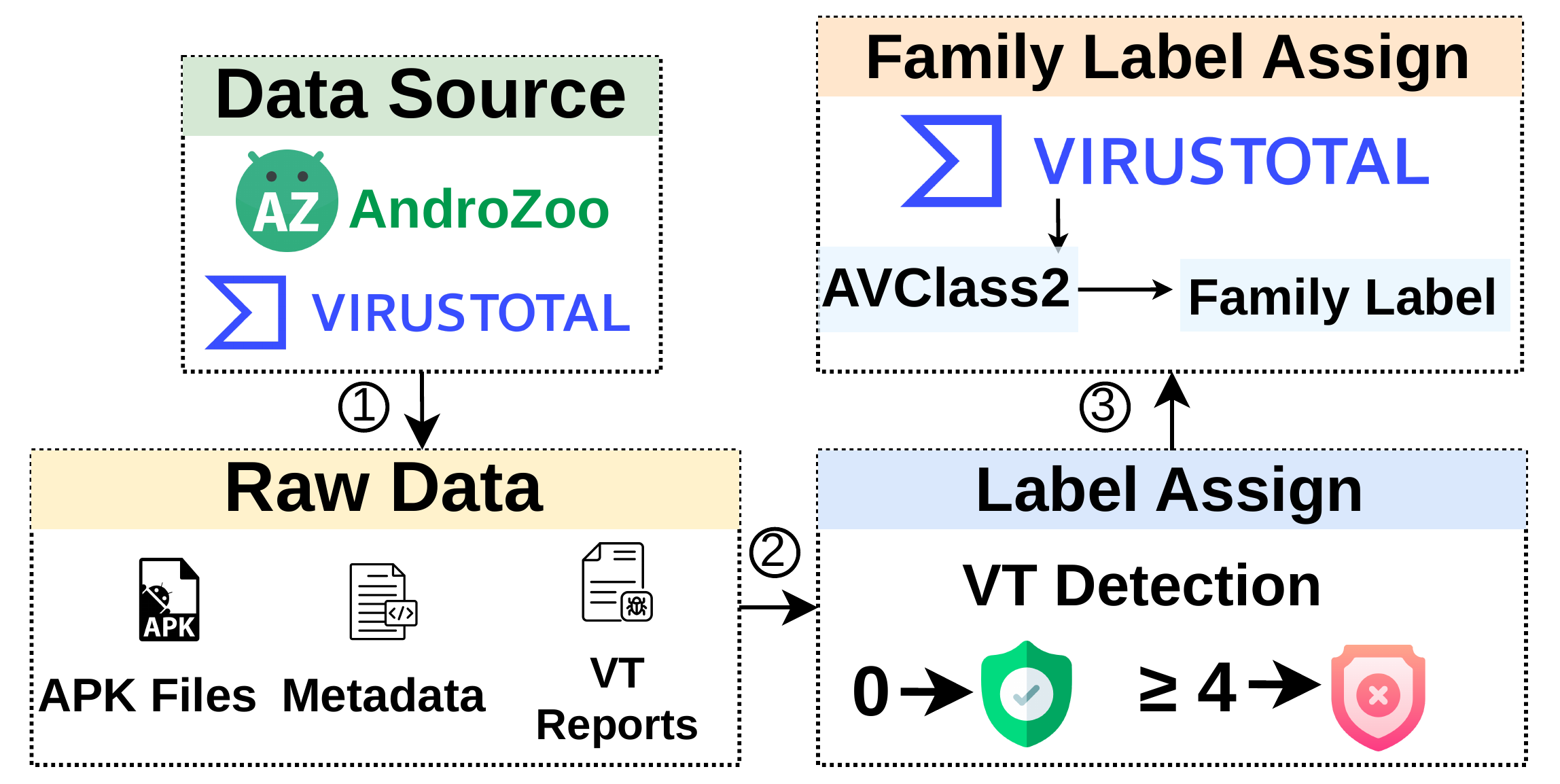}
    \caption{Data collection strategy.}
    \label{fig:mcndroid-Creation}
\end{subfigure}
\hfill
\begin{subfigure}{0.54\textwidth}
    \centering
    \includegraphics[width=\linewidth]{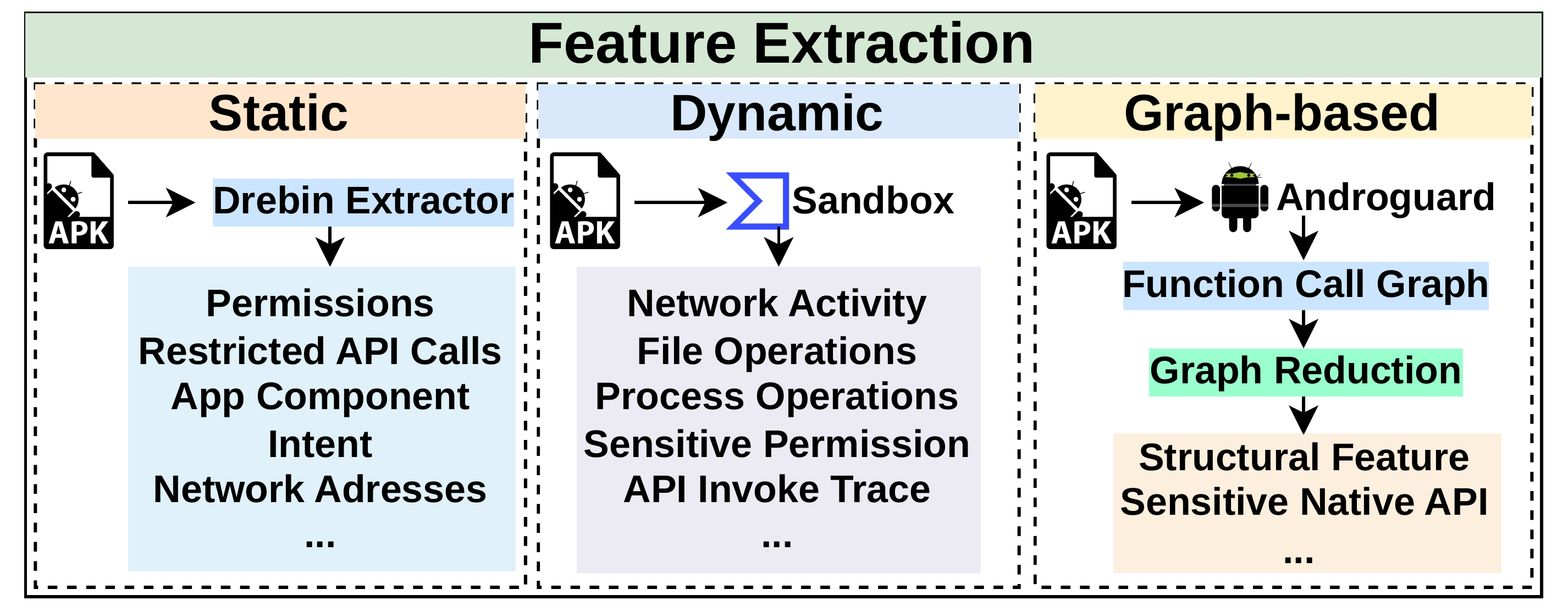}
    \caption{Feature extraction pipeline.}
    \label{fig:feature-extraction-pipeline}
\end{subfigure}
\caption{Overview of the McNdroid dataset creation process.}
\end{figure}


\subsection{Dataset Collection Strategy}
We construct {\system} from AndroZoo~\cite{androzoo}, which provides APK SHA-256 hashes, VirusTotal (VT) metadata, and submission timestamps. These metadata allow us to build a longitudinal dataset for studying Android malware evolution under temporal distribution shift. A schematic representation of our data collection stragegy is depicted in Figure~\ref{fig:mcndroid-Creation}.


\BfPara{Data Collection and Temporal Coverage}
We collect APKs from 2013 to 2025. We exclude 2015 because AndroZoo contains too few valid entries for that year~\cite{androzoo}. Removing this sparse year avoids distorted temporal comparisons and maintains a more consistent timeline.


\BfPara{Label Assignment and Noise Reduction}
We use \texttt{vt\_detection} as a weak-label proxy, following prior malware benchmarking practice~\cite{haque2025lamda,tesseract}. Applications with \texttt{vt\_detection = 0} are labeled benign, and those with \texttt{vt\_detection >= 4} are labeled malware. A threshold of 4 enforces stronger agreement among antivirus vendors, reducing labeling noise and improving reliability. Samples with intermediate counts, i.e., 1--3, are discarded as ambiguous. 
We release raw \texttt{vt\_detection} values so researchers can vary thresholds and test sensitivity to alternative label definitions.


\BfPara{Deduplication and Integrity Checks}
Following prior work~\cite{liu2026unraveling}, we perform strict deduplication using SHA-256 hashes to eliminate redundant samples and prevent data leakage. 

\BfPara{Cross-Modality Consistency Filtering}
To support fair multimodal evaluation, we retain only APKs with valid static features, dynamic outputs, and function-call graphs. Samples are removed if any extraction pipeline fails, including APKTool errors, incomplete sandbox reports, or graph-construction failures. This intersection filtering reduces the dataset from approximately 1.5M APKs to 858,859 samples, trading coverage for complete and aligned multimodal views.



\BfPara{Family Label Allocation}
To facilitate fine-grained analysis beyond binary classification, we assign malware family labels using AVClass2~\cite{avclass2}, a widely adopted tool for normalizing heterogeneous antivirus vendor outputs. AVClass2 aggregates noisy and often inconsistent VirusTotal~\cite{virustotal} labels into a unified and standardized family taxonomy. For each sample, we retrieve the corresponding VirusTotal report, and extract the inferred family label.

\BfPara{Evaluation Splits}
We adopt an 80–20 train–test split with stratification to preserve the original malware-to-benign class distribution~\cite{haque2025lamda}. This ensures unbiased evaluation under realistic class imbalance without introducing distributional skew. 
Detailed dataset statistics and malware family distributions are provided in Appendix~\ref{app:dataset_stats}.

\subsection{Feature Extraction}
Figure~\ref{fig:feature-extraction-pipeline} summarizes the feature extraction pipeline, from APK collection to static, dynamic, and FCG-based representations.

\BfPara{Static Feature}
Our first modality uses static features, which support efficient application profiling without execution. 
Following DREBIN~\cite{arp2014drebin}, we unpack each APK and parse \texttt{AndroidManifest.xml} and the disassembled \texttt{classes.dex} bytecode. We extract hardware components, requested and used permissions, app components, intents, suspicious API calls, and network addresses. These indicators are encoded as a high-dimensional binary bag-of-features vector, where each dimension records the presence or absence of a static attribute. To prevent temporal leakage, we construct the global vocabulary only from the 2013 training split. We then apply low-variance filtering with \texttt{VarianceThreshold} $(0.001)$ to remove rare or uninformative attributes, reducing the static feature space from 501,525 to 2,390 features per application~\cite{haque2025lamda,arp2014drebin,madar2025,continual-learning-malware}.

\BfPara{Dynamic Feature}
Our second modality captures runtime behavior from structured VirusTotal sandbox reports~\cite{virustotal}. The extracted signals cover network activity, file-system operations, process execution, inter-process communication, and API-call traces. We transform raw telemetry to a hybrid feature space--(i) explicit features for stable behavioral patterns, and (ii) hashed features for high-entropy or open-vocabulary artifacts. Explicit features include system-interaction counts, protocol usage statistics, telephony access, location access, reflection, privilege-escalation attempts, and persistence indicators. High-cardinality artifacts, such as domains, file paths, command strings, and textual fields, are mapped to fixed-dimensional vectors using feature hashing~\cite{hashingtrick,ember,ember2024}. This design preserves common behavioral signals while bounding the dimensionality of open-vocabulary artifacts. The resulting dynamic representation contains 17,483 features per application.

\BfPara{Function Call Graph (FCG)}
Our third modality captures program structure through function-call graphs (FCGs). Following prior work~\cite{mariconti2017mamadroid,sdac,higraph}, we extract FCGs with Androguard~\cite{androguard} and model each application as a directed graph of function calls. Because raw FCGs are large, noisy, and API-heavy~\cite{shokouhinejad2025recent,mpdroid}, we reduce complexity in two steps: retaining local developer-defined functions, where malicious logic is often embedded~\cite{higraph}, and preserving edges incident to security-critical APIs identified by MalScan~\cite{malscan}. From the reduced graph, we extract structural statistics, sensitive-interaction ratios, and binary indicators for sensitive API presence. As with static features, the sensitive-API vocabulary is built only from training data, and each application is projected into a fixed-dimensional feature space. The resulting FCG-derived representation contains 2,793 features per application. We also release caller--callee extraction scripts to support custom graph representations.

\section{Benchmark Design}
\label{sec:Benchmark}

\begin{figure}[!t]
    \centering
    \includegraphics[width=\linewidth]{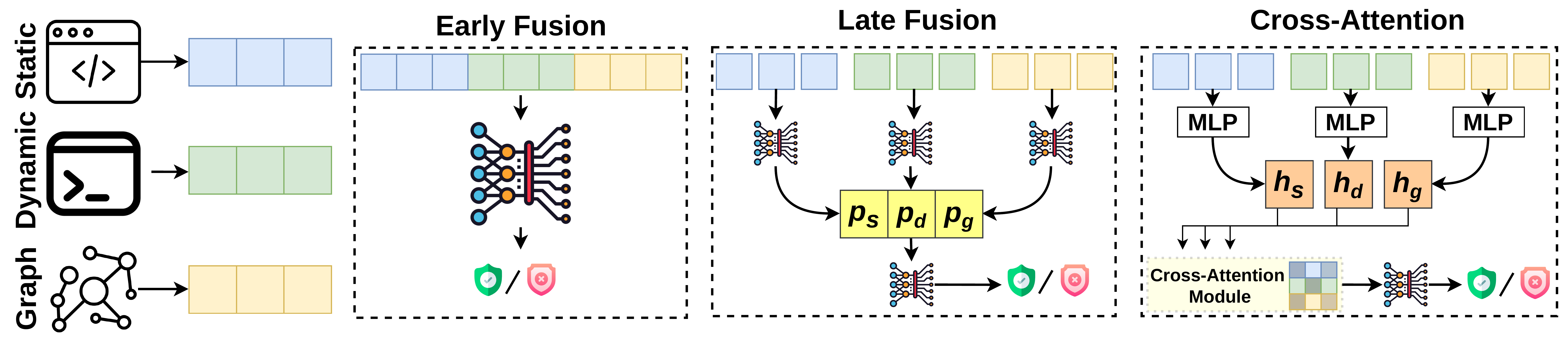}
\caption{Overview of the multimodal fusion strategies evaluated in \system.}
    \label{fig:benchmarkDesign}
\end{figure}

{\system} provides a large-scale multimodal dataset and reproducible benchmark for Android malware detection under temporal distribution shift. {\system} standardizes data curation, weak-label assignment, feature extraction, and evaluation splits, enabling fair comparison across methods~\cite{arp2014drebin,haque2025lamda,higraph,mariconti2017mamadroid,apigraph}, modality-specific analysis, and robustness evaluation under realistic temporal shift.


\BfPara{Problem Formulation: Malware Detection under Distribution Shift}
We formulate Android malware detection as supervised binary classification. Each application $i$ is represented by three aligned modalities,
$x_i = (x_i^{s}, x_i^{d}, x_i^{g})$, corresponding to static, dynamic, and graph-based features. The goal is to learn a detector $f(x_i) \rightarrow y_i$, where $y_i \in \{0,1\}$ denotes benign or malicious behavior. Unlike IID evaluation, {\system} evaluates models under temporal distribution shift: detectors are trained on historical samples and tested on future applications~\cite{haque2025lamda,anoshift}. This setting measures whether models generalize as applications, malware families, APIs, and attacker behaviors evolve.

\BfPara{Temporal Generalization}
To simulate deployment, {\system} uses chronologically ordered splits following prior temporal malware evaluation protocols~\cite{haque2025lamda,cade,chen2023continuous}. Applications are partitioned by year, and models are trained only on earlier years and evaluated on later years. Increasing the train--test gap creates stronger temporal shift, allowing us to measure how performance degrades over time. This design prevents temporal leakage; unless otherwise stated, all temporal experiments preserve chronological ordering and avoid random mixing across years.

\BfPara{Modality-Controlled Evaluation}
Each retained application has aligned static, dynamic, and graph-based views. This alignment allows evaluation of (i) single-modality, (ii) pairwise-modality, and (iii) full multimodal settings under identical sample sets and temporal splits. As a result, performance differences can be attributed to modality choice or fusion strategy rather than changes in the underlying data.

\BfPara{Multimodal Fusion Protocols}
{\system} supports evaluation of three canonical multimodal fusion strategies, summarized in Figure~\ref{fig:benchmarkDesign}: (i) \emph{early fusion}, where features from different modalities are concatenated before classification to learn joint feature interactions~\cite{perez2019early,boulahia2021early}; (ii) \emph{late fusion}, where modality-specific models are trained independently and their outputs are combined at the decision level~\cite{franco2020multimodal}; and (iii) \emph{cross-attention fusion}, where modality-specific embeddings are integrated through learnable attention mechanisms to capture fine-grained cross-modal interactions~\cite{zhumultimodal,kim2018multimodal}. sIn early fusion, static, dynamic, and graph-based feature matrices are horizontally concatenated; we also evaluate pairwise variants---static--dynamic, static--graph-based, and dynamic--graph-based---to identify complementary modality pairs. In late fusion, modality-specific learners produce probability estimates that are used as meta-features for a second-stage learner~\cite{franco2020multimodal}. In cross-attention fusion, each modality is projected into a shared embedding space using a dedicated MLP encoder~\cite{kim2018multimodal}; modality tokens then attend to one another through multi-head attention~\cite{zhumultimodal}, and the resulting context-aware representations are normalized, concatenated, and used for prediction. Before training and evaluation, we verify hash and label consistency across modalities so that all fusion strategies use matched samples.

\begin{table}[!t]
\vspace{-0.3cm}
\centering
\tiny
\caption{Comparison of unimodal and best multimodal performances across three temporal splits.\\ 
$^\dagger$ Multimodal denotes the best fusion configuration per model--split pair.}
\label{tab:comparison_unimodal}
\begin{tabular}{l|l|cc|cc|cc|cc}
\toprule
\multirow{2}{*}{\textbf{Split}} & \multirow{2}{*}{\textbf{Model}} 
& \multicolumn{2}{c|}{\textbf{Static}} 
& \multicolumn{2}{c|}{\textbf{Dynamic}} 
& \multicolumn{2}{c|}{\textbf{Graph}} 
& \multicolumn{2}{c}{\textbf{Multimodal}$^\dagger$} \\
\cmidrule(lr){3-4} \cmidrule(lr){5-6} \cmidrule(lr){7-8} \cmidrule(lr){9-10}
& & F1 & ROC-AUC 
  & F1 & ROC-AUC 
  & F1 & ROC-AUC
  & F1 & ROC-AUC \\
\midrule

\multirow{6}{*}{IID}
  & MLP         & \meanstd{90.6}{6.5} & \meanstd{97.4}{2.0} & \meanstd{82.9}{7.8} & \meanstd{92.0}{5.0} & \meanstd{82.1}{5.3} & \meanstd{93.3}{2.8} & \meanstd{90.7}{5.5} & \meanstd{98.3}{1.0} \\
  & LightGBM    & \meanstd{91.6}{5.5} & \meanstd{98.4}{1.2} & \meanstd{85.2}{7.7} & \meanstd{94.0}{4.2} & \meanstd{91.5}{5.1} & \meanstd{98.1}{1.3} & \meanstd{92.6}{4.5} & \meanstd{98.7}{0.9} \\
  & XGBoost     & \meanstd{91.5}{5.8} & \meanstd{98.2}{1.4} & \meanstd{85.2}{8.0} & \meanstd{93.6}{4.6} & \meanstd{92.1}{5.1} & \meanstd{98.8}{0.8} & \meanstd{92.1}{5.0} & \meanstd{98.8}{0.8} \\
  & SVM         & \meanstd{90.0}{5.3} & \meanstd{97.1}{2.0} & \meanstd{76.9}{6.0} & \meanstd{89.3}{3.9} & \meanstd{86.2}{5.5} & \meanstd{95.0}{2.7} & \meanstd{90.5}{4.9} & \meanstd{98.1}{0.9} \\
  & DetectBERT  & \meanstd{89.6}{5.6} & \meanstd{97.9}{1.2} & \meanstd{81.3}{7.7} & \meanstd{90.4}{5.6} & \meanstd{0.0}{0.0} & \meanstd{50.0}{0.0} & \meanstd{90.7}{5.8} & \meanstd{98.1}{1.3} \\
  & Transformer & \meanstd{88.5}{6.5} & \meanstd{97.7}{1.4} & \meanstd{80.4}{7.6} & \meanstd{91.1}{4.9} & \meanstd{41.8}{10.3} & \meanstd{66.9}{4.0} & \meanstd{91.0}{4.1} & \meanstd{97.9}{1.0} \\
\midrule

\multirow{6}{*}{NEAR}
  & MLP         & \meanstd{43.1}{30.2} & \meanstd{77.3}{13.4} & \meanstd{38.7}{24.5} & \meanstd{69.9}{10.9} & \meanstd{36.9}{24.7} & \meanstd{65.2}{8.3} & \meanstd{42.7}{26.5} & \meanstd{77.9}{13.4} \\
  & LightGBM    & \meanstd{40.3}{31.8} & \meanstd{82.3}{11.0} & \meanstd{34.7}{28.3} & \meanstd{74.5}{12.4} & \meanstd{32.6}{28.6} & \meanstd{81.5}{8.2} & \meanstd{39.7}{28.6} & \meanstd{78.9}{14.6} \\
  & XGBoost     & \meanstd{39.4}{32.1} & \meanstd{84.6}{7.6} & \meanstd{35.5}{26.7} & \meanstd{73.3}{12.7} & \meanstd{34.6}{28.2} & \meanstd{84.4}{6.2} & \meanstd{39.6}{29.3} & \meanstd{81.7}{12.3} \\
  & SVM         & \meanstd{43.9}{30.4} & \meanstd{73.9}{16.5} & \meanstd{40.2}{19.9} & \meanstd{71.2}{9.4} & \meanstd{31.1}{26.6} & \meanstd{77.4}{6.3} & \meanstd{46.2}{26.6} & \meanstd{78.0}{12.1} \\
  & DetectBERT  & \meanstd{42.1}{30.9} & \meanstd{70.5}{22.4} & \meanstd{39.7}{23.4} & \meanstd{67.0}{13.9} & \meanstd{0.0}{0.0} & \meanstd{49.9}{0.1} & \meanstd{43.9}{24.4} & \meanstd{78.2}{11.4} \\
  & Transformer & \meanstd{39.5}{30.2} & \meanstd{68.4}{24.3} & \meanstd{39.2}{25.1} & \meanstd{68.4}{14.8} & \meanstd{29.4}{17.5} & \meanstd{67.5}{9.7} & \meanstd{44.8}{27.2} & \meanstd{80.8}{9.91} \\
\midrule

\multirow{6}{*}{FAR}
  & MLP         & \meanstd{28.2}{16.9} & \meanstd{81.2}{12.4} & \meanstd{32.2}{22.9} & \meanstd{63.7}{6.3} & \meanstd{15.4}{13.3} & \meanstd{53.3}{13.4} & \meanstd{29.0}{21.0} & \meanstd{79.4}{12.3} \\
  & LightGBM    & \meanstd{35.7}{24.3} & \meanstd{84.0}{12.8} & \meanstd{25.3}{20.6} & \meanstd{65.1}{16.6} & \meanstd{29.8}{18.1} & \meanstd{76.4}{11.8} & \meanstd{47.9}{27.1} & \meanstd{79.1}{17.9} \\
  & XGBoost     & \meanstd{31.5}{23.4} & \meanstd{85.3}{12.2} & \meanstd{27.7}{22.2} & \meanstd{65.6}{14.0} & \meanstd{27.1}{15.0} & \meanstd{77.1}{11.9} & \meanstd{42.7}{25.7} & \meanstd{80.3}{16.3} \\
  & SVM         & \meanstd{33.3}{18.5} & \meanstd{79.8}{11.3} & \meanstd{29.1}{20.0} & \meanstd{56.0}{8.7} & \meanstd{13.7}{11.8} & \meanstd{83.0}{9.1} & \meanstd{41.6}{21.9} & \meanstd{79.8}{10.3} \\
  & DetectBERT  & \meanstd{34.8}{21.0} & \meanstd{80.9}{16.9} & \meanstd{32.5}{23.7} & \meanstd{62.3}{7.2} & \meanstd{10.6}{15.0} & \meanstd{54.5}{7.0} & \meanstd{38.1}{25.4} & \meanstd{84.0}{11.9} \\
  & Transformer & \meanstd{28.9}{21.3} & \meanstd{82.4}{17.2} & \meanstd{29.4}{20.9} & \meanstd{60.4}{7.0} & \meanstd{18.3}{13.8} & \meanstd{35.8}{17.2} & \meanstd{43.4}{26.1} & \meanstd{69.2}{16.3} \\

\bottomrule
\end{tabular}%

\vspace{-0.3cm}
\end{table}

\section{Benchmark Validation under Temporal Distribution Shift}
\label{sec:Benchmarkvalid}

\BfPara{Experimental Setup}
Following prior temporal evaluation studies~\cite{anoshift,haque2025lamda}, we train models on 2013 samples and evaluate them on progressively later years to measure robustness under temporal distribution shift. We define three evaluation regimes: IID, using temporally adjacent 2014 samples as a near-distribution baseline; NEAR, using samples from 2016--2017; and FAR, using samples from 2018--2025. This protocol fixes the training distribution and increases the temporal distance of the test distribution, allowing us to measure how detector performance changes as software ecosystems, malware families, and behaviors evolve.

We first evaluate each modality independently to establish unimodal baselines. For static, dynamic, and graph-based representations, we train widely used malware detection models, including Linear SVM, LightGBM, MLP, XGBoost, DetectBERT~\cite{xiong2021nystromformer}, and Transformer-based architectures~\citep{haque2025lamda,arp2014drebin,ember}. We then apply the same temporal protocol to multimodal models that fuse static, dynamic, and graph-based representations. This ensures that unimodal and multimodal methods are compared under identical samples, labels, and temporal splits. All experiments are repeated over five random seeds, and results are reported as \textit{mean$\pm$std}, averaged over runs and years within each split. Table~\ref{tab:comparison_unimodal} compares unimodal baselines with multimodal variants.

\BfPara{Results}
Table~\ref{tab:comparison_unimodal} shows substantial temporal degradation. All models perform best on the IID split, while F1-scores drop sharply on NEAR and FAR, indicating poor generalization from 2013 to later malware distributions. Among unimodal representations, static features are the most stable, dynamic features provide moderate robustness, and graph-based features are more variable, with severe failures for some model--modality pairs. FAR performance is not always worse than NEAR, suggesting that drift is not strictly monotonic with calendar distance; nevertheless, both temporal regimes remain far below IID.

Multimodal fusion provides clearer gains under long-term drift. For example, FAR LightGBM improves from \meanstd{35.7}{24.3} with the best unimodal static representation to \meanstd{47.9}{27.1} with multimodal fusion, while FAR Transformer improves from \meanstd{29.4}{20.9} with the best unimodal dynamic representation to \meanstd{43.4}{26.1}. Figure~\ref{fig:fusion_f1_score} further shows that multimodal variants are more robust than the strongest unimodal baseline in later years, although temporal degradation remains. The full unimodal and multimodal results are reported in Appendix~\ref{app:full_temporal_results}. Overall, these findings suggest that static, dynamic, and graph-based views capture complementary signals of Android malware evolution.

\begin{figure}[!t]
    \centering

    \begin{minipage}[t]{0.47\linewidth}
        \centering
        \resizebox{\linewidth}{!}{%
            \begin{tikzpicture}
\begin{axis}[
    width=8cm,
    height=4cm,
    ybar,
    xtick pos=bottom,
    ytick pos=left,
    bar width=3.5pt,
    enlargelimits=0.08,
    legend style={
        at={(0.9999,.9905)},
        anchor=north east,
        legend columns=3,
        font=\tiny,
        inner xsep=1pt,
        inner ysep=1pt,
        row sep=0pt,
        cells={anchor=west},
    },
legend image code/.code={
    \draw[#1] (0cm,-0.04cm) rectangle (0.07cm,0.07cm);
},
    legend image code/.code={
        \draw[#1] (0cm,-0.08cm) rectangle (0.1cm,0.10cm);
    },
    ylabel={F1 Score},
    xlabel={Year},
    xlabel style={font=\bfseries\footnotesize, yshift=5pt},
    ylabel style={font=\bfseries\footnotesize, xshift=2pt, yshift=-5pt},
    symbolic x coords={2013,2014,2016,2017,2018,2019,2020,2021,2022,2023,2024,2025},
    xtick=data,
    x tick label style={rotate=45, anchor=east, font=\bfseries\tiny},
    y tick label style={font=\bfseries\tiny},
    ytick={0,0.2,0.4,0.6,0.8,1.0},
    ymin=0,
    ymax=1.0,
]

\addplot[
    fill=blue!30,
    draw=blue!30!black,
    bar shift=-4.5pt,
    error bars/.cd,
    y dir=both,
    y explicit,
    error bar style={draw=black, line width=0.35pt},
    error mark options={draw=black, line width=0.35pt, mark size=1pt}
] table[
    x=year,
    y=f1_score,
    y error=f1_score_std,
    col sep=comma,
    restrict expr to domain={\pdfstrcmp{\thisrow{method}}{stage2_base_data}}{0:0}
] {tikz/yearly_bar_top1_unimodal_top2_multimodal_f1_score.csv};

\addplot[
    fill=red!60,
    draw=red!30!black,
    bar shift=0pt,
    error bars/.cd,
    y dir=both,
    y explicit,
    error bar style={draw=black, line width=0.35pt},
    error mark options={draw=black, line width=0.35pt, mark size=1pt}
] table[
    x=year,
    y=f1_score,
    y error=f1_score_std,
    col sep=comma,
    restrict expr to domain={\pdfstrcmp{\thisrow{method}}{stage3_feature_fusion}}{0:0}
] {tikz/yearly_bar_top1_unimodal_top2_multimodal_f1_score.csv};

\addplot[
    fill=green!30,
    draw=green!30!black,
    bar shift=4.5pt,
    error bars/.cd,
    y dir=both,
    y explicit,
    error bar style={draw=black, line width=0.35pt},
    error mark options={draw=black, line width=0.35pt, mark size=1pt}
] table[
    x=year,
    y=f1_score,
    y error=f1_score_std,
    col sep=comma,
    restrict expr to domain={\pdfstrcmp{\thisrow{method}}{stage1_data_gml_pairwise}}{0:0}
] {tikz/yearly_bar_top1_unimodal_top2_multimodal_f1_score.csv};

\legend{Static only,Feature fusion,Static + Graph-based}
\end{axis}
\end{tikzpicture}
        }
        \vspace{-0.75em}
        \caption{Year-wise F1-score comparison of the best unimodal baseline, feature-fusion model, and static--graph multimodal model. 
        }
        \label{fig:fusion_f1_score}
    \end{minipage}
    \hfill
    \begin{minipage}[t]{0.48\linewidth}
        \centering
        \resizebox{\linewidth}{!}{%
            \begin{tikzpicture}
\begin{axis}[
    width=8cm,
    height=4cm,
    ybar,
    xtick pos=bottom,
    ytick pos=left,
    bar width=3.5pt,
    enlargelimits=0.08,
    legend style={
        at={(0.75,0.98)},
        anchor=north east,
        legend columns=3,
        fill=white,
        draw=black,
        line width=0.3pt,
        font=\bfseries\tiny
    },
    legend image code/.code={
        \draw[#1] (0cm,-0.08cm) rectangle (0.12cm,0.12cm);
    },
    ylabel={Mean feature-wise KS},
    xlabel={Year},
    xlabel style={font=\bfseries\footnotesize, yshift=5pt},
    ylabel style={font=\bfseries\footnotesize, xshift=2pt, yshift=-5pt},
    symbolic x coords={2013,2014,2016,2017,2018,2019,2020,2021,2022,2023,2024,2025},
    xtick=data,
    x tick label style={rotate=45, anchor=east, font=\bfseries\tiny},
    scaled y ticks=false,
    yticklabel style={
        /pgf/number format/fixed,
        /pgf/number format/precision=2,
        font=\bfseries\tiny
    },
    ymin=0,
    ymax=0.16,
    ytick={0,0.02,0.04,0.06,0.08,0.10,0.12,0.14,0.16},
]

\addplot[
    fill=cyan!30,
    draw=cyan!30!black,
    bar shift=-4.5pt,
    error bars/.cd,
    y dir=both,
    y explicit,
    error bar style={draw=black, line width=0.35pt},
    error mark options={draw=black, line width=0.35pt, mark size=1pt}
] table[
    x=year,
    y=data_mean,
    y error=data_std,
    col sep=comma
] {tikz/ks-test/ks-test.csv};

\addplot[
    fill=purple!60,
    draw=purple!30!black,
    bar shift=0pt,
    error bars/.cd,
    y dir=both,
    y explicit,
    error bar style={draw=black, line width=0.35pt},
    error mark options={draw=black, line width=0.35pt, mark size=1pt}
] table[
    x=year,
    y=gml_mean,
    y error=gml_std,
    col sep=comma
] {tikz/ks-test/ks-test.csv};

\addplot[
    fill=orange!30,
    draw=orange!30!black,
    bar shift=4.5pt,
    error bars/.cd,
    y dir=both,
    y explicit,
    error bar style={draw=black, line width=0.35pt},
    error mark options={draw=black, line width=0.35pt, mark size=1pt}
] table[
    x=year,
    y=json_mean,
    y error=json_std,
    col sep=comma
] {tikz/ks-test/ks-test.csv};

\legend{Static,Graph-based,Dynamic}
\end{axis}
\end{tikzpicture}
        }
        \vspace{-0.75em}
        \caption{Temporal distribution shift across static, graph-based, and dynamic features measured by the mean feature-wise KS statistic. 
        }
        \label{fig:ks-drift}
    \end{minipage}

\end{figure}

\section{Comprehensive Multimodal Drift Analysis}
\label{sec:DriftAnalysis}

Having established substantial temporal degradation in {\system}, we next analyze how drift appears across modalities, features, and malware families.





\subsection{Distributional Shift Across Time}

\BfPara{Experimental Setting}
We quantify temporal distribution shift by comparing each test year with the 2013 training reference using feature-wise two-sample Kolmogorov--Smirnov (KS) statistics~\cite{gao2024comprehensive,ks-testneurips,kstest2014}. KS is a non-parametric measure of the maximum difference between two empirical distributions, making it suitable for measuring covariate shift across feature modalities.

\BfPara{Analysis}
Figure~\ref{fig:ks-drift} shows that all modalities diverge from the 2013 reference distribution over time, indicating sustained temporal covariate shift. The graph-based representation exhibits the largest drift, increasing from $0.0006$ in the 2013 baseline comparison to $0.1405$ in 2025. This is consistent with the evolving Android ecosystem: operating-system updates and API changes can deprecate, modify, or introduce APIs over time~\cite{bavota2014impact,yang2018android,fazzini2019automated}, which may alter function-call structure and graph-derived features. The high KS divergence for graph features aligns with our earlier results, where graph-based models show greater performance variability under temporal shift.

\subsection{Feature Stability Across Modalities and Class Distributions}

\BfPara{Setting}
While the KS analysis measures overall covariate shift, we further examine whether feature stability differs between malware and benign samples. For each modality and class, we estimate yearly feature distributions using normalized histograms and compare each year against the 2013 reference with Jeffreys divergence~\cite{Jeffreysdivergence}:
\(
D_J(p, q) = D_{KL}(p \parallel q) + D_{KL}(q \parallel p).
\)






Jeffreys divergence is symmetric and non-negative, and captures discrepancies in both directions between two distributions. 
The divergence is computed per feature and averaged across features to obtain one statistic per modality, class, and year~\cite{anoshift,haque2025lamda}. In our analysis, each year is compared to a fixed reference year 2013, allowing us to capture both long-term drift and short-term changes. 
By evaluating malware and benign distributions separately, this setup enables a more precise characterization of class-specific feature instability over time.

\begin{figure}[!t]
    \centering

    \begin{subfigure}[t]{0.48\linewidth}
        \centering
        \resizebox{\linewidth}{!}{%
            \input{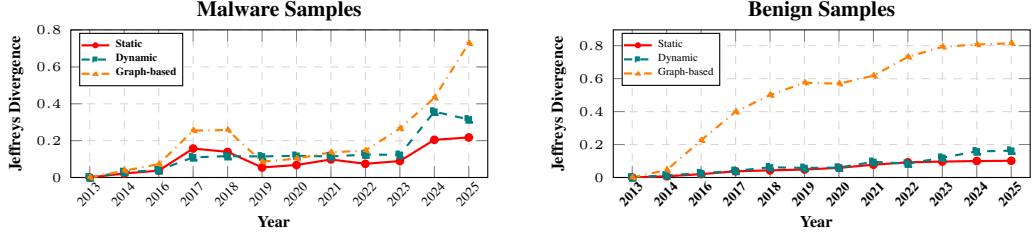}
        }
        \label{fig:jd-malware}
    \end{subfigure}
    \hfill
    \begin{subfigure}[t]{0.48\linewidth}
        \centering
        \resizebox{\linewidth}{!}{%
            \begin{tikzpicture}
\begin{axis}[
   width=7cm,
   height=3.5cm,
   title={Benign Samples},
   title style={font=\bfseries\fontsize{8}{6}\selectfont, yshift=-6pt},
    xlabel={Year},
    ylabel={Jeffreys Divergence},
    xlabel style={font=\bfseries\fontsize{6}{6}\selectfont, xshift=-2,yshift=5},
    ylabel style={font=\bfseries\fontsize{6}{6}\selectfont, yshift=-2},
    xtick=data,
    symbolic x coords={2013,2014,2016,2017,2018,2019,2020,2021,2022,2023,2024,2025},
    x tick label style={
        rotate=45,
        anchor=east,
        font=\fontsize{5}{5}\selectfont\bfseries,
        xshift=4pt,
        yshift=-4pt
    },
    y tick label style={font=\fontsize{5}{5}\selectfont,},
    enlarge x limits=0.05,
    ymin=0,
    grid=both,
    grid style={dashed, gray!30},
    legend columns=1,
legend style={
    at={(0.02,0.98)},
    anchor=north west,
    fill=white,
    draw=black,
    font=\fontsize{4}{4}\selectfont,
    inner sep=0.3pt,
    row sep=-3pt,
},
legend image post style={
    scale=0.6  
},  
    legend cell align={left},
]

\addplot[color=red, mark=*, mark size=1.2pt, thick]
    table[x=year, y=mean_divergence, col sep=comma]
    {tikz/stability/benign_drift_data.csv};

\addplot[color=teal, mark=square*, mark size=1.2pt, dashed, thick]
    table[x=year, y=mean_divergence, col sep=comma]
    {tikz/stability/benign_drift_json.csv};

\addplot[color=orange, mark=triangle*, mark size=1.2pt, dash dot, thick]
    table[x=year, y=mean_divergence, col sep=comma]
    {tikz/stability/benign_drift_gml.csv};

\legend{Static, Dynamic, Graph-based}

\end{axis}
\end{tikzpicture}
        }
        \label{fig:jd-benign}
    \end{subfigure}
    \vspace{-0.6em}
    \caption{Temporal stability of malware and benign samples measured by Jeffreys divergence.}
    \label{fig:jd-aio}
\end{figure}

\BfPara{Analysis}
Figure~\ref{fig:jd-aio} shows the temporal evolution of Jeffreys divergence for malware and benign samples across static, dynamic, and graph-based modalities. For malware, all modalities show low divergence in the initial years, indicating stability relative to the 2013 baseline. Static and dynamic features remain relatively stable, while graph-based features increase after the mid-period, suggesting that structural representations become less stable as malware evolves. Benign samples show stronger drift. Static and dynamic features increase gradually, indicating moderate distributional change, whereas graph-based features rise sharply with temporal distance and reach the highest divergence among all modalities. This suggests that benign application structure changes substantially over time, likely reflecting shifts in APIs, frameworks, and development practices. Overall, static features are the most stable across both classes, dynamic features show moderate drift, and graph-based representations exhibit the strongest temporal evolution, especially within the benign distribution.

\subsection{Cross-Modal Disagreement Analysis}

\begin{figure}[t]
    \centering
    \hspace{-1.75em}%
    \begin{minipage}[t]{0.49\linewidth}
        \centering
    \begin{tikzpicture}
    \begin{axis}[
        width=7cm,
        height=3.5cm,
        xtick pos=bottom,
        ytick pos=left,
        enlargelimits=0.04,
        grid=both,
        grid style={line width=0.15pt, draw=gray!25},
        major grid style={line width=0.2pt, draw=gray!35},
        legend style={
            at={(1.0,1.49)},
            anchor=north east,
            legend columns=2,
            fill=white,
            draw=black,
            line width=0.1pt,
            font=\bfseries\tiny
        },
        legend image post style={scale=0.8},
        ylabel={Kappa ($\kappa$)},
        xlabel={Test Year},
        xlabel style={font=\bfseries\tiny, yshift=6pt}, 
        ylabel style={font=\bfseries\tiny, xshift=2pt, yshift=-12pt},
        symbolic x coords={2014,2016,2017,2018,2019,2020,2021,2022,2023,2024,2025},
        xtick=data,
        x tick label style={
            rotate=45,
            anchor=east,
            font=\bfseries\tiny
        },
        scaled y ticks=false,
        y tick label style={
            font=\bfseries\tiny,
            /pgf/number format/fixed,
            /pgf/number format/precision=2
        },
        ymin=-0.15,
        ymax=0.85,
        ytick={-0.10,0.00,0.20,0.40,0.60,0.80}
    ]

    \addplot[
        thick,
        blue!70!black,
        solid,
        mark=*,
        mark size=1pt,
        mark options={fill=blue!70!black, draw=blue!70!black},
        error bars/.cd,
            y dir=both,
            y explicit,
            error bar style={line width=0.65pt},
            error mark options={line width=0.45pt, mark size=1.4pt}
    ] coordinates {
        (2014,0.7685) +- (0,0.0008)
        (2016,0.5864) +- (0,0.0020)
        (2017,0.1454) +- (0,0.0086)
        (2018,0.2288) +- (0,0.0041)
        (2019,0.4207) +- (0,0.0034)
        (2020,0.2918) +- (0,0.0066)
        (2021,0.1384) +- (0,0.0042)
        (2022,0.2445) +- (0,0.0028)
        (2023,0.0635) +- (0,0.0122)
        (2024,-0.0988) +- (0,0.0028)
        (2025,-0.0826) +- (0,0.0122)
    };
    \addlegendentry{Fleiss' Kappa (mean)}

    \addplot[
        red!70,
        dotted,
        line width=1.5pt
    ] coordinates {
        (2014,0) (2025,0)
    };
    \addlegendentry{Random chance ($\kappa=0$)}

    \addplot[
        green!60!black,
        dotted,
        line width=1.5pt
    ] coordinates {
        (2014,0.8) (2025,0.8)
    };
    \addlegendentry{Strong agreement ($\kappa=0.8$)}

    \end{axis}
    \end{tikzpicture}
        \vspace{-0.3em}
        \captionof{figure}{Year-wise Fleiss' Kappa with standard deviation shown as error bars.}
       \label{fig:kappa}
    \end{minipage}%
    \hspace{1em}%
    \begin{minipage}[t]{0.49\linewidth}
        \centering
    \begin{tikzpicture}
    \begin{axis}[
        width=7cm,
        height=3.5cm,
        xtick pos=bottom,
        ytick pos=left,
        enlargelimits=0.03,
        grid=both,
        grid style={line width=0.15pt, draw=gray!25},
        major grid style={line width=0.2pt, draw=gray!35},
        legend style={
            at={(0.003,1.30)},
            anchor=north west,
            legend columns=3,
            fill=white,
            draw=black,
            line width=0.2pt,
            font=\bfseries\tiny
        },
        legend image post style={scale=0.8},
        ylabel={Dissent Rate},
        xlabel={Test Year},
        xlabel style={font=\bfseries\tiny, yshift=6pt}, 
        ylabel style={font=\bfseries\tiny, xshift=2pt, yshift=-5pt},
        symbolic x coords={2014,2016,2017,2018,2019,2020,2021,2022,2023,2024,2025},
        xtick=data,
        x tick label style={
            rotate=45,
            anchor=east,
            font=\bfseries\tiny
        },
        scaled y ticks=false,
        y tick label style={
            font=\bfseries\tiny,
            /pgf/number format/fixed,
            /pgf/number format/precision=2
        },
        ymin=0.00,
        ymax=0.35,
        ytick={0.00,0.05,0.10,0.15,0.20,0.25,0.30,0.35}
    ]

    \addplot[name path=dataupper, draw=none, forget plot] coordinates {
        (2014,0.0353)
        (2016,0.0633)
        (2017,0.0142)
        (2018,0.0186)
        (2019,0.0960)
        (2020,0.1186)
        (2021,0.0983)
        (2022,0.0884)
        (2023,0.0158)
        (2024,0.0084)
        (2025,0.0057)
    };

    \addplot[name path=datalower, draw=none, forget plot] coordinates {
        (2014,0.0345)
        (2016,0.0605)
        (2017,0.0116)
        (2018,0.0156)
        (2019,0.0846)
        (2020,0.1058)
        (2021,0.0909)
        (2022,0.0830)
        (2023,0.0140)
        (2024,0.0034)
        (2025,0.0031)
    };

    \addplot[
        blue!25,
        fill opacity=0.18,
        draw=none,
        forget plot
    ] fill between[of=dataupper and datalower];

    \addplot[name path=gmlupper, draw=none, forget plot] coordinates {
        (2014,0.0338)
        (2016,0.0627)
        (2017,0.0102)
        (2018,0.0132)
        (2019,0.1092)
        (2020,0.1083)
        (2021,0.0511)
        (2022,0.0942)
        (2023,0.0360)
        (2024,0.0115)
        (2025,0.0076)
    };

    \addplot[name path=gmllower, draw=none, forget plot] coordinates {
        (2014,0.0330)
        (2016,0.0609)
        (2017,0.0084)
        (2018,0.0118)
        (2019,0.1060)
        (2020,0.1033)
        (2021,0.0477)
        (2022,0.0902)
        (2023,0.0348)
        (2024,0.0103)
        (2025,0.0050)
    };

    \addplot[
        green!20,
        fill opacity=0.18,
        draw=none,
        forget plot
    ] fill between[of=gmlupper and gmllower];

    \addplot[name path=jsonupper, draw=none, forget plot] coordinates {
        (2014,0.0648)
        (2016,0.0961)
        (2017,0.0289)
        (2018,0.0339)
        (2019,0.1081)
        (2020,0.1327)
        (2021,0.1119)
        (2022,0.1501)
        (2023,0.1681)
        (2024,0.3236)
        (2025,0.2781)
    };

    \addplot[name path=jsonlower, draw=none, forget plot] coordinates {
        (2014,0.0638)
        (2016,0.0949)
        (2017,0.0257)
        (2018,0.0315)
        (2019,0.1013)
        (2020,0.1245)
        (2021,0.1083)
        (2022,0.1419)
        (2023,0.1431)
        (2024,0.2724)
        (2025,0.2433)
    };

    \addplot[
        red!25,
        fill opacity=0.18,
        draw=none,
        forget plot
    ] fill between[of=jsonupper and jsonlower];

    \addplot[
        thick,
        blue!75!black,
        solid,
        mark=*,
        mark size=1.8pt,
        mark options={fill=blue!75!black, draw=blue!75!black},
        error bars/.cd,
            y dir=both,
            y explicit,
            error bar style={line width=0.45pt},
            error mark options={line width=0.45pt, mark size=1.3pt}
    ] coordinates {
        (2014,0.0349) +- (0,0.0004)
        (2016,0.0619) +- (0,0.0014)
        (2017,0.0129) +- (0,0.0013)
        (2018,0.0171) +- (0,0.0015)
        (2019,0.0903) +- (0,0.0057)
        (2020,0.1122) +- (0,0.0064)
        (2021,0.0946) +- (0,0.0037)
        (2022,0.0857) +- (0,0.0027)
        (2023,0.0149) +- (0,0.0009)
        (2024,0.0059) +- (0,0.0025)
        (2025,0.0044) +- (0,0.0013)
    };
    \addlegendentry{Static}

    \addplot[
        thick,
        green!60!black,
        dashed,
        mark=triangle*,
        mark size=2.0pt,
        mark options={fill=green!60!black, draw=green!60!black},
        error bars/.cd,
            y dir=both,
            y explicit,
            error bar style={line width=0.45pt},
            error mark options={line width=0.45pt, mark size=1.3pt}
    ] coordinates {
        (2014,0.0334) +- (0,0.0004)
        (2016,0.0618) +- (0,0.0009)
        (2017,0.0093) +- (0,0.0009)
        (2018,0.0125) +- (0,0.0007)
        (2019,0.1076) +- (0,0.0016)
        (2020,0.1058) +- (0,0.0025)
        (2021,0.0494) +- (0,0.0017)
        (2022,0.0922) +- (0,0.0020)
        (2023,0.0354) +- (0,0.0006)
        (2024,0.0109) +- (0,0.0006)
        (2025,0.0063) +- (0,0.0013)
    };
    \addlegendentry{Graph-based}

    \addplot[
        thick,
        red!75!black,
        dash dot,
        mark=square*,
        mark size=1.8pt,
        mark options={fill=red!75!black, draw=red!75!black},
        error bars/.cd,
            y dir=both,
            y explicit,
            error bar style={line width=0.45pt},
            error mark options={line width=0.45pt, mark size=1.3pt}
    ] coordinates {
        (2014,0.0643) +- (0,0.0005)
        (2016,0.0955) +- (0,0.0006)
        (2017,0.0273) +- (0,0.0016)
        (2018,0.0327) +- (0,0.0012)
        (2019,0.1047) +- (0,0.0034)
        (2020,0.1286) +- (0,0.0041)
        (2021,0.1101) +- (0,0.0018)
        (2022,0.1460) +- (0,0.0041)
        (2023,0.1556) +- (0,0.0125)
        (2024,0.2980) +- (0,0.0256)
        (2025,0.2607) +- (0,0.0174)
    };
    \addlegendentry{Dynamic}

    \end{axis}
    \end{tikzpicture}
        \vspace{-0.3em}
        \captionof{figure}{Year-wise dissenter rate comparison for static, dynamic and graph-based feature.}
        \label{fig:dissenter}
    \end{minipage}

\end{figure}

\BfPara{Setting}
To test whether modalities make consistent predictions over time, we train three XGBoost classifiers, one per modality, on 2013 samples and evaluate them on yearly test sets from 2014 to 2025. We measure inter-modality agreement with Fleiss' Kappa~\cite{fleiss1971measuring}, treating each classifier as an independent rater. 
For each test sample, we define the dissenting modality as the one whose prediction differs from the other two, then aggregate dissent rates by modality and year.
All experiments are repeated with three random seeds, and results are reported as \textit{mean$\pm$std}.

\BfPara{Analysis}
Figure~\ref{fig:kappa} shows that cross-modal agreement is highest in 2014 ($\kappa = 0.768$) and declines over time, becoming negative in 2024. This indicates that, for recent samples, modality-specific classifiers agree less often than expected by chance. The dissenter analysis in Figure~\ref{fig:dissenter} identifies which modality most often disagrees with the other two. Dynamic features are the main source of disagreement, acting as the conflicting modality in approximately 30\% of samples in 2024--2025, while static and graph-based classifiers remain more consistent with each other. These results suggest that temporal drift disrupts both individual feature spaces and cross-modal agreement.

\section{Concept Drift Adaptation Across Modalities}
\label{sec:cda-main}
\BfPara{Setting}
We evaluate concept-drift adaptation (CDA) on {\system} using two recent adaptation methods: Chen et al.~\cite{chen2023continuous} and CADE~\cite{cade}. Our goal is to assess how existing CDA methods handle temporal drift across static, dynamic, graph-based, and fused representations.

\begin{wraptable}[10]{r}{0.60\textwidth}
\vspace{-5pt}
\centering
\tiny
\setlength{\tabcolsep}{1pt}
\caption{Performance of CDA methods on {\system} across IID, NEAR, and FAR settings using three modalities and feature fusion (budget = 400).}
\resizebox{\linewidth}{!}{
\label{tab:adaptation-cd}
\centering
\scriptsize
\setlength{\tabcolsep}{4pt}
\renewcommand{\arraystretch}{1.25}

\begin{tabular}{p{2.2cm}|c|c|c|c|c}
\hline
\textbf{Method}
& \textbf{Setting}
& \textbf{Static}
& \textbf{Dynamic}
& \textbf{Graph-based}
& \textbf{Feature Fusion} \\
\hline

\multirow{3}{*}{\centering Chen et al.~\cite{chen2023continuous}}
& IID  
& \meanstd{95.85}{0.25} 
& \meanstd{73.91}{3.70}
& \meanstd{90.31}{1.55}
& \meanstd{93.85}{1.32} \\

& NEAR 
& \meanstd{87.04}{1.24}
& \meanstd{70.74}{1.07}
& \meanstd{73.01}{1.70}
& \meanstd{82.93}{0.65} \\

& FAR  
& \meanstd{70.54}{2.31}
& \meanstd{46.12}{2.22}
& \meanstd{62.47}{1.49}
& \meanstd{70.02}{0.53} \\
\hline

\multirow{3}{*}{\centering{CADE~\cite{cade}}}
& IID
& \meanstd{92.42}{0.32}
& \meanstd{76.86}{0.88}
& \meanstd{86.05}{1.10}
& \meanstd{82.03}{3.95} \\

& NEAR
& \meanstd{52.24}{7.87}
& \meanstd{36.22}{4.87}
& \meanstd{29.37}{4.66}
& \meanstd{41.56}{3.93} \\

& FAR
& \meanstd{20.56}{0.81}
& \meanstd{14.41}{1.14}
& \meanstd{6.31}{1.36}
& \meanstd{14.30}{4.16} \\
\hline

\end{tabular}
}
\vspace{-10pt}
\label{tab:adaptation-cd}
\end{wraptable}

We follow the active-learning framework of Chen et al.~\cite{chen2023continuous}, which adapts models in monthly cycles under labeling budgets of 50, 100, 200, and 400 samples. In each cycle, a subset of test samples is selected for labeling, and the model is retrained to reduce drift-induced degradation. We report the 400-sample budget in the main text and provide results for smaller budgets in Appendix~\ref{app:extra-cda}.

\BfPara{Analysis}
Table~\ref{tab:adaptation-cd} shows that Chen et al.~\cite{chen2023continuous} outperforms CADE across most settings, but both methods degrade substantially from IID to NEAR and FAR. This suggests that existing CDA methods remain limited under long-term multimodal malware drift. {\system} exposes this limitation by evaluating adaptation across static, dynamic, graph-based, and fused representations, each of which exhibits distinct drift behavior. For example, Chen et al. achieves comparable FAR performance with static features and feature fusion, while CADE collapses across modalities, with graph-based features dropping to \meanstd{6.31}{1.36}. These results highlight the need for modality-aware CDA methods that can adapt not only to evolving feature distributions but also to shifting cross-modal relationships.

\section{Discussion and Limitation}
\label{sec:discuss}

Unlike single-modality benchmarks~\cite{malnetCIKM,haque2025lamda,higraph,arp2014drebin,mariconti2017mamadroid}, {\system} aligns three complementary Android malware modalities: static features, dynamic behavioral features, and graph-based structural features. This enables controlled analysis of how modalities evolve independently and jointly under temporal distribution shift. Our results show substantial degradation across modalities, while multimodal fusion improves robustness in several long-term settings, suggesting complementary rather than redundant signals. Drift analyses further show that graph-based features exhibit the strongest temporal divergence, feature importance changes over time, and agreement among modality-specific classifiers declines in later years. These findings make {\system} a diagnostic benchmark for studying modality-specific drift, cross-modal inconsistency, malware-family evolution, explanation drift, and adaptation under temporal shift.


{\system} also has a few limitations. First, we retain only applications with valid static, dynamic, and graph-based representations. This reduces coverage but ensures sample-level modality alignment for fair unimodal and multimodal evaluation. Second, like many malware datasets, {\system} relies on VirusTotal-derived weak labels and AVClass2 family labels, which may contain noise despite conservative filtering. Finally, {\system} uses a more balanced sampling strategy than some real deployment settings to emphasize malware-family diversity and maintain sufficient malicious samples across years. This may underrepresent benign-heavy deployment environments, but supports controlled evaluation of multimodal drift, family evolution, and robustness under temporal shift.




\section{Conclusion}
We present McNdroid, a longitudinal multimodal Android malware benchmark spanning 12 years. It enables systematic analysis of how detection performance evolves under shifts in malware behavior, benign application distributions, malware-family composition, and modality-specific feature spaces. Our evaluations show that long-term malware evolution degrades robustness, shifts feature distributions, reduces cross-modal agreement, changes explanations, and challenges adaptation methods. With broad temporal coverage, diverse malware families, and aligned multimodal features, McNdroid provides a reproducible benchmark for robust, interpretable, and adaptive malware detection under non-stationary security conditions.

\bibliographystyle{plainnat}
\bibliography{main_references}

\begin{thebibliography}{92}
\providecommand{\natexlab}[1]{#1}
\providecommand{\url}[1]{\texttt{#1}}
\expandafter\ifx\csname urlstyle\endcsname\relax
  \providecommand{\doi}[1]{doi: #1}\else
  \providecommand{\doi}{doi: \begingroup \urlstyle{rm}\Url}\fi

\bibitem[Allix et~al.(2016)Allix, Bissyand{\'e}, Klein, and Le~Traon]{androzoo}
Kevin Allix, Tegawend{\'e}~F. Bissyand{\'e}, Jacques Klein, and Yves Le~Traon.
\newblock Androzoo: Collecting millions of android apps for the research community.
\newblock In \emph{International Conference on Mining Software Repositories (MSR)}, 2016.

\bibitem[Anderson and Roth(2018)]{ember}
Hyrum~S Anderson and Phil Roth.
\newblock {EMBER}: an open dataset for training static pe malware machine learning models.
\newblock \emph{arXiv preprint}, 2018.

\bibitem[Aonzo et~al.(2020)Aonzo, Georgiu, Verderame, and Merlo]{aonzo2020obfuscapk}
Simone Aonzo, Gabriel~Claudiu Georgiu, Luca Verderame, and Alessio Merlo.
\newblock Obfuscapk: An open-source black-box obfuscation tool for android apps.
\newblock \emph{(SoftwareX)}, 2020.

\bibitem[Arevalo et~al.(2017)Arevalo, Solorio, Montes-y G{\'o}mez, and Gonz{\'a}lez]{arevalo2017gated}
John Arevalo, Thamar Solorio, Manuel Montes-y G{\'o}mez, and Fabio~A Gonz{\'a}lez.
\newblock Gated multimodal units for information fusion.
\newblock \emph{arXiv preprint}, 2017.

\bibitem[Arp et~al.(2014)Arp, Spreitzenbarth, Hubner, Gascon, Rieck, and Siemens]{arp2014drebin}
Daniel Arp, Michael Spreitzenbarth, Malte Hubner, Hugo Gascon, Konrad Rieck, and CERT Siemens.
\newblock {Drebin}: Effective and explainable detection of android malware in your pocket.
\newblock In \emph{Network and Distributed System Security Symposium (NDSS)}, 2014.

\bibitem[Arp et~al.(2022)Arp, Quiring, Pendlebury, Warnecke, Pierazzi, Wressnegger, Cavallaro, and Rieck]{dosanddont}
Daniel Arp, Erwin Quiring, Feargus Pendlebury, Alexander Warnecke, Fabio Pierazzi, Christian Wressnegger, Lorenzo Cavallaro, and Konrad Rieck.
\newblock Dos and don{\textquoteright}ts of machine learning in computer security.
\newblock In \emph{USENIX Security Symposium}, 2022.

\bibitem[Baier et~al.(2021)Baier, Schl{\"o}r, Sch{\"o}ffer, and K{\"u}hl]{baier2021detecting}
Lucas Baier, Tim Schl{\"o}r, Jakob Sch{\"o}ffer, and Niklas K{\"u}hl.
\newblock Detecting concept drift with neural network model uncertainty.
\newblock \emph{arXiv preprint arXiv:2107.01873}, 2021.

\bibitem[Bavota et~al.(2014)Bavota, Linares-Vasquez, Bernal-Cardenas, Di~Penta, Oliveto, and Poshyvanyk]{bavota2014impact}
Gabriele Bavota, Mario Linares-Vasquez, Carlos~Eduardo Bernal-Cardenas, Massimiliano Di~Penta, Rocco Oliveto, and Denys Poshyvanyk.
\newblock The impact of api change-and fault-proneness on the user ratings of android apps.
\newblock \emph{IEEE Transactions on Software Engineering (TSE)}, 2014.

\bibitem[Berger and Zhou(2014)]{kstest2014}
Vance~W Berger and YanYan Zhou.
\newblock Kolmogorov--smirnov test: Overview.
\newblock \emph{Wiley statsref: Statistics reference online}, 2014.

\bibitem[Boulahia et~al.(2021)Boulahia, Amamra, Madi, and Daikh]{boulahia2021early}
Said~Yacine Boulahia, Abdenour Amamra, Mohamed~Ridha Madi, and Said Daikh.
\newblock Early, intermediate and late fusion strategies for robust deep learning-based multimodal action recognition.
\newblock \emph{Machine Vision and Applications}, 2021.

\bibitem[Cai(2020)]{cai2020assessing}
Haipeng Cai.
\newblock Assessing and improving malware detection sustainability through app evolution studies.
\newblock \emph{ACM Transactions on Software Engineering and Methodology (TOSEM)}, 2020.

\bibitem[Campello et~al.(2013)Campello, Moulavi, and Sander]{hdbscan}
Ricardo J. G.~B. Campello, Davoud Moulavi, and J{\"o}rg Sander.
\newblock Density-based clustering based on hierarchical density estimates.
\newblock In \emph{Pacific-Asia Conference on Knowledge Discovery and Data Mining (PA-KDD)}, 2013.

\bibitem[Castro et~al.(2019)Castro, Hazarika, P{\'e}rez-Rosas, Zimmermann, Mihalcea, and Poria]{castro2019towards}
Santiago Castro, Devamanyu Hazarika, Ver{\'o}nica P{\'e}rez-Rosas, Roger Zimmermann, Rada Mihalcea, and Soujanya Poria.
\newblock Towards multimodal sarcasm detection (an \_obviously\_ perfect paper).
\newblock In \emph{Annual Meeting of the Association for Computational Linguistics (ACL)}, 2019.

\bibitem[Chen et~al.(2018)Chen, Carvalho, Baracaldo, Ludwig, Edwards, Lee, Molloy, and Srivastava]{chen2018detecting}
Bryant Chen, Wilka Carvalho, Nathalie Baracaldo, Heiko Ludwig, Benjamin Edwards, Taesung Lee, Ian Molloy, and Biplav Srivastava.
\newblock Detecting backdoor attacks on deep neural networks by activation clustering.
\newblock \emph{arXiv preprint arXiv:1811.03728}, 2018.

\bibitem[Chen et~al.(2025)Chen, Wang, Chen, Zhang, Qin, and Zhang]{higraph}
Han Chen, Hanchen Wang, Hongmei Chen, Ying Zhang, Lu~Qin, and Wenjie Zhang.
\newblock Higraph: A large-scale hierarchical graph dataset for malware analysis.
\newblock \emph{arXiv preprint arXiv:2509.02113}, 2025.

\bibitem[Chen and Guestrin(2016)]{chen2016xgboost}
Tianqi Chen and Carlos Guestrin.
\newblock Xgboost: A scalable tree boosting system.
\newblock In \emph{ACM SIGKDD International Conference on Knowledge Discovery and Data Mining (SIGKDD)}, 2016.

\bibitem[Chen et~al.(2020)Chen, Wang, She, and Jana]{chen2020training}
Yizheng Chen, Shiqi Wang, Dongdong She, and Suman Jana.
\newblock On training robust $\{$PDF$\}$ malware classifiers.
\newblock In \emph{USENIX Security Symposium}, 2020.

\bibitem[Chen et~al.(2023)Chen, Ding, and Wagner]{chen2023continuous}
Yizheng Chen, Zhoujie Ding, and David Wagner.
\newblock Continuous learning for android malware detection.
\newblock In \emph{USENIX Security Symposium}, 2023.

\bibitem[Chow et~al.(2025)Chow, D'Onghia, Linhardt, Kan, Arp, Cavallaro, and Pierazzi]{chow2025breaking}
Theo Chow, Mario D'Onghia, Lorenz Linhardt, Zeliang Kan, Daniel Arp, Lorenzo Cavallaro, and Fabio Pierazzi.
\newblock Breaking out from the tesseract: Reassessing ml-based malware detection under spatio-temporal drift.
\newblock \emph{arXiv preprint arXiv:2506.23814}, 2025.

\bibitem[Desnos(2011)]{androguard}
Anthony Desnos.
\newblock Androguard: Reverse engineering, malware and goodware analysis of android applications.
\newblock \url{https://github.com/androguard/androguard}, 2011.

\bibitem[Dragoi et~al.(2022)Dragoi, Burceanu, Haller, Manolache, and Brad]{anoshift}
Marius Dragoi, Elena Burceanu, Emanuela Haller, Andrei Manolache, and Florin Brad.
\newblock Anoshift: A distribution shift benchmark for unsupervised anomaly detection.
\newblock \emph{Advances in Neural Information Processing Systems (NeurIPS)}, 2022.

\bibitem[Enck et~al.(2014)Enck, Gilbert, Han, Tendulkar, Chun, Cox, Jung, McDaniel, and Sheth]{enck2014taintdroid}
William Enck, Peter Gilbert, Seungyeop Han, Vasant Tendulkar, Byung-Gon Chun, Landon~P Cox, Jaeyeon Jung, Patrick McDaniel, and Anmol~N Sheth.
\newblock Taintdroid: an information-flow tracking system for realtime privacy monitoring on smartphones.
\newblock \emph{ACM Transactions on Computer Systems (TOCS)}, 2014.

\bibitem[{European Union Agency for Cybersecurity (ENISA)}(2025)]{ENISAThreatLandscape}
{European Union Agency for Cybersecurity (ENISA)}.
\newblock Threat landscape.
\newblock \url{https://www.enisa.europa.eu/topics/cyber-threats/threat-landscape}, 2025.
\newblock URL \url{https://www.enisa.europa.eu/topics/cyber-threats/threat-landscape}.
\newblock Accessed 2025-11-27.

\bibitem[Fazzini et~al.(2019)Fazzini, Xin, and Orso]{fazzini2019automated}
Mattia Fazzini, Qi~Xin, and Alessandro Orso.
\newblock Automated api-usage update for android apps.
\newblock In \emph{Proceedings of the 28th ACM SIGSOFT international symposium on software testing and analysis (SIGSOFT)}, 2019.

\bibitem[Feng et~al.(2023)Feng, Bose, Zhang, Hebbar, Ramakrishna, Gupta, Zhang, Avestimehr, and Narayanan]{feng2023fedmultimodal}
Tiantian Feng, Digbalay Bose, Tuo Zhang, Rajat Hebbar, Anil Ramakrishna, Rahul Gupta, Mi~Zhang, Salman Avestimehr, and Shrikanth Narayanan.
\newblock Fedmultimodal: A benchmark for multimodal federated learning.
\newblock In \emph{ACM SIGKDD International Conference on Knowledge Discovery and Data Mining}, 2023.

\bibitem[Fleiss(1971)]{fleiss1971measuring}
Joseph~L Fleiss.
\newblock Measuring nominal scale agreement among many raters.
\newblock \emph{American Psychological Association Psychological bulletin}, 1971.

\bibitem[Franco et~al.(2020)Franco, Magnani, and Maio]{franco2020multimodal}
Annalisa Franco, Antonio Magnani, and Dario Maio.
\newblock A multimodal approach for human activity recognition based on skeleton and rgb data.
\newblock \emph{Pattern Recognition Letters}, 2020.

\bibitem[Freitas et~al.(2021)Freitas, Dong, Neil, and Chau]{malnetNeurIPS}
Scott Freitas, Yuxiao Dong, Joshua Neil, and Duen~Horng Chau.
\newblock A large-scale database for graph representation learning.
\newblock \emph{Advances in Neural Information Processing Systems (NeurIPS)}, 2021.

\bibitem[Freitas et~al.(2022)Freitas, Duggal, and Chau]{malnetCIKM}
Scott Freitas, Rahul Duggal, and Duen~Horng Chau.
\newblock Malnet: A large-scale image database of malicious software.
\newblock In \emph{ACM International Conference on Information \& Knowledge Management (CIKM)}, 2022.

\bibitem[Gao et~al.(2024)Gao, Huang, Li, Wu, Wu, and Yuan]{gao2024comprehensive}
Cuiying Gao, Gaozhun Huang, Heng Li, Bang Wu, Yueming Wu, and Wei Yuan.
\newblock A comprehensive study of learning-based android malware detectors under challenging environments.
\newblock In \emph{IEEE/ACM International Conference on Software Engineering (ICSE)}, 2024.

\bibitem[Hamilton et~al.(2017)Hamilton, Ying, and Leskovec]{hamilton2017inductive-graph}
Will Hamilton, Zhitao Ying, and Jure Leskovec.
\newblock Inductive representation learning on large graphs.
\newblock \emph{Advances in Neural Information Processing Systems (NeurIPS)}, 2017.

\bibitem[Haque et~al.(2026)Haque, Hossain, Kamol, Alam, Amalapuram, Talukder, and Rahman]{haque2025lamda}
Md~Ahsanul Haque, Ismail Hossain, Md~Mahmuduzzaman Kamol, Md~Jahangir Alam, Suresh~Kumar Amalapuram, Sajedul Talukder, and Mohammad~Saidur Rahman.
\newblock {LAMDA}: A longitudinal android malware benchmark for concept drift analysis.
\newblock In \emph{International Conference on Learning Representations (ICLR)}, 2026.

\bibitem[Hasan et~al.(2019)Hasan, Rahman, Zadeh, Zhong, Tanveer, Morency, and Hoque]{hasan2019ur}
Md~Kamrul Hasan, Wasifur Rahman, AmirAli~Bagher Zadeh, Jianyuan Zhong, Md~Iftekhar Tanveer, Louis-Philippe Morency, and Mohammed~Ehsan Hoque.
\newblock Ur-funny: A multimodal language dataset for understanding humor.
\newblock In \emph{Conference on Empirical Methods in Natural Language Processing and the International Joint Conference on Natural Language Processing (EMNLP-IJCNLP)}, 2019.

\bibitem[He et~al.(2023{\natexlab{a}})He, Xia, Zhang, and Ji]{he2023efficient}
Ping He, Yifan Xia, Xuhong Zhang, and Shouling Ji.
\newblock Efficient query-based attack against ml-based android malware detection under zero knowledge setting.
\newblock In \emph{ACM SIGSAC Conference on Computer and Communications Security (CCS)}, 2023{\natexlab{a}}.

\bibitem[He et~al.(2023{\natexlab{b}})He, Liu, Wu, Yang, Ren, and Qin]{msdroid}
Yiling He, Yiping Liu, Lei Wu, Ziqi Yang, Kui Ren, and Zhan Qin.
\newblock Msdroid: Identifying malicious snippets for android malware detection.
\newblock \emph{IEEE Transactions on Dependable and Secure Computing (TDSC)}, 2023{\natexlab{b}}.

\bibitem[Hou et~al.(2017)Hou, Ye, Song, and Abdulhayoglu]{hou2017hindroid}
Shifu Hou, Yanfang Ye, Yangqiu Song, and Melih Abdulhayoglu.
\newblock Hindroid: An intelligent android malware detection system based on structured heterogeneous information network.
\newblock In \emph{ACM SIGKDD International Conference on Knowledge Discovery and Data Mining}, 2017.

\bibitem[Jeffreys(1946)]{Jeffreysdivergence}
Harold Jeffreys.
\newblock An invariant form for the prior probability in estimation problems.
\newblock \emph{Proceedings of the Royal Society of London. Series A. Mathematical and Physical Sciences}, 1946.

\bibitem[Johnson et~al.(2016)Johnson, Pollard, Shen, Lehman, Feng, Ghassemi, Moody, Szolovits, Anthony~Celi, and Mark]{johnson2016mimic}
Alistair~EW Johnson, Tom~J Pollard, Lu~Shen, Li-wei~H Lehman, Mengling Feng, Mohammad Ghassemi, Benjamin Moody, Peter Szolovits, Leo Anthony~Celi, and Roger~G Mark.
\newblock Mimic-iii, a freely accessible critical care database.
\newblock \emph{Scientific data}, 2016.

\bibitem[Joyce et~al.(2025)Joyce, Miller, Roth, Zak, Zaresky-Williams, Anderson, Raff, and Holt]{ember2024}
Robert~J Joyce, Gideon Miller, Phil Roth, Richard Zak, Elliott Zaresky-Williams, Hyrum Anderson, Edward Raff, and James Holt.
\newblock {EMBER2024--A Benchmark Dataset for Holistic Evaluation of Malware Classifiers}.
\newblock \emph{arXiv preprint arXiv:2506.05074}, 2025.

\bibitem[Kay et~al.(2017)Kay, Carreira, Simonyan, Zhang, Hillier, Vijayanarasimhan, Viola, Green, Back, Natsev, et~al.]{kay2017kinetics}
Will Kay, Joao Carreira, Karen Simonyan, Brian Zhang, Chloe Hillier, Sudheendra Vijayanarasimhan, Fabio Viola, Tim Green, Trevor Back, Paul Natsev, et~al.
\newblock The kinetics human action video dataset.
\newblock \emph{arXiv preprint arXiv:1705.06950}, 2017.

\bibitem[Ke et~al.(2017)Ke, Meng, Finley, Wang, Chen, Ma, Ye, and Liu]{lightgbm}
Guolin Ke, Qi~Meng, Thomas Finley, Taifeng Wang, Wei Chen, Weidong Ma, Qiwei Ye, and Tie-Yan Liu.
\newblock Lightgbm: A highly efficient gradient boosting decision tree.
\newblock \emph{Advances in Neural Information Processing Systems (NeurIPS)}, 2017.

\bibitem[Kim et~al.(2018)Kim, Kang, Rho, Sezer, and Im]{kim2018multimodal}
TaeGuen Kim, BooJoong Kang, Mina Rho, Sakir Sezer, and Eul~Gyu Im.
\newblock A multimodal deep learning method for android malware detection using various features.
\newblock \emph{IEEE Transactions on Information Forensics and Security (TIFS)}, 2018.

\bibitem[Kondracki et~al.(2022)Kondracki, Azad, Miramirkhani, and Nikiforakis]{kondracki2022droid}
Brian Kondracki, Babak~Amin Azad, Najmeh Miramirkhani, and Nick Nikiforakis.
\newblock The droid is in the details: Environment-aware evasion of android sandboxes.
\newblock In \emph{Network and Distributed System Security Symposium (NDSS)}, 2022.

\bibitem[Kulinski et~al.(2020)Kulinski, Bagchi, and Inouye]{ks-testneurips}
Sean Kulinski, Saurabh Bagchi, and David~I Inouye.
\newblock Feature shift detection: Localizing which features have shifted via conditional distribution tests.
\newblock \emph{Advances in Neural Information Processing Systems (NeurIPS)}, 2020.

\bibitem[Lakshmanan(2024)]{lakshmanan2024darkgate}
Ravie Lakshmanan.
\newblock Darkgate malware exploits samba file shares in short-lived campaign.
\newblock \emph{The Hacker News}, 2024.
\newblock URL \url{https://thehackernews.com/2024/07/darkgate-malware-exploits-samba-file.html}.

\bibitem[Lashkari et~al.(2018)Lashkari, Kadir, Taheri, and Ghorbani]{lashkari2018toward}
Arash~Habibi Lashkari, Andi Fitriah~A Kadir, Laya Taheri, and Ali~A Ghorbani.
\newblock Toward developing a systematic approach to generate benchmark android malware datasets and classification.
\newblock In \emph{2018 International Carnahan conference on security technology (ICCST)}, 2018.

\bibitem[Lee et~al.(2020{\natexlab{a}})Lee, Yi, Mart{\'\i}n-Mart{\'\i}n, Savarese, and Bohg]{lee2020multimodal}
Michelle~A Lee, Brent Yi, Roberto Mart{\'\i}n-Mart{\'\i}n, Silvio Savarese, and Jeannette Bohg.
\newblock Multimodal sensor fusion with differentiable filters.
\newblock In \emph{2020 IEEE/RSJ International Conference on Intelligent Robots and Systems (IROS)}. IEEE, 2020{\natexlab{a}}.

\bibitem[Lee et~al.(2020{\natexlab{b}})Lee, Zhu, Zachares, Tan, Srinivasan, Savarese, Fei-Fei, Garg, and Bohg]{lee2020making}
Michelle~A Lee, Yuke Zhu, Peter Zachares, Matthew Tan, Krishnan Srinivasan, Silvio Savarese, Li~Fei-Fei, Animesh Garg, and Jeannette Bohg.
\newblock Making sense of vision and touch: Learning multimodal representations for contact-rich tasks.
\newblock \emph{IEEE Transactions on Robotics (T-RO)}, 2020{\natexlab{b}}.

\bibitem[Lei et~al.(2019)Lei, Qin, Wang, Li, and Ye]{lei2019evedroid}
Tao Lei, Zhan Qin, Zhibo Wang, Qi~Li, and Dengpan Ye.
\newblock Evedroid: Event-aware android malware detection against model degrading for iot devices.
\newblock \emph{IEEE Internet of Things Journal (IoT-J)}, 2019.

\bibitem[Leiva et~al.(2020)Leiva, Hota, and Oulasvirta]{leiva2020enrico}
Luis~A Leiva, Asutosh Hota, and Antti Oulasvirta.
\newblock Enrico: A dataset for topic modeling of mobile ui designs.
\newblock In \emph{International Conference on Human-Computer Interaction with Mobile Devices and Services}, 2020.

\bibitem[Li et~al.(2025)Li, Iyengar, Kundu, and Bertino]{li2025revisiting}
Adrian~Shuai Li, Arun Iyengar, Ashish Kundu, and Elisa Bertino.
\newblock Revisiting concept drift in windows malware detection: Adaptation to real drifted malware with minimal samples.
\newblock In \emph{Network and Distributed System Security Symposium (NDSS)}, 2025.

\bibitem[Li et~al.(2021)Li, Zhou, Yuan, Luo, Gao, and Chen]{ramda}
Heng Li, Shiyao Zhou, Wei Yuan, Xiapu Luo, Cuiying Gao, and Shuiyan Chen.
\newblock Robust android malware detection against adversarial example attacks.
\newblock In \emph{Web Conference (WWW)}, 2021.

\bibitem[Liang et~al.(2021)Liang, Lyu, Fan, Wu, Cheng, Wu, Chen, Wu, Lee, Zhu, et~al.]{liang2021multibench}
Paul~Pu Liang, Yiwei Lyu, Xiang Fan, Zetian Wu, Yun Cheng, Jason Wu, Leslie Chen, Peter Wu, Michelle~A Lee, Yuke Zhu, et~al.
\newblock Multibench: Multiscale benchmarks for multimodal representation learning.
\newblock \emph{Advances in Neural Information Processing Systems (NeurIPS)}, 2021.

\bibitem[Liang et~al.(2023)Liang, Lyu, Fan, Agarwal, Cheng, Morency, and Salakhutdinov]{liang2023multizoo}
Paul~Pu Liang, Yiwei Lyu, Xiang Fan, Arav Agarwal, Yun Cheng, Louis-Philippe Morency, and Ruslan Salakhutdinov.
\newblock Multizoo and multibench: A standardized toolkit for multimodal deep learning.
\newblock \emph{Journal of Machine Learning Research}, 2023.

\bibitem[Liu et~al.(2026)Liu, Zeng, Pierazzi, Yang, Cavallaro, and Liang]{liu2026unraveling}
Jiahao Liu, Jun Zeng, Fabio Pierazzi, Ziqi Yang, Lorenzo Cavallaro, and Zhenkai Liang.
\newblock Unraveling the key of machine learning-based android malware detection.
\newblock \emph{ACM Transactions on Software Engineering and Methodology (TOSEM)}, 2026.

\bibitem[Lundberg and Lee(2017)]{shap}
Scott~M Lundberg and Su-In Lee.
\newblock A unified approach to interpreting model predictions.
\newblock In \emph{Advances in Neural Information Processing Systems (NeurIPS)}, 2017.

\bibitem[Mahdavifar et~al.(2020)Mahdavifar, Kadir, Fatemi, Alhadidi, and Ghorbani]{mahdavifar2020dynamic}
Samaneh Mahdavifar, Andi Fitriah~Abdul Kadir, Rasool Fatemi, Dima Alhadidi, and Ali~A Ghorbani.
\newblock Dynamic android malware category classification using semi-supervised deep learning.
\newblock In \emph{2020 IEEE Intl Conf on Dependable, Autonomic and Secure Computing, Intl Conf on Pervasive Intelligence and Computing, Intl Conf on Cloud and Big Data Computing, Intl Conf on Cyber Science and Technology Congress (DASC/PiCom/CBDCom/CyberSciTech)}, 2020.

\bibitem[Mariconti et~al.(2017)Mariconti, Onwuzurike, Andriotis, De~Cristofaro, Ross, and Stringhini]{mariconti2017mamadroid}
Enrico Mariconti, Lucky Onwuzurike, Panagiotis Andriotis, Emiliano De~Cristofaro, Gianluca Ross, and Gianluca Stringhini.
\newblock Mamadroid: Detecting android malware by building markov chains of behavioral models.
\newblock In \emph{Network and Distributed System Security Symposium (NDSS)}, 2017.

\bibitem[Nguyen et~al.(2024)Nguyen, Tran, Johnson, and Leach]{nguyen2024pbp}
Dung~Thuy Nguyen, Ngoc~N Tran, Taylor~T Johnson, and Kevin Leach.
\newblock Pbp: Post-training backdoor purification for malware classifiers.
\newblock \emph{arXiv preprint arXiv:2412.03441}, 2024.

\bibitem[Ortega and Pratapagiri(2025)]{godfather-casestudy}
Fernando Ortega and Vishnu Pratapagiri.
\newblock Your mobile app, their playground: The dark side of virtualization, 2025.
\newblock URL \url{https://zimperium.com/blog/your-mobile-app-their-playground-the-dark-side-of-the-virtualization}.

\bibitem[Park et~al.(2025)Park, Ji, Park, Rahman, and Oh]{malcl}
Jimin Park, AHyun Ji, Minji Park, Mohammad~Saidur Rahman, and Se~Eun Oh.
\newblock {MalCL: Leveraging gan-based generative replay to combat catastrophic forgetting in malware classification}.
\newblock In \emph{AAAI Conference on Artificial Intelligence}, 2025.

\bibitem[Pendlebury et~al.(2019)Pendlebury, Pierazzi, Jordaney, Kinder, and Cavallaro]{tesseract}
Feargus Pendlebury, Fabio Pierazzi, Roberto Jordaney, Johannes Kinder, and Lorenzo Cavallaro.
\newblock {TESSERACT}: {Eliminating} experimental bias in malware classification across space and time.
\newblock In \emph{USENIX Security Symposium}, 2019.

\bibitem[P{\'e}rez-R{\'u}a et~al.(2019)P{\'e}rez-R{\'u}a, Vielzeuf, Pateux, Baccouche, and Jurie]{perez2019early}
Juan-Manuel P{\'e}rez-R{\'u}a, Valentin Vielzeuf, St{\'e}phane Pateux, Moez Baccouche, and Fr{\'e}d{\'e}ric Jurie.
\newblock Mfas: Multimodal fusion architecture search.
\newblock In \emph{Proceedings of the IEEE/CVF Conference on Computer Vision and Pattern Recognition (CVPRW)}, 2019.

\bibitem[Rahman et~al.(2022)Rahman, Coull, and Wright]{continual-learning-malware}
Mohammad~Saidur Rahman, Scott~E. Coull, and Matthew Wright.
\newblock {On the Limitations of Continual Learning for Malware Classification}.
\newblock In \emph{Conference on Lifelong Learning Agents (CoLLAs)}, 2022.

\bibitem[Rahman et~al.(2025)Rahman, Coull, Yu, and Wright]{madar2025}
Mohammad~Saidur Rahman, Scott Coull, Qi~Yu, and Matthew Wright.
\newblock {MADAR: Efficient continual learning for malware analysis with distribution-aware replay}.
\newblock In \emph{Conference on Applied Machine Learning in Information Security (CAMLIS)}, 2025.

\bibitem[Rastogi et~al.(2013)Rastogi, Chen, and Jiang]{rastogi2013droidchameleon}
Vaibhav Rastogi, Yan Chen, and Xuxian Jiang.
\newblock Droidchameleon: Evaluating android anti-malware against transformation attacks.
\newblock In \emph{ACM SIGSAC Conference on Computer and Communications Security (CCS)}, 2013.

\bibitem[Rolnick et~al.(2019)Rolnick, Ahuja, Schwarz, Lillicrap, and Wayne]{er}
David Rolnick, Arun Ahuja, Jonathan Schwarz, Timothy Lillicrap, and Gregory Wayne.
\newblock Experience replay for continual learning.
\newblock In \emph{Neural Information Processing Systems (NeurIPS)}, 2019.

\bibitem[Sebasti{\'a}n and Caballero(2020)]{avclass2}
Silvia Sebasti{\'a}n and Juan Caballero.
\newblock Avclass2: Massive malware tag extraction from av labels.
\newblock In \emph{Annual Computer Security Applications Conference (ACSAC)}, 2020.

\bibitem[Shijo and Salim(2015)]{dynamic-malware}
PV~Shijo and AJPCS Salim.
\newblock Integrated static and dynamic analysis for malware detection.
\newblock \emph{Procedia Computer Science}, 2015.

\bibitem[Shokouhinejad et~al.(2025)Shokouhinejad, Razavi-Far, Mohammadian, Rabbani, Ansong, Higgins, and Ghorbani]{shokouhinejad2025recent}
Hossein Shokouhinejad, Roozbeh Razavi-Far, Hesamodin Mohammadian, Mahdi Rabbani, Samuel Ansong, Griffin Higgins, and Ali~A Ghorbani.
\newblock Recent advances in malware detection: Graph learning and explainability.
\newblock \emph{arXiv preprint arXiv:2502.10556}, 2025.

\bibitem[{\v{S}}rndic and Laskov(2013)]{vsrndic2013detection}
Nedim {\v{S}}rndic and Pavel Laskov.
\newblock Detection of malicious pdf files based on hierarchical document structure.
\newblock In \emph{Network and Distributed System Security Symposium (NDSS)}, 2013.

\bibitem[Tam et~al.(2015)Tam, Khan, Fattori, and Cavallaro]{tam2015copperdroid}
Kimberly Tam, Salahuddin~J Khan, Aristide Fattori, and Lorenzo Cavallaro.
\newblock Copperdroid: automatic reconstruction of android malware behaviors.
\newblock In \emph{Network and Distributed System Security Symposium (NDSS)}, 2015.

\bibitem[van~de Ven et~al.(2024)van~de Ven, Soures, and Kudithipudi]{van2024continual}
Gido~M van~de Ven, Nicholas Soures, and Dhireesha Kudithipudi.
\newblock Continual learning and catastrophic forgetting.
\newblock \emph{arXiv preprint arXiv:2403.05175}, 2024.

\bibitem[Vielzeuf et~al.(2018)Vielzeuf, Lechervy, Pateux, and Jurie]{vielzeuf2018centralnet}
Valentin Vielzeuf, Alexis Lechervy, St{\'e}phane Pateux, and Fr{\'e}d{\'e}ric Jurie.
\newblock Centralnet: a multilayer approach for multimodal fusion.
\newblock In \emph{European Conference on Computer Vision (ECCV) workshops}, 2018.

\bibitem[Virustotal()]{virustotalstats}
Virustotal.
\newblock {V}irus{T}otal --- virustotal.com.
\newblock \url{https://www.virustotal.com/gui/intelligence-overview}.
\newblock [Accessed 21-10-2025].

\bibitem[VirusTotal(2025)]{virustotal}
VirusTotal.
\newblock {VirusTotal -- Stats}, 2025.
\newblock https://www.virustotal.com/gui/stats.

\bibitem[Weinberger et~al.(2009)Weinberger, Dasgupta, Langford, Smola, and Attenberg]{hashingtrick}
Kilian Weinberger, Anirban Dasgupta, John Langford, Alex Smola, and Josh Attenberg.
\newblock Feature hashing for large scale multitask learning.
\newblock In \emph{International Conference on Machine Learning (ICML)}, 2009.

\bibitem[Wu et~al.(2021{\natexlab{a}})Wu, Chen, Gao, Fan, Liu, Wen, and Lyu]{xmal}
Bozhi Wu, Sen Chen, Cuiyun Gao, Lingling Fan, Yang Liu, Weiping Wen, and Michael~R. Lyu.
\newblock Why an android app is classified as malware: Toward malware classification interpretation.
\newblock \emph{ACM Transactions on Software Engineering and Methodology (TOSEM)}, 2021{\natexlab{a}}.

\bibitem[Wu et~al.(2019)Wu, Li, Zou, Yang, Zhang, and Jin]{malscan}
Yueming Wu, Xiaodi Li, Deqing Zou, Wei Yang, Xin Zhang, and Hai Jin.
\newblock Malscan: Fast market-wide mobile malware scanning by social-network centrality analysis.
\newblock In \emph{IEEE/ACM International Conference on Automated Software Engineering (ASE)}. IEEE, 2019.

\bibitem[Wu et~al.(2021{\natexlab{b}})Wu, Zou, Yang, Li, and Jin]{homdroid}
Yueming Wu, Deqing Zou, Wei Yang, Xiang Li, and Hai Jin.
\newblock Homdroid: detecting android covert malware by social-network homophily analysis.
\newblock In \emph{Proceedings of the 30th ACM SIGSOFT International Symposium on Software Testing and Analysis (ISSTA)}, 2021{\natexlab{b}}.

\bibitem[Xiong et~al.(2021)Xiong, Zeng, Chakraborty, Tan, Fung, Li, and Singh]{xiong2021nystromformer}
Yunyang Xiong, Zhanpeng Zeng, Rudrasis Chakraborty, Mingxing Tan, Glenn Fung, Yin Li, and Vikas Singh.
\newblock Nystr{\"o}mformer: A nystr{\"o}m-based algorithm for approximating self-attention.
\newblock In \emph{AAAI conference on artificial intelligence (AAAI)}, 2021.

\bibitem[Xu et~al.(2024)Xu, Yao, Zhang, Dawoud, Park, and Saltaformaggio]{xu2024dva}
Haichuan Xu, Mingxuan Yao, Runze Zhang, Mohamed~Moustafa Dawoud, Jeman Park, and Brendan Saltaformaggio.
\newblock $\{$DVa$\}$: Extracting victims and abuse vectors from android accessibility malware.
\newblock In \emph{USENIX Security Symposium}, 2024.

\bibitem[Xu et~al.()Xu, Li, Deng, and Xu]{sdac}
Jiayun Xu, Yingjiu Li, Robert~H. Deng, and Ke~Xu.
\newblock Sdac: A slow-aging solution for android malware detection using semantic distance based api clustering.
\newblock \emph{IEEE Transactions on Dependable and Secure Computing (TDSC)}.

\bibitem[Yan and Yin(2012)]{yan2012droidscope}
Lok-Kwong Yan and Heng Yin.
\newblock Droidscope: Seamlessly reconstructing the os and dalvik semantic views for dynamic android malware analysis.
\newblock In \emph{USENIX Security Symposium}, 2012.

\bibitem[Yang et~al.(2018)Yang, Jones, Moninger, and Che]{yang2018android}
Guowei Yang, Jeffrey Jones, Austin Moninger, and Meiru Che.
\newblock How do android operating system updates impact apps?
\newblock In \emph{International Conference on Mobile Software Engineering and Systems (MOBILESoft)}, 2018.

\bibitem[Yang et~al.(2021)Yang, Guo, Hao, Ciptadi, Ahmadzadeh, Xing, and Wang]{cade}
Limin Yang, Wenbo Guo, Qingying Hao, Arridhana Ciptadi, Ali Ahmadzadeh, Xinyu Xing, and Gang Wang.
\newblock {CADE: Detecting} and explaining concept drift samples for security applications.
\newblock In \emph{USENIX Security Symposium}, 2021.

\bibitem[Zadeh et~al.(2016)Zadeh, Zellers, Pincus, and Morency]{zadeh2016mosi}
Amir Zadeh, Rowan Zellers, Eli Pincus, and Louis-Philippe Morency.
\newblock Mosi: multimodal corpus of sentiment intensity and subjectivity analysis in online opinion videos.
\newblock \emph{arXiv preprint arXiv:1606.06259}, 2016.

\bibitem[Zadeh et~al.(2018)Zadeh, Liang, Poria, Cambria, and Morency]{zadeh2018multimodal}
AmirAli~Bagher Zadeh, Paul~Pu Liang, Soujanya Poria, Erik Cambria, and Louis-Philippe Morency.
\newblock Multimodal language analysis in the wild: Cmu-mosei dataset and interpretable dynamic fusion graph.
\newblock In \emph{Annual Meeting of the Association for Computational Linguistics (ACL)}, 2018.

\bibitem[Zhang et~al.(2025)Zhang, Su, Liu, and Yang]{mpdroid}
Sanfeng Zhang, Heng Su, Hongxian Liu, and Wang Yang.
\newblock Mpdroid: A multimodal pre-training android malware detection method with static and dynamic features.
\newblock \emph{Computers \& Security}, 2025.

\bibitem[Zhang et~al.(2020)Zhang, Zhang, Zhong, Ding, Cao, Zhang, Zhang, and Yang]{apigraph}
Xiaohan Zhang, Yuan Zhang, Ming Zhong, Daizong Ding, Yinzhi Cao, Yukun Zhang, Mi~Zhang, and Min Yang.
\newblock Enhancing state-of-the-art classifiers with api semantics to detect evolved android malware.
\newblock In \emph{ACM SIGSAC Conference on Computer and Communications Security (CCS)}, 2020.

\bibitem[Zhao et~al.(2021)Zhao, Zhou, Zhu, Zhan, Zhou, Li, Yu, Yuan, and Luo]{zhao2021structural}
Kaifa Zhao, Hao Zhou, Yulin Zhu, Xian Zhan, Kai Zhou, Jianfeng Li, Le~Yu, Wei Yuan, and Xiapu Luo.
\newblock Structural attack against graph based android malware detection.
\newblock In \emph{ACM SIGSAC Conference on Computer and Communications Security (CCS)}, 2021.

\bibitem[Zhu et~al.(2019)Zhu, Xi, Jing, Wu, Xia, and Zhang]{zhumultimodal}
Dali Zhu, Tong Xi, Pengfei Jing, Di~Wu, Qing Xia, and Yiming Zhang.
\newblock A transparent and multimodal malware detection method for android apps.
\newblock In \emph{ACM Conference on Modeling, Analysis and Simulation of Wireless and Mobile Systems (MSWIM)}, 2019.

\end{thebibliography}

\clearpage

\appendix


\section*{Overview of Appendix}
The following appendices provide further information:

\begin{itemize}
    \item \hyperref[app:dataset_stats]{\textbf{A Dataset Statistics}}
    \item \hyperref[app:prior-multimodal-datasets]{\textbf{B Comparison with Prior Multimodal Datasets}}
    \item \hyperref[app:godfather_case_study]{\textbf{C Motivating Case Study: Recent GodFather Variants}}
    \item \hyperref[app:model-descriptions]{\textbf{D Model Descriptions}}
    \item \hyperref[app:fusion_strategies]{\textbf{E Multimodal Fusion Strategies}}
    \item \hyperref[app:full_temporal_results]{\textbf{F Extended Temporal Results}}
    \item \hyperref[app:feature-space-visualization]{\textbf{G Feature-Space Visualization}}
    \item \hyperref[app:feature_importance]{\textbf{H Feature Importance Across Modalities}}
    \item \hyperref[app:missing_modal]{\textbf{I Sensitivity to Missing Modalities}}
    \item \hyperref[app:label_drift]{\textbf{J Prediction Uncertainty Under Label Drift}}
    \item \hyperref[app:extra-cda]{\textbf{K Additional Concept-Drift Adaptation Results}}
    \item \hyperref[app:unsupervised-family-classification]{\textbf{L Unsupervised Family Structure Analysis}}
    \item \hyperref[app:continual-learning]{\textbf{M Continual Learning on \system}}
    \item \hyperref[app:family-temporal-stability]{\textbf{N Malware Family Temporal Stability}}
    \item \hyperref[app:computation]{\textbf{O Computational Resources}}
    \item \hyperref[app:broaderImpacts]{\textbf{P Broader Impacts}}

    \item \hyperref[app:datasetdocument]{\textbf{Q Dataset Documentation}}
\end{itemize}

\setcounter{section}{0}

\section{Dataset Statistics}
\label{app:dataset_stats}

Table~\ref{tab:yearwise_compact_distribution} summarizes the temporal composition of the dataset, reporting the number of malware and benign samples per year along with family-level statistics. Overall, the dataset exhibits substantial variation in both class distribution and family diversity across years, reflecting the evolving nature of the Android malware ecosystem. Earlier years (e.g., 2013–2014) show moderate family diversity, while later years (e.g., 2019–2022) exhibit a significant increase in the number of distinct malware families, indicating higher behavioral heterogeneity. The number of newly observed families also varies considerably across years, highlighting periods of rapid emergence of previously unseen malware variants.

A notable characteristic of the dataset is the large proportion of singleton families—families represented by a single sample within a given year—which suggests a long-tailed distribution of malware. This reflects the presence of many rare or short-lived variants, a common phenomenon in real-world malware dynamics~\cite{lakshmanan2024darkgate}. In contrast, the number of samples labeled as \textit{unknown} remains relatively small, indicating that the family labeling process is generally reliable.

Table~\ref{tab:top10_families} provides a fine-grained view of the temporal dynamics of the ten most prevalent malware families. The distribution reveals strong temporal shifts in family dominance: certain families (e.g., \textit{dowgin}, \textit{airpush}, \textit{kuguo}) are prominent in earlier years, while others exhibit transient spikes or gradual decline over time. This non-stationary behavior highlights the presence of concept drift at the family level, where the relative prevalence of malware families changes significantly across years. Such dynamics underscore the importance of evaluating models under temporally evolving conditions rather than assuming a static data distribution.

\begin{table}[t]
\centering
\caption{Year-wise distribution of malware and benign samples with corresponding malware family number.}
\label{tab:yearwise_compact_distribution}
\begin{tabular}{lrrrrrr}
\toprule
\textbf{Year} & \textbf{Malware} & \textbf{Benign} & \textbf{Valid Families} & \textbf{New Families} & \textbf{Singleton} & \textbf{Unknown} \\
\midrule
2013 & 38,324 & 58,226 & 230 & 230 & 1,411 & 27 \\
2014 & 37,061 & 77,108 & 227 & 94 & 1,950 & 322 \\
2016 & 30,683 & 39,516 & 350 & 159 & 4,545 & 154 \\
2017 & 19,781 & 48,620 & 202 & 87 & 8,894 & 1,068 \\
2018 & 34,254 & 45,688 & 345 & 132 & 19,059 & 1,182 \\
2019 & 46,553 & 32,578 & 624 & 261 & 18,370 & 21 \\
2020 & 41,086 & 25,164 & 558 & 123 & 25,919 & 14 \\
2021 & 24,727 & 28,707 & 286 & 50 & 19,626 & 16 \\
2022 & 44,133 & 36,649 & 641 & 159 & 24,558 & 3 \\
2023 & 5,526 & 40,220 & 199 & 31 & 3,992 & 15 \\
2024 & 760 & 54,869 & 61 & 14 & 604 & 0 \\
2025 & 612 & 48,010 & 73 & 14 & 347 & 2 \\
\bottomrule
\end{tabular}
\end{table}

\begin{table}[t]
\centering
\caption{Year-wise distribution of the top 10 malware families by total sample count, excluding benign, unknown, and singleton samples.}
\label{tab:top10_families}
\scriptsize
\begin{tabular}{lrrrrrrrrrr}
\toprule
\textbf{Year} & \textbf{dowgin} & \textbf{airpush} & \textbf{kuguo} & \textbf{smsreg} & \textbf{dnotua} & \textbf{gappusin} & \textbf{revmob} & \textbf{leadbolt} & \textbf{adwo} & \textbf{youmi} \\
\midrule
2013 & 6,839 \nochange & 5,615 \nochange & 5,295 \nochange & 1,041 \nochange & 2 \nochange & 3,247 \nochange & 296 \nochange & 1,807 \nochange & 2,822 \nochange & 1,004 \nochange \\
2014 & 11,824 \inc & 4,135 \dec & 3,304 \dec & 1,358 \inc & 5 \inc & 931 \dec & 2,391 \inc & 2,040 \inc & 606 \dec & 853 \dec \\
2016 & 2,259 \dec & 2,227 \dec & 4,025 \inc & 1,089 \dec & 35 \inc & 1,659 \inc & 1,869 \dec & 930 \dec & 947 \inc & 1,168 \inc \\
2017 & 32 \dec & 1,151 \dec & 3 \dec & 67 \dec & 3,547 \inc & 3 \dec & 281 \dec & 373 \dec & 6 \dec & 34 \dec \\
2018 & 17 \dec & 1,178 \inc & 8 \inc & 142 \inc & 5,053 \inc & 5 \inc & 170 \dec & 527 \inc & 6 \nochange & 35 \inc \\
2019 & 980 \inc & 988 \dec & 679 \inc & 5,069 \inc & 451 \dec & 676 \inc & 1,552 \inc & 464 \dec & 402 \inc & 321 \inc \\
2020 & 619 \dec & 522 \dec & 533 \dec & 2,245 \dec & 616 \inc & 303 \dec & 67 \dec & 137 \dec & 123 \dec & 347 \inc \\
2021 & 100 \dec & 101 \dec & 118 \dec & 538 \dec & 85 \dec & 118 \dec & 48 \dec & 55 \dec & 39 \dec & 96 \dec \\
2022 & 570 \inc & 159 \inc & 774 \inc & 2,874 \inc & 772 \inc & 336 \inc & 43 \dec & 62 \inc & 96 \inc & 372 \inc \\
2023 & 54 \dec & 50 \dec & 33 \dec & 132 \dec & 7 \dec & 48 \dec & 22 \dec & 18 \dec & 30 \dec & 48 \dec \\
2024 & 0 \dec & 3 \dec & 0 \dec & 7 \dec & 1 \dec & 0 \dec & 0 \dec & 0 \dec & 0 \dec & 0 \dec \\
2025 & 0 \nochange & 6 \inc & 0 \nochange & 5 \dec & 0 \dec & 0 \nochange & 0 \nochange & 2 \inc & 0 \nochange & 0 \nochange \\
\bottomrule
\end{tabular}
\end{table}

\section{\system vs Prior Multimodal Datasets}
\label{app:prior-multimodal-datasets}

\begin{table}[t]
\centering
\tiny
\caption{Representative multimodal datasets across different application sectors.}
\label{tab:multimodal_datasets_sectors}
\resizebox{\textwidth}{!}{
\begin{tabular}{llll|p{4.5cm}}
\hline
\textbf{Sector} & \textbf{Dataset} & \textbf{Modalities / Data Types} & \textbf{\# Samples} & \textbf{Prediction Task} \\
\hline
Cybersecurity 
& \textbf{McNdroid (Ours)} & Static, Dynamic, and Graph structure & 858,859 & Malware detection, temporal generalization, concept-drift detection, distributional shift analysis, intra-family shift, inter-family shift and family classification \\
& MalNet~\cite{malnetCIKM,malnetNeurIPS} & Graph structure, node features & 1,262,024 & Malware detection and classification \\

\hline
Affective Computing & MUSTARD~\cite{castro2019towards} & Text, Video, Audio & 690 & Sarcasm detection \\
 & CMU-MOSI~\cite{zadeh2016mosi} & Text, Video, Audio & 2,199 & Sentiment analysis \\
 & UR-FUNNY~\cite{hasan2019ur} & Text, Video, Audio & 16,514 & Humor detection \\
 & CMU-MOSEI~\cite{zadeh2018multimodal} & Text, Video, Audio & 22,777 & Sentiment and emotion recognition \\
\hline
Healthcare & MIMIC~\cite{johnson2016mimic} & Time-series, Tabular & 36,212 & Mortality and ICD-9 code prediction \\
\hline
Robotics & MuJoCo Push~\cite{lee2020multimodal} & Image, Force, Proprioception & 37,990 & Object pose estimation \\
 & Vision\&Touch~\cite{lee2020making} & Image, Force, Proprioception & 147,000 & Contact and robot pose estimation \\
\hline
Finance & Stocks-F\&B~\cite{liang2021multibench} & Time-series financial indicators & 5,218 & Stock price and volatility prediction \\
 & Stocks-Health~\cite{liang2021multibench} & Time-series financial indicators & 5,218 & Stock price and volatility prediction \\
 & Stocks-Tech~\cite{liang2021multibench} & Time-series financial indicators & 5,218 & Stock price and volatility prediction \\
\hline
Human-Computer Interaction & Enrico~\cite{leiva2020enrico} & Image, Structured layout & 1,460 & Interface design understanding \\
\hline
Multimedia & Kinetics400-S~\cite{kay2017kinetics} & Video, Audio, Optical flow & 2,624 & Human action recognition \\
 & MM-IMDb~\cite{arevalo2017gated} & Text, Image & 25,959 & Movie genre classification \\
 & AV-MNIST~\cite{vielzeuf2018centralnet} & Image, Audio & 70,000 & Digit classification \\
 & Kinetics400-L~\cite{kay2017kinetics} & Video, Audio, Optical flow & 306,245 & Human action recognition \\
\hline
\end{tabular}
}
\end{table}

Table~\ref{tab:multimodal_datasets_sectors} highlights the limitations of existing multimodal datasets for studying temporal multimodal generalization. Prior benchmarks span diverse sectors, including affective computing, healthcare, robotics, finance, human-computer interaction, multimedia, and cybersecurity. However, their evaluation tasks are primarily conventional in-distribution prediction problems, such as sarcasm detection, sentiment and emotion recognition, mortality and ICD-9 code prediction, object/contact/robot pose estimation, stock price and volatility forecasting, interface understanding, action recognition, movie genre classification, digit recognition, and malware classification. These datasets are valuable for studying multimodal fusion and representation learning, but they are not designed to evaluate model aging, concept drift, or distribution shift under naturally evolving deployment conditions.

\system addresses these limitations by extending multimodal benchmarking to Android malware evolution. Unlike prior datasets in Table~\ref{tab:multimodal_datasets_sectors}, \system provides aligned static, dynamic, and graph-structured representations together with temporally ordered tasks for malware detection, family classification, temporal generalization, model aging, concept-drift detection, distribution shift analysis, and intra-/inter-family shift analysis.



\section{Motivating Case Study: Recent GodFather Variants}
\label{app:godfather_case_study}

We use the recent virtualization-based GodFather Android malware reported by Zimperium zLabs~\cite{godfather-casestudy} as an external motivating case study, not as an part of this paper's contribution. This campaign highlights why Android malware detection benefits from multimodal representations. Unlike conventional overlay-based banking malware~\cite{xu2024dva}, recent GodFather variants execute legitimate banking and cryptocurrency applications inside an attacker-controlled virtualized environment. This preserves the appearance of the original app while allowing the malware to intercept user inputs, alter runtime behavior, and exfiltrate sensitive information.

This design exposes complementary evidence across three modalities. Static Drebin-like features reveal the declared attack surface, including suspicious permissions, accessibility-service declarations, manifest components, encoded configuration strings, embedded payloads, target package names, C2 indicators, and virtualization or hooking artifacts. However, static analysis is weakened by anti-analysis techniques such as APK packaging manipulation, JADX-blocking fields, manifest obfuscation, and dynamically shifted logic. Dynamic analysis captures the executed attack workflow, including accessibility abuse, virtualized app launch, user redirection, runtime hooking, C2 communication, payload download, credential interception, and remote-control behavior. Yet these traces may be incomplete when behavior depends on target apps, granted permissions, live C2 infrastructure, regional conditions, delayed execution, or attacker-issued commands. Graph-based features provide the missing structural view by modeling how the host malware, accessibility service, virtual container, proxy activity manager, hooking framework, C2 handler, target-app interceptors, and sensitive source-to-sink paths are coordinated.

Thus, the GodFather case demonstrates the brittleness of single-modality detection. Static features describe what suspicious capabilities are present, dynamic features describe what behavior is observed at runtime, and graph-based features describe how malicious components are organized. Their fusion provides a more complete and explainable representation for detecting advanced Android banking malware under obfuscation, sandbox evasion, and modular attack design.

\section{Model Descriptions}
\label{app:model-descriptions}

\BfPara{Multi-Layer Perceptron (MLP)}
The MLP model is implemented as a four-layer fully connected neural network for malware classification, adapted from prior work~\cite{continual-learning-malware,madar2025}. The hidden layers have dimensions 1024, 512, 256, and 128, respectively. Each hidden layer is followed by batch normalization, ReLU activation, and dropout with a rate of 0.5. For binary classification, the final classifier consists of a linear layer followed by a sigmoid activation, while for multi-class classification, a linear output layer is used. The model is trained using the Adam optimizer with a learning rate of 0.001 and a batch size of 512.

\BfPara{LightGBM}
In addition to MLP, we employ LightGBM~\cite{lightgbm}, a gradient-boosted decision tree ensemble, for both binary and multi-class classification. The model is trained with up to 5000 estimators and a learning rate of 0.02. Each tree is allowed up to 256 leaves to provide sufficient model capacity. To mitigate overfitting, we apply 80\% subsampling of both rows and features, along with L1 and L2 regularization. Early stopping is used based on validation performance, halting training if no improvement is observed for 100 rounds. For binary classification, the Area Under the Curve (AUC) metric is used, while multi-class classification uses multi-class log loss. These configurations follow standard practices of prior works in malware detection such as EMBER~\cite{ember} and TESSERACT~\cite{tesseract}.

\BfPara{XGBoost}
The XGBoost model is configured with a maximum tree depth of 12 and a learning rate of 0.05. We use the histogram-based tree construction method with GPU acceleration for efficient training. The objective function is binary logistic loss for binary classification and soft-probability output for multi-class classification~\cite{chen2016xgboost}. The model is trained for up to 3000 boosting rounds, with additional regularization through row and feature subsampling.

\BfPara{Support Vector Machine (SVM)}
A linear Support Vector Machine is implemented using \texttt{LinearSVC}. The model is trained with a maximum of 20{,}000 iterations and a tolerance of $10^{-3}$ ~\cite{chen2023continuous}. To obtain probabilistic outputs required for ranking-based evaluation metrics, the classifier is calibrated using \texttt{CalibratedClassifierCV} with sigmoid calibration. A \texttt{MaxAbsScaler} is applied within a pipeline to ensure compatibility with high-dimensional sparse features.

\BfPara{DetectBERT}
DetectBERT is implemented as a lightweight transformer-based model for malware classification. The input feature vector is first projected into a hidden representation using a fully connected layer, followed by a learnable \texttt{[CLS]} token. The resulting sequence is processed by two stacked transformer layers utilizing Nyström-based self-attention~\cite{xiong2021nystromformer}, with residual connections and LayerNorm. The model supports multiple aggregation strategies, including \texttt{[CLS]}-based representation, summation, averaging, and random token selection. The final representation is passed through a linear classification head.

\BfPara{Transformer-based}
We utilize a Vision Transformer (ViT)-based model adapted for malware classification using static feature vectors~\cite{haque2025lamda}. Each input sample is projected into a hidden-dimensional token via a linear embedding layer. A learnable \texttt{[CLS]} token is prepended, and positional embeddings are added. The resulting sequence is processed through a stack of transformer encoder blocks, each consisting of LayerNorm, multi-head self-attention, and a feed-forward network with GELU activation and residual connections. The final \texttt{[CLS]} representation is used for classification through a LayerNorm and linear output layer, enabling both binary and fine-grained multi-class malware classification.

\section{Multimodal Fusion Strategies}
\label{app:fusion_strategies}

We consider four model-level strategies for integrating the three malware feature modalities: static, graph-based, and dynamic features. The first three strategies use conventional classifiers over fused feature spaces or prediction scores, while the fourth learns modality-specific embeddings and explicitly models cross-modal interactions through attention.

\subsection{Pairwise Feature Fusion}
\label{app:pairwise_fusion}

Pairwise fusion evaluates whether two modalities provide complementary information when modeled jointly. For each modality pair \((m_i, m_j)\), the corresponding feature vectors are concatenated into a single input representation:
\(
x^{(i,j)} = [x^{(i)} \Vert x^{(j)}].
\)
We instantiate pairwise fusion for Static+Dynamic, Static+Graph-based, and Dynamic+Graph-based, training a separate classifier on each concatenated feature space. This provides a controlled way to assess complementary information between two modalities without requiring all three simultaneously. The selected feature are horizontally stacked, the classifier is fit on the fused representation.

\subsection{Late Fusion}
\label{app:late_fusion}

Late fusion combines modalities at the decision level by training one base classifier per modality and concatenating their predicted probabilities into a three-dimensional meta-representation:
\[
z = [p_{\text{Static}}, p_{\text{Graph-based}}, p_{\text{Dynamic}}].
\]
A stacker is then trained on out-of-fold predictions, which prevents reliance on inflated in-sample scores. At test time, it maps the three modality probabilities to the final prediction.

\subsection{Cross-Attention Fusion}
\label{app:cross_attention_fusion}

Cross-attention fusion replaces direct feature concatenation with a learned representation-level interaction mechanism~\cite{kim2018multimodal}. Each modality is first processed by an MLP encoder~\cite{continual-learning-malware}:
\(
h_m = E_m(x^{(m)}),
\)
where \(E_m\) denotes the modality-specific encoder for modality \(m\), and \(h_m\) is the corresponding dense modality token. Static, Graph-based, and Dynamic modalities are projected into a shared latent space using separate feed-forward encoders composed of linear transformations, normalization, nonlinear activation, and dropout. A learned modality embedding is added to each projected token to preserve modality identity in the shared representation space.

The fusion module performs directional cross-attention. Each modality token serves as a query and attends to the other two modality tokens as keys and values:
\[
\tilde{h}_{\text{Static}} =
\text{Attn}_{\text{Static}}(h_{\text{Static}}, [h_{\text{Graph-based}}, h_{\text{Dynamic}}]),
\]
\[
\tilde{h}_{\text{Graph-based}} =
\text{Attn}_{\text{Graph-based}}(h_{\text{Graph-based}}, [h_{\text{Static}}, h_{\text{Dynamic}}]),
\]
\[
\tilde{h}_{\text{Dynamic}} =
\text{Attn}_{\text{Dynamic}}(h_{\text{Dynamic}}, [h_{\text{Static}}, h_{\text{Graph-based}}]).
\]
This design allows each modality to selectively incorporate complementary evidence from the other modalities while maintaining its own query-specific representation. The attended context is combined with the original modality token using a residual connection and then normalized. The three resulting modality-specific representations are concatenated to form the final fused representation, which is normalized once more before being passed to the classifier.

\section{Extended Temporal Results}
\label{app:full_temporal_results}

Tables~\ref{tab:f1_all_multimodal} and~\ref{tab:rocauc_all_multimodal} report the complete F1-score and ROC-AUC results for all multimodal fusion strategies across IID, NEAR, and FAR temporal splits. A subset of these results is analyzed in the main paper; here, we provide the full model- and metric-level comparison. Under IID evaluation, multimodal methods achieve uniformly strong performance, indicating that all modality combinations are effective when the test distribution is temporally close to the training distribution. In contrast, performance degrades substantially under NEAR and FAR splits. Among the fusion strategies, Static+Graph-based provides the clearest gains under long-term drift, achieving the strongest FAR F1-score for several model families, including LightGBM, XGBoost, DetectBERT, and ViT. However, this trend is not universal, as Static+Dynamic and late fusion remain competitive for specific classifiers. These results indicate that multimodal fusion utilizes complementary information from static, dynamic, and graph-based representations, but its effectiveness depends on both the modality pairing and the downstream model.

\begin{table*}[!t]
\centering
\scriptsize
\caption{Comparison of multimodal strategies across temporal splits using F1-score.}
\label{tab:f1_all_multimodal}
\resizebox{\textwidth}{!}{%
\begin{tabular}{ll|cccccc}
\toprule
\textbf{Split} & \textbf{Model} & \textbf{Static + Dynamic} 
& \textbf{Static + GML} 
& \textbf{Dynamic + Graph-based} 
& \textbf{Late Fusion} 
& \textbf{Feature Fusion} & \textbf{Cross-Attention} \\
\midrule
\multirow{7}{*}{IID} & MLP & \meanstd{89.3}{8.3} & \meanstd{83.9}{6.7} & \meanstd{80.6}{11.6} & \meanstd{90.7}{7.8} & \meanstd{84.2}{9.3} & -- \\
 & LightGBM & \meanstd{90.6}{8.7} & \meanstd{91.9}{7.4} & \meanstd{91.4}{7.2} & \meanstd{92.6}{6.4} & \meanstd{91.6}{7.6} & -- \\
 & XGBoost & \meanstd{90.7}{8.9} & \meanstd{92.1}{7.1} & \meanstd{91.5}{7.2} & \meanstd{91.7}{6.6} & \meanstd{91.6}{7.8} & -- \\
 & SVM & \meanstd{89.8}{7.6} & \meanstd{89.5}{8.9} & \meanstd{87.1}{8.7} & \meanstd{90.5}{7.0} & \meanstd{89.2}{9.4} & -- \\
 & DetectBERT & \meanstd{89.4}{7.9} & \meanstd{89.9}{5.7} & \meanstd{87.1}{7.2} & \meanstd{90.7}{8.3} & \meanstd{89.9}{5.6} & -- \\
 & ViT & \meanstd{89.0}{8.2} & \meanstd{91.0}{5.8} & \meanstd{87.6}{6.9} & \meanstd{90.5}{7.9} & \meanstd{90.0}{6.3} & -- \\
 & Cross-Attention & -- & -- & -- & -- & -- & \meanstd{90.0}{8.6} \\
\midrule
\multirow{7}{*}{NEAR} & MLP & \meanstd{42.6}{37.5} & \meanstd{31.4}{41.9} & \meanstd{29.5}{25.9} & \meanstd{41.5}{38.7} & \meanstd{30.1}{37.2} & -- \\
 & LightGBM & \meanstd{39.7}{40.5} & \meanstd{39.5}{44.2} & \meanstd{36.9}{39.5} & \meanstd{36.5}{42.9} & \meanstd{39.4}{42.8} & -- \\
 & XGBoost & \meanstd{39.2}{39.6} & \meanstd{39.4}{42.5} & \meanstd{37.3}{37.6} & \meanstd{39.5}{40.2} & \meanstd{39.6}{41.5} & -- \\
 & SVM & \meanstd{46.2}{37.7} & \meanstd{39.3}{39.8} & \meanstd{33.3}{35.4} & \meanstd{41.3}{42.3} & \meanstd{39.3}{41.2} & -- \\
 & DetectBERT & \meanstd{44.0}{34.5} & \meanstd{36.1}{43.8} & \meanstd{41.1}{36.1} & \meanstd{42.8}{41.7} & \meanstd{38.8}{42.1} & -- \\
 & ViT & \meanstd{44.8}{38.5} & \meanstd{36.4}{42.6} & \meanstd{39.5}{36.3} & \meanstd{41.9}{41.1} & \meanstd{38.6}{44.1} & -- \\
 & Cross-Attention & -- & -- & -- & -- & -- & \meanstd{40.8}{40.7} \\
\midrule
\multirow{7}{*}{FAR} & MLP & \meanstd{28.2}{18.2} & \meanstd{15.6}{13.1} & \meanstd{15.0}{11.2} & \meanstd{29.0}{22.5} & \meanstd{20.3}{16.0} & -- \\
 & LightGBM & \meanstd{37.7}{27.3} & \meanstd{48.0}{29.0} & \meanstd{41.3}{21.6} & \meanstd{31.9}{17.1} & \meanstd{45.4}{30.4} & -- \\
 & XGBoost & \meanstd{33.2}{22.4} & \meanstd{42.7}{27.5} & \meanstd{27.2}{20.2} & \meanstd{33.6}{20.3} & \meanstd{42.0}{30.3} & -- \\
 & SVM & \meanstd{41.6}{23.5} & \meanstd{34.1}{19.2} & \meanstd{29.5}{18.0} & \meanstd{31.8}{20.4} & \meanstd{40.2}{22.6} & -- \\
 & DetectBERT & \meanstd{36.7}{23.6} & \meanstd{38.1}{27.1} & \meanstd{27.9}{19.6} & \meanstd{28.7}{21.7} & \meanstd{35.3}{24.2} & -- \\
 & ViT & \meanstd{34.2}{21.7} & \meanstd{43.4}{27.9} & \meanstd{25.6}{16.1} & \meanstd{31.1}{18.9} & \meanstd{33.7}{21.6} & -- \\
 & Cross-Attention & -- & -- & -- & -- & -- & \meanstd{33.0}{19.9} \\
\bottomrule
\end{tabular}%
}
\end{table*}

\begin{table*}[!t]
\centering
\scriptsize
\caption{Comparison of multimodal strategies across temporal splits using ROC-AUC.}
\label{tab:rocauc_all_multimodal}
\resizebox{\textwidth}{!}{%
\begin{tabular}{ll|cccccc}
\toprule
\textbf{Split} & \textbf{Model} & \textbf{Static + Dynamic} 
& \textbf{Static + Graph-based} 
& \textbf{Dynamic + Graph-based} 
& \textbf{Late Fusion} 
& \textbf{Feature Fusion} & \textbf{Cross-Attention} \\
\midrule
\multirow{7}{*}{IID} & MLP & \meanstd{97.3}{2.4} & \meanstd{95.9}{2.0} & \meanstd{93.4}{4.3} & \meanstd{98.3}{1.4} & \meanstd{95.8}{2.9} & -- \\
 & LightGBM & \meanstd{98.3}{1.8} & \meanstd{98.7}{1.2} & \meanstd{98.1}{1.9} & \meanstd{98.7}{1.2} & \meanstd{98.6}{1.4} & -- \\
 & XGBoost & \meanstd{98.1}{2.0} & \meanstd{98.8}{1.1} & \meanstd{98.0}{2.1} & \meanstd{97.8}{2.0} & \meanstd{98.5}{1.6} & -- \\
 & SVM & \meanstd{97.0}{2.8} & \meanstd{96.4}{3.8} & \meanstd{95.5}{3.8} & \meanstd{98.1}{1.3} & \meanstd{96.3}{4.0} & -- \\
 & DetectBERT & \meanstd{96.3}{3.4} & \meanstd{96.1}{3.5} & \meanstd{94.5}{4.3} & \meanstd{98.1}{1.8} & \meanstd{95.6}{3.7} & -- \\
 & ViT & \meanstd{97.4}{2.2} & \meanstd{97.9}{1.4} & \meanstd{95.3}{3.6} & \meanstd{98.2}{1.5} & \meanstd{96.7}{2.7} & -- \\
 & Cross-Attention & -- & -- & -- & -- & -- & \meanstd{97.8}{2.1} \\
\midrule
\multirow{7}{*}{NEAR} & MLP & \meanstd{77.9}{18.9} & \meanstd{66.0}{26.6} & \meanstd{66.0}{17.0} & \meanstd{80.0}{16.7} & \meanstd{64.4}{22.5} & -- \\
 & LightGBM & \meanstd{78.9}{20.7} & \meanstd{73.9}{28.3} & \meanstd{86.1}{7.9} & \meanstd{82.4}{16.0} & \meanstd{80.7}{18.9} & -- \\
 & XGBoost & \meanstd{81.1}{17.0} & \meanstd{79.3}{20.8} & \meanstd{83.3}{11.3} & \meanstd{80.8}{13.4} & \meanstd{81.7}{17.4} & -- \\
 & SVM & \meanstd{78.0}{17.1} & \meanstd{75.3}{16.0} & \meanstd{78.8}{8.5} & \meanstd{79.1}{17.8} & \meanstd{77.5}{14.4} & -- \\
 & DetectBERT & \meanstd{78.2}{16.2} & \meanstd{82.4}{7.8} & \meanstd{80.1}{8.2} & \meanstd{80.7}{16.9} & \meanstd{82.6}{7.9} & -- \\
 & ViT & \meanstd{80.8}{14.0} & \meanstd{74.6}{23.0} & \meanstd{72.7}{11.4} & \meanstd{77.9}{19.8} & \meanstd{76.6}{16.3} & -- \\
 & Cross-Attention & -- & -- & -- & -- & -- & \meanstd{73.1}{25.9} \\
\midrule
\multirow{7}{*}{FAR} & MLP & \meanstd{76.7}{10.1} & \meanstd{72.8}{10.3} & \meanstd{69.1}{6.8} & \meanstd{79.3}{13.1} & \meanstd{69.8}{10.8} & -- \\
 & LightGBM & \meanstd{79.7}{15.7} & \meanstd{79.1}{19.2} & \meanstd{82.0}{14.1} & \meanstd{81.9}{14.3} & \meanstd{80.9}{17.1} & -- \\
 & XGBoost & \meanstd{80.1}{14.6} & \meanstd{80.3}{17.4} & \meanstd{77.6}{14.5} & \meanstd{79.6}{13.8} & \meanstd{80.1}{16.8} & -- \\
 & SVM & \meanstd{79.8}{11.0} & \meanstd{83.6}{11.9} & \meanstd{82.7}{9.2} & \meanstd{82.8}{11.1} & \meanstd{83.7}{11.9} & -- \\
 & DetectBERT & \meanstd{75.9}{12.4} & \meanstd{84.0}{12.8} & \meanstd{71.5}{9.1} & \meanstd{79.9}{12.9} & \meanstd{77.0}{9.2} & -- \\
 & ViT & \meanstd{76.6}{9.8} & \meanstd{69.2}{17.4} & \meanstd{56.1}{5.5} & \meanstd{79.1}{13.4} & \meanstd{62.6}{6.2} & -- \\
 & Cross-Attention & -- & -- & -- & -- & -- & \meanstd{75.6}{13.7} \\
\bottomrule
\end{tabular}%
}
\end{table*}

\section{Feature-Space Visualization}
\label{app:feature-space-visualization}
We use UMAP visualizations to assess the class-wise organization of learned representations across years and modalities. For each modality, we train an MLP using 2013 samples and extract 128-dimensional penultimate-layer embeddings from the fixed 2013-trained model for each year-specific test set. To obtain a consistent visualization space, we fit PCA with 50 retained components followed by UMAP on the 2013 training embeddings and apply the resulting fixed PCA--UMAP transformation to all test samples. This reference-fixed protocol avoids fitting a separate visualization for each year, ensuring that the observed year-wise malware--benign overlap, separation, and density patterns are measured in a common representation space induced by the source-year model. These results indicate that temporal drift affects each modality differently, motivating multimodal evaluation under modality-specific distribution shift. They also suggest a promising direction for future work: incorporating semantic API similarity, as explored by APIGraph~\cite{apigraph}, into graph-based representations to improve robustness against API-level and framework-level evolution.



\begin{figure*}[t]
\centering

\begin{subfigure}[t]{0.155\textwidth}
    \centering
    \includegraphics[width=\linewidth]{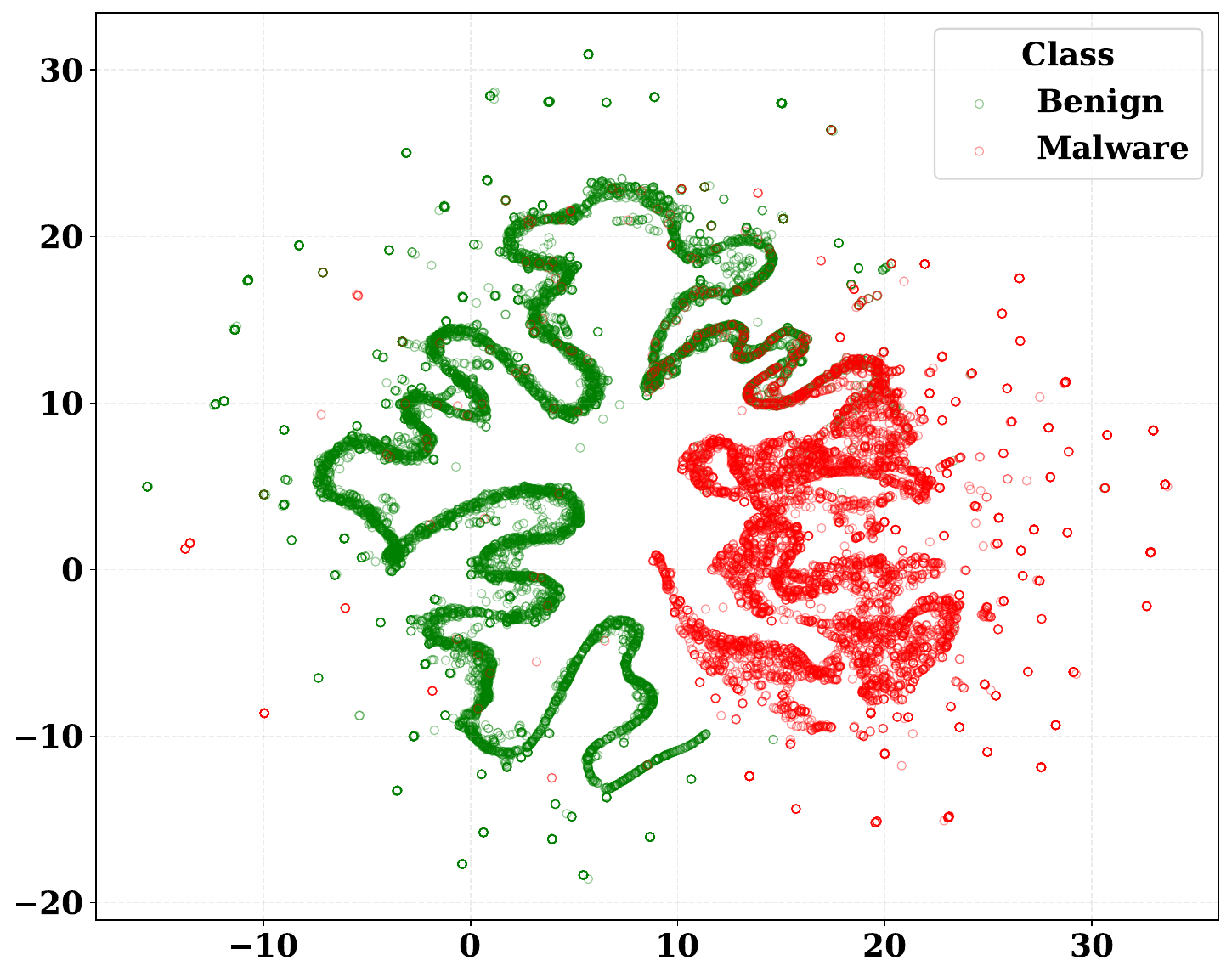}
    \caption{2013}
    \label{fig:data-year-2013}
\end{subfigure}
\begin{subfigure}[t]{0.155\textwidth}
    \centering
    \includegraphics[width=\linewidth]{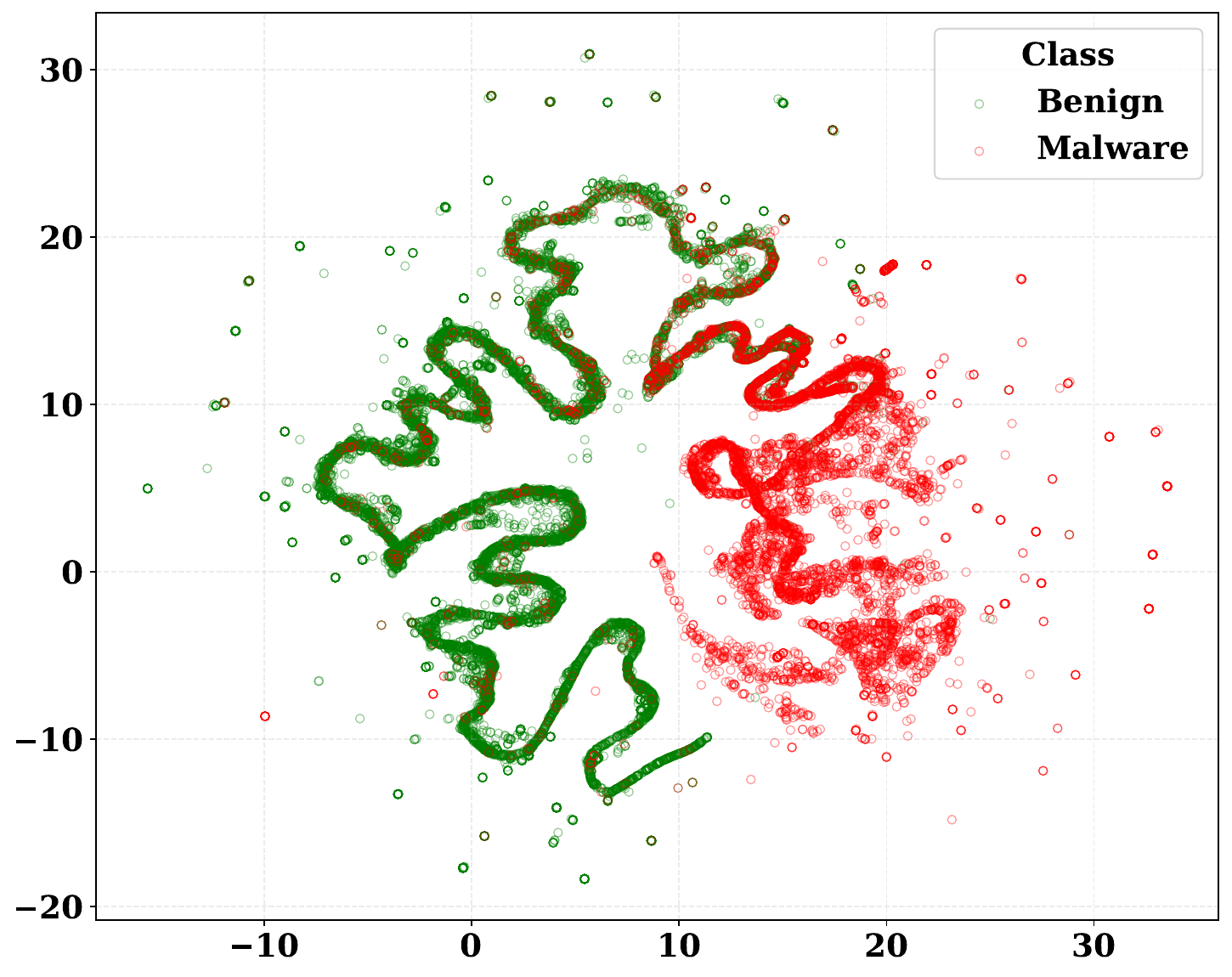}
    \caption{2014}
    \label{fig:data-year-2014}
\end{subfigure}
\begin{subfigure}[t]{0.155\textwidth}
    \centering
    \includegraphics[width=\linewidth]{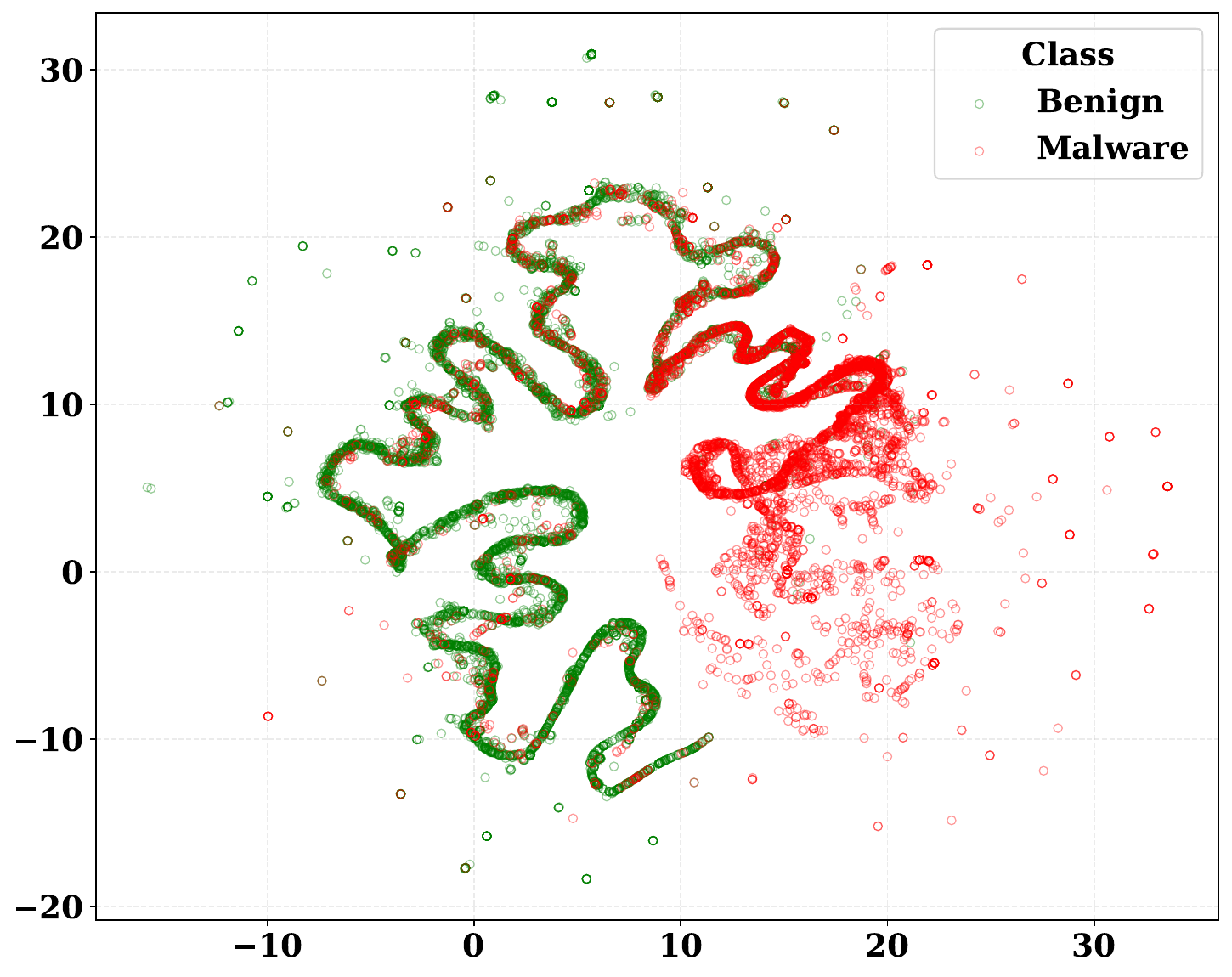}
    \caption{2016}
    \label{fig:data-year-2016}
\end{subfigure}
\begin{subfigure}[t]{0.155\textwidth}
    \centering
    \includegraphics[width=\linewidth]{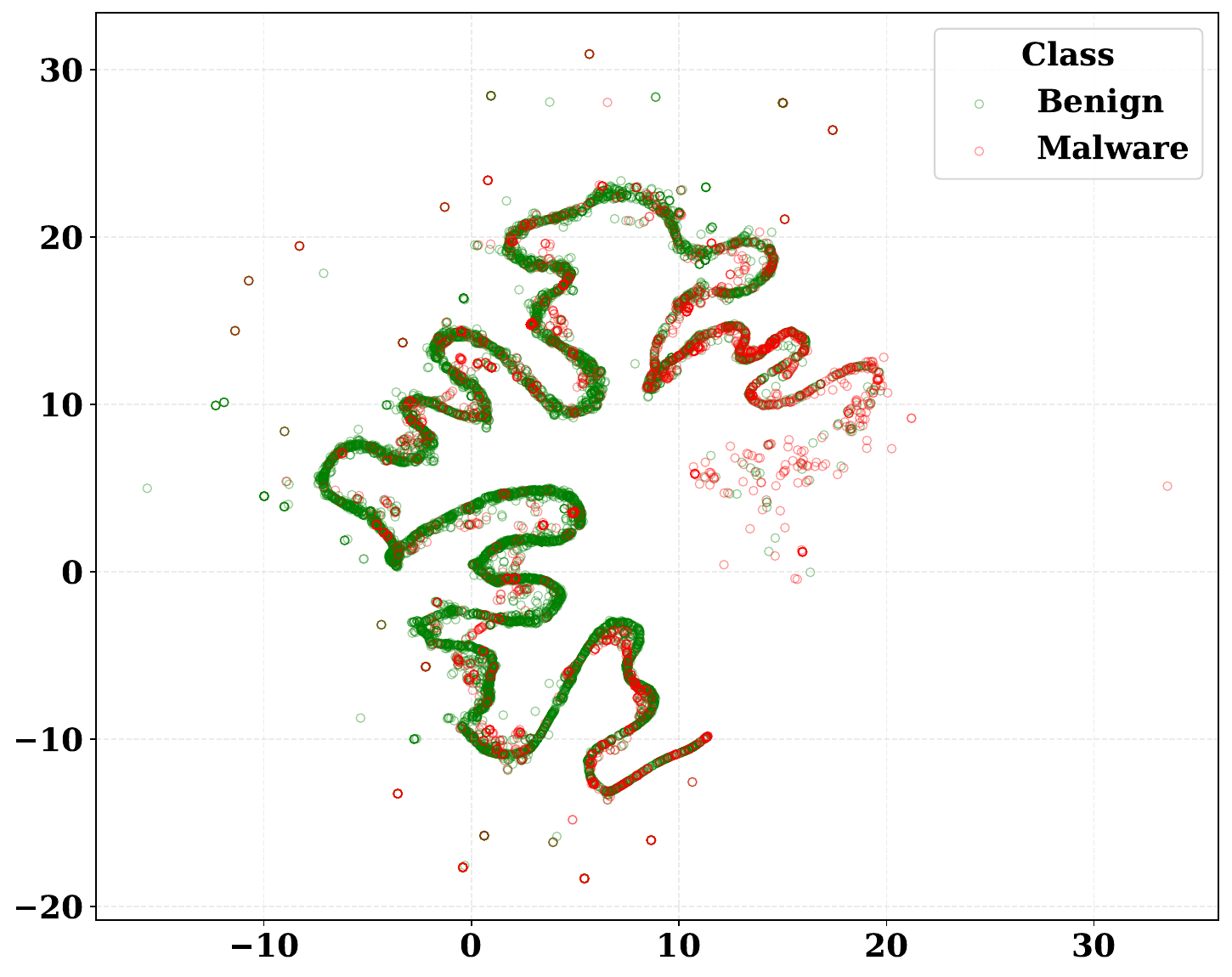}
    \caption{2017}
    \label{fig:data-year-2017}
\end{subfigure}
\begin{subfigure}[t]{0.155\textwidth}
    \centering
    \includegraphics[width=\linewidth]{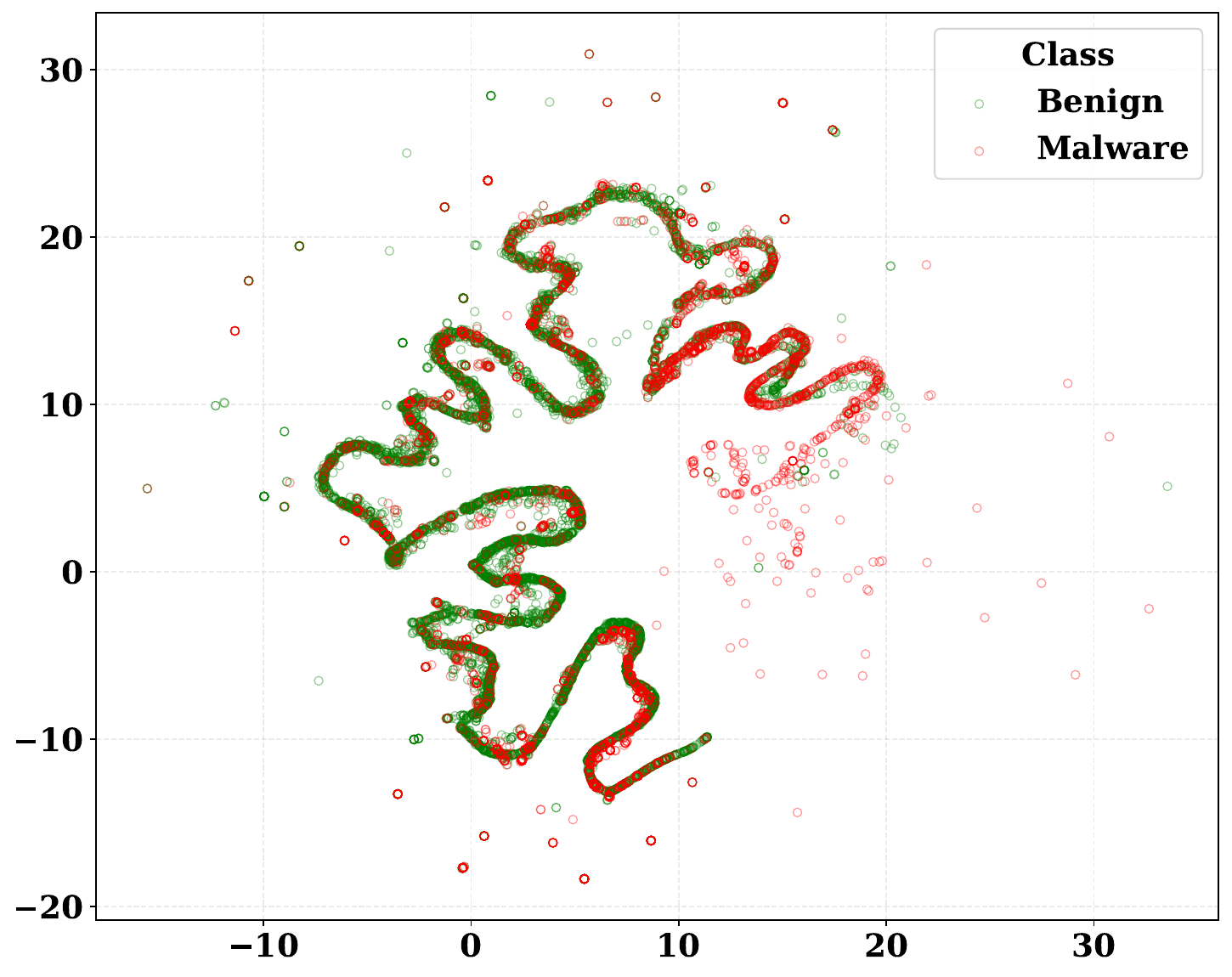}
    \caption{2018}
    \label{fig:data-year-2018}
\end{subfigure}
\begin{subfigure}[t]{0.155\textwidth}
    \centering
    \includegraphics[width=\linewidth]{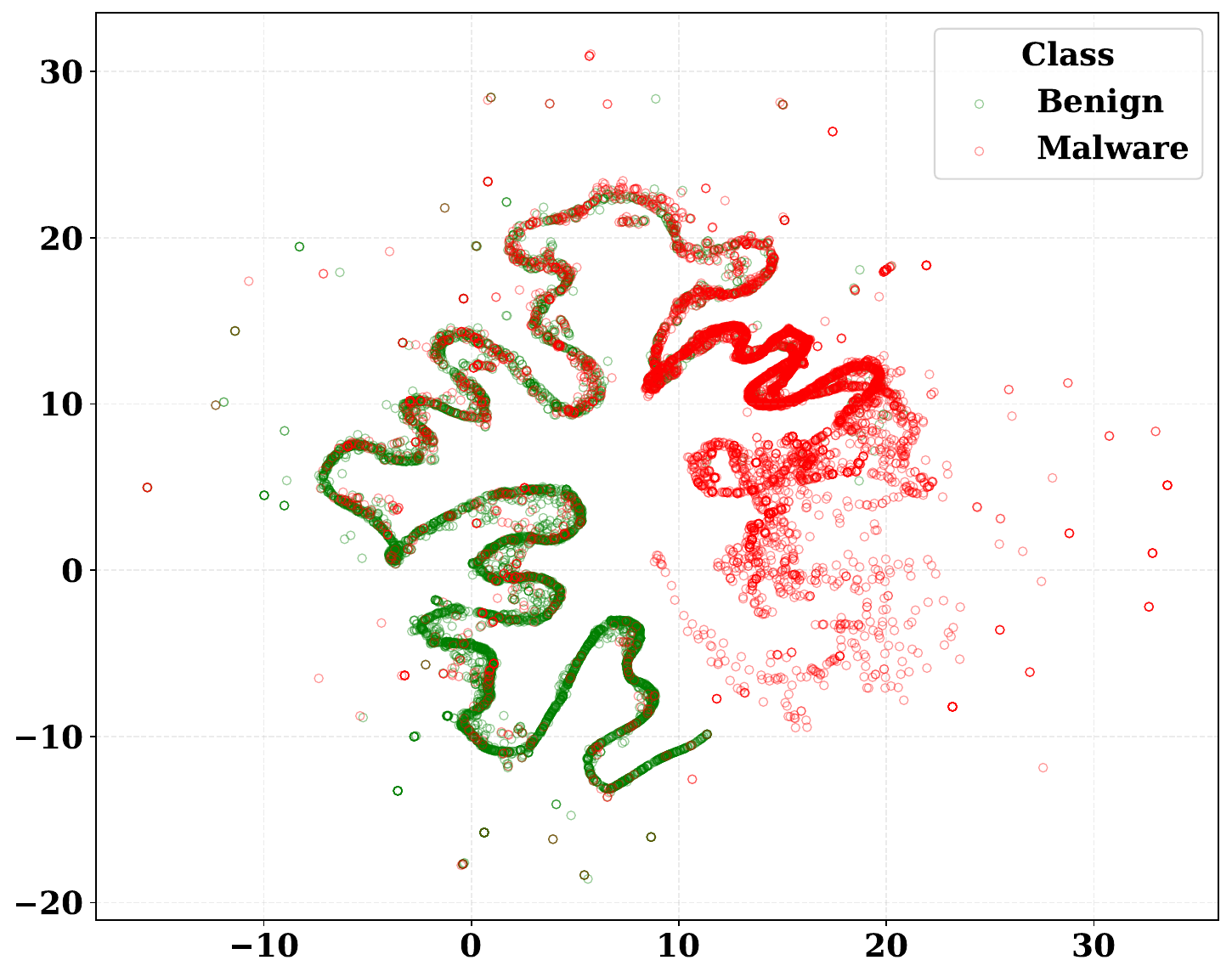}
    \caption{2019}
    \label{fig:data-year-2019}
\end{subfigure}

\medskip

\begin{subfigure}[t]{0.155\textwidth}
    \centering
    \includegraphics[width=\linewidth]{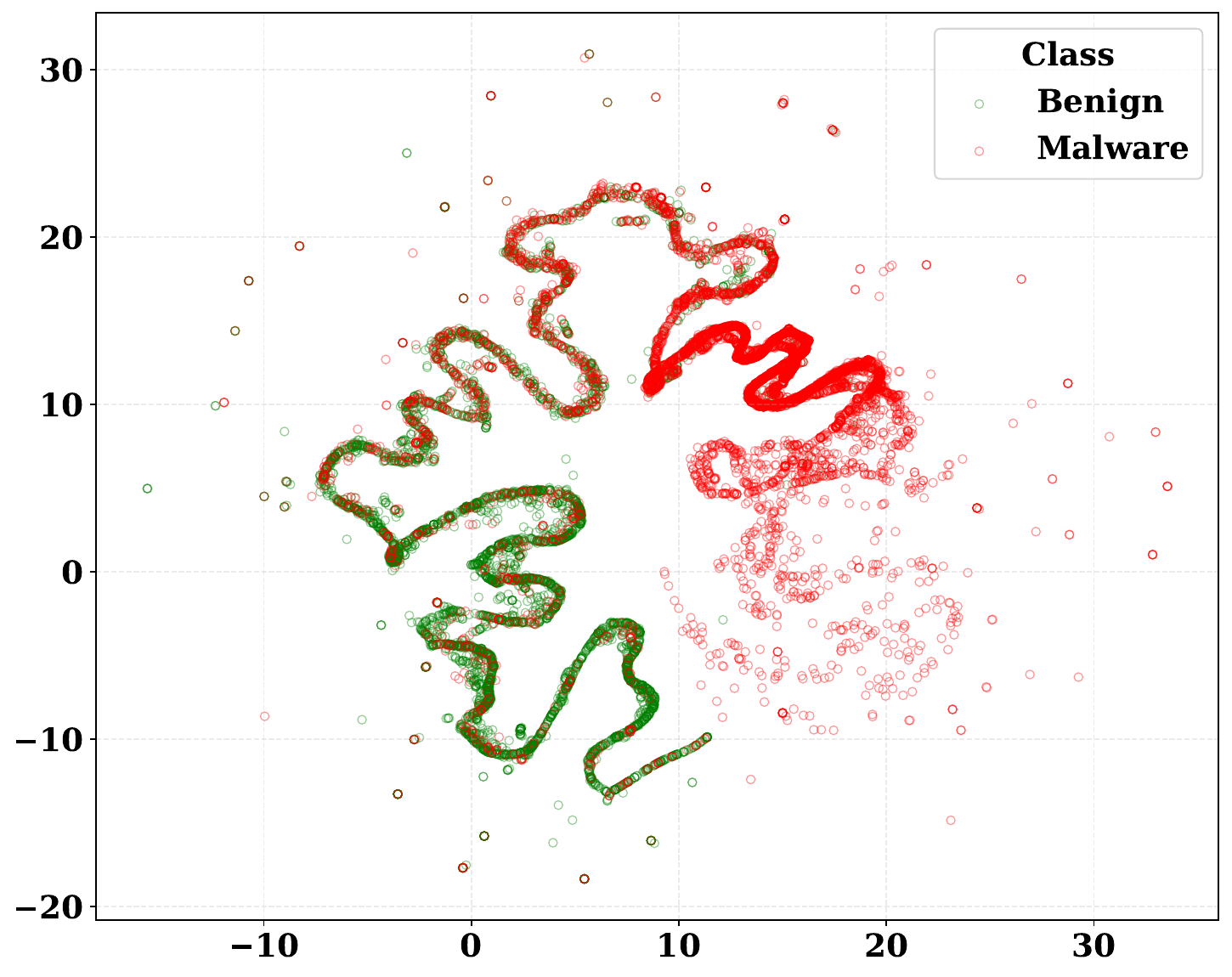}
    \caption{2020}
    \label{fig:data-year-2020}
\end{subfigure}
\begin{subfigure}[t]{0.155\textwidth}
    \centering
    \includegraphics[width=\linewidth]{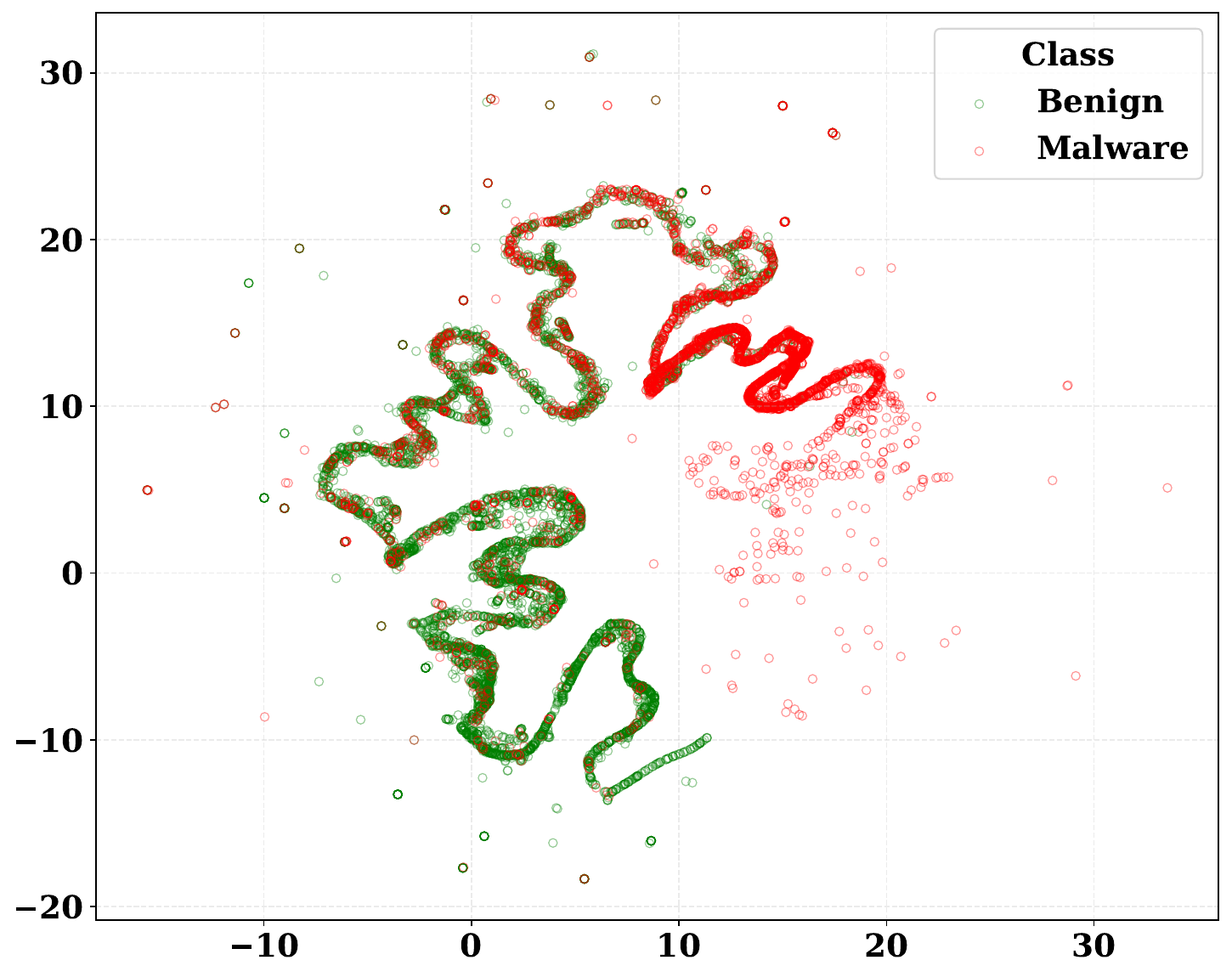}
    \caption{2021}
    \label{fig:data-year-2021}
\end{subfigure}
\begin{subfigure}[t]{0.155\textwidth}
    \centering
    \includegraphics[width=\linewidth]{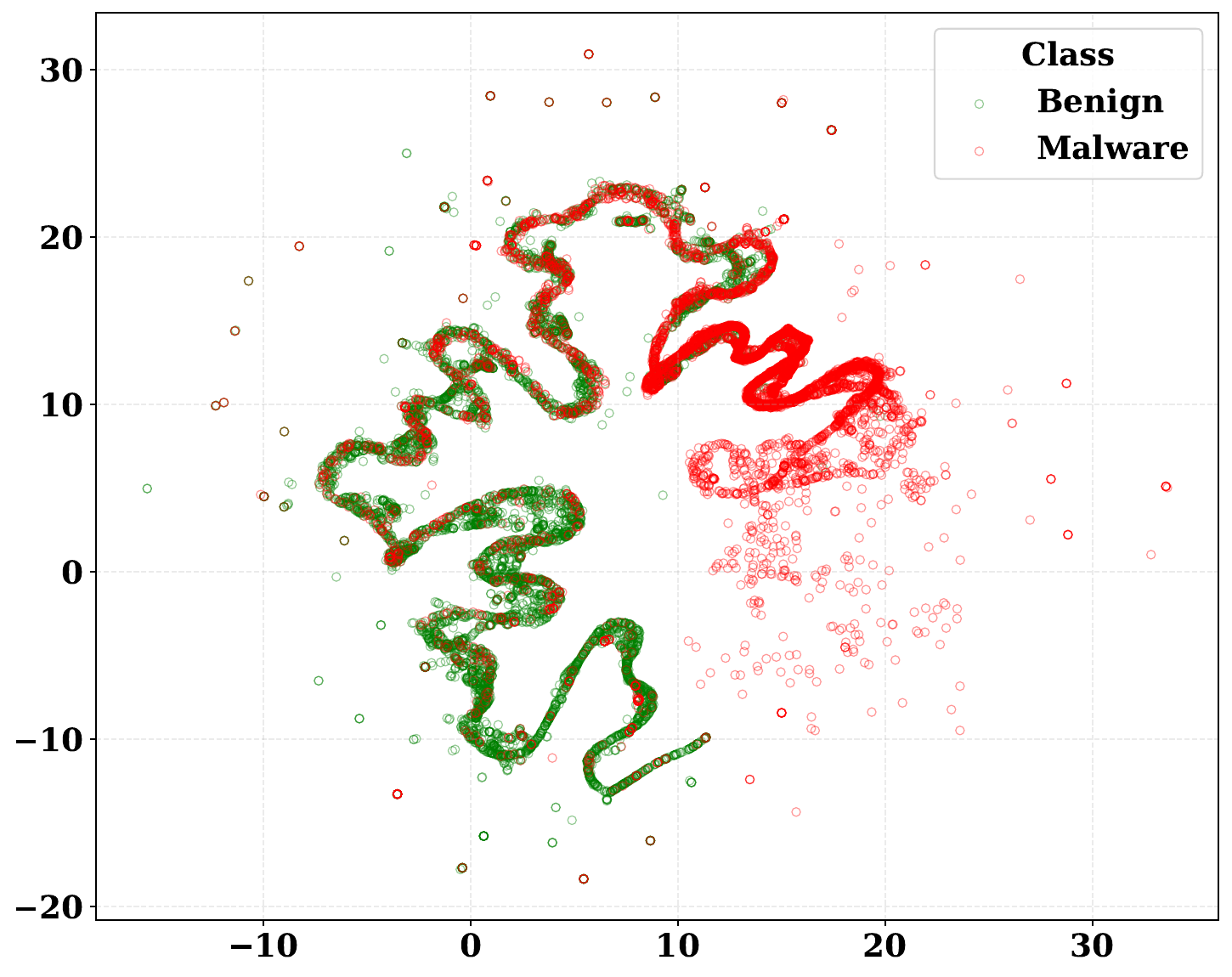}
    \caption{2022}
    \label{fig:data-year-2022}
\end{subfigure}
\begin{subfigure}[t]{0.155\textwidth}
    \centering
    \includegraphics[width=\linewidth]{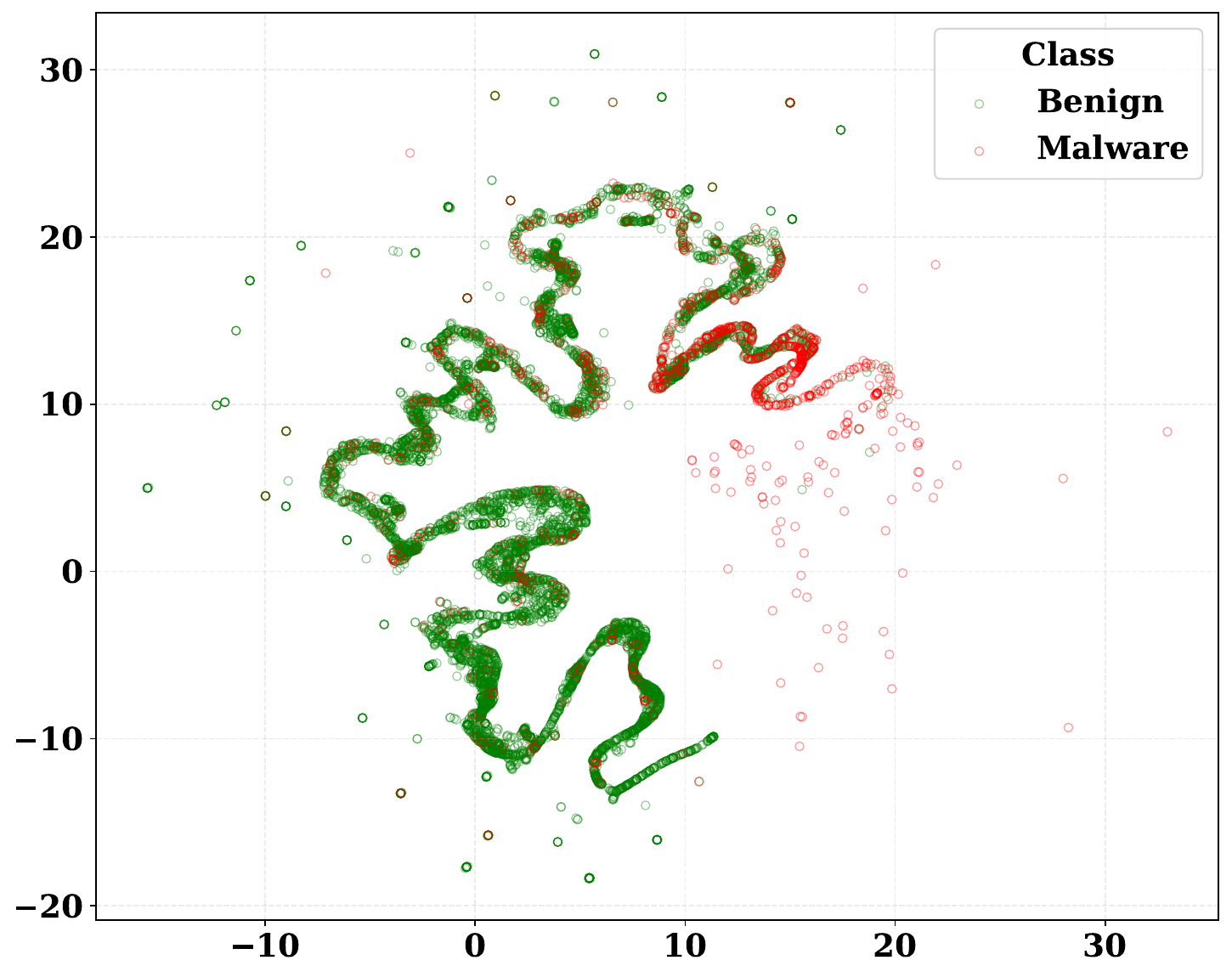}
    \caption{2023}
    \label{fig:data-year-2023}
\end{subfigure}
\begin{subfigure}[t]{0.155\textwidth}
    \centering
    \includegraphics[width=\linewidth]{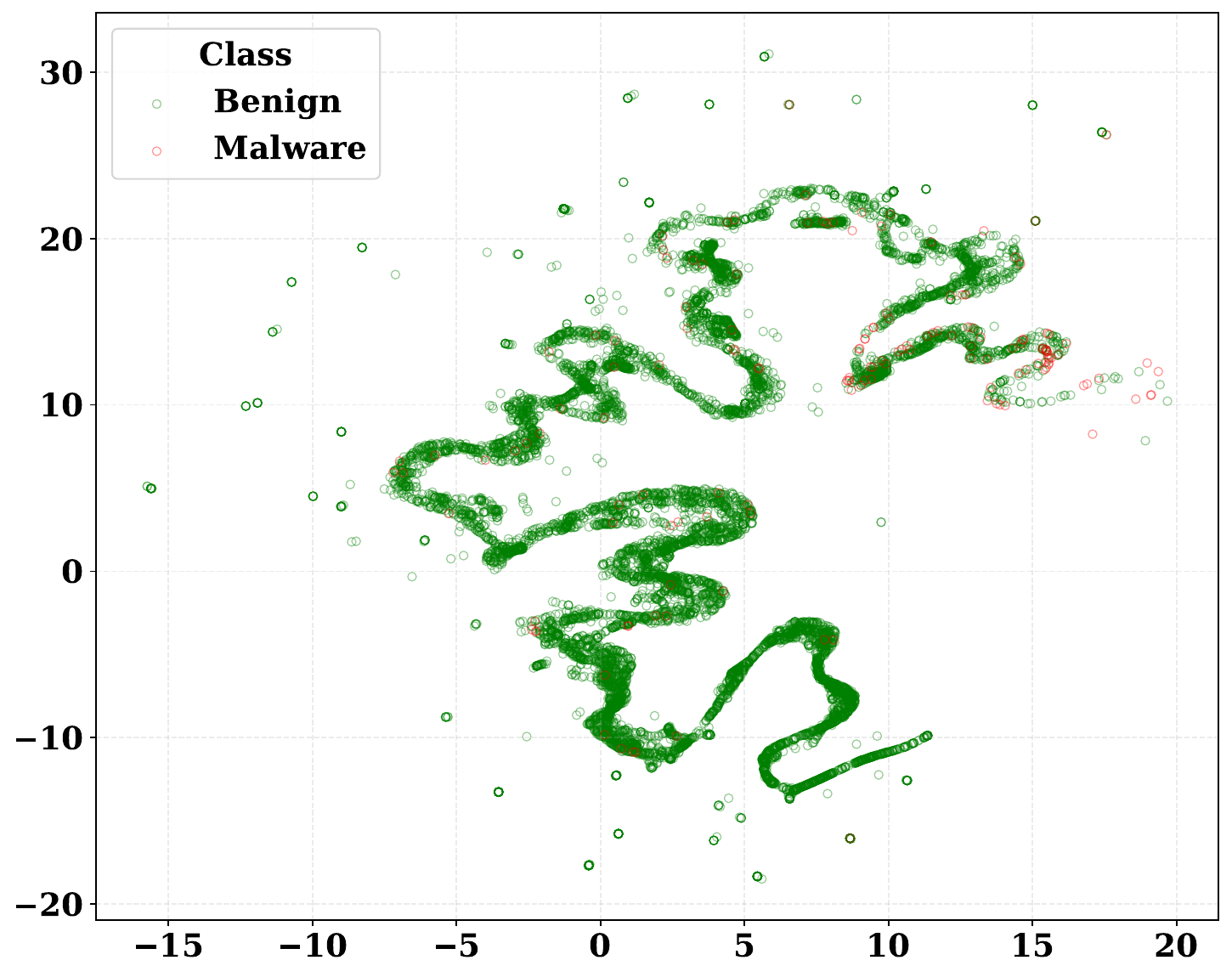}
    \caption{2024}
    \label{fig:data-year-2024}
\end{subfigure}
\begin{subfigure}[t]{0.155\textwidth}
    \centering
    \includegraphics[width=\linewidth]{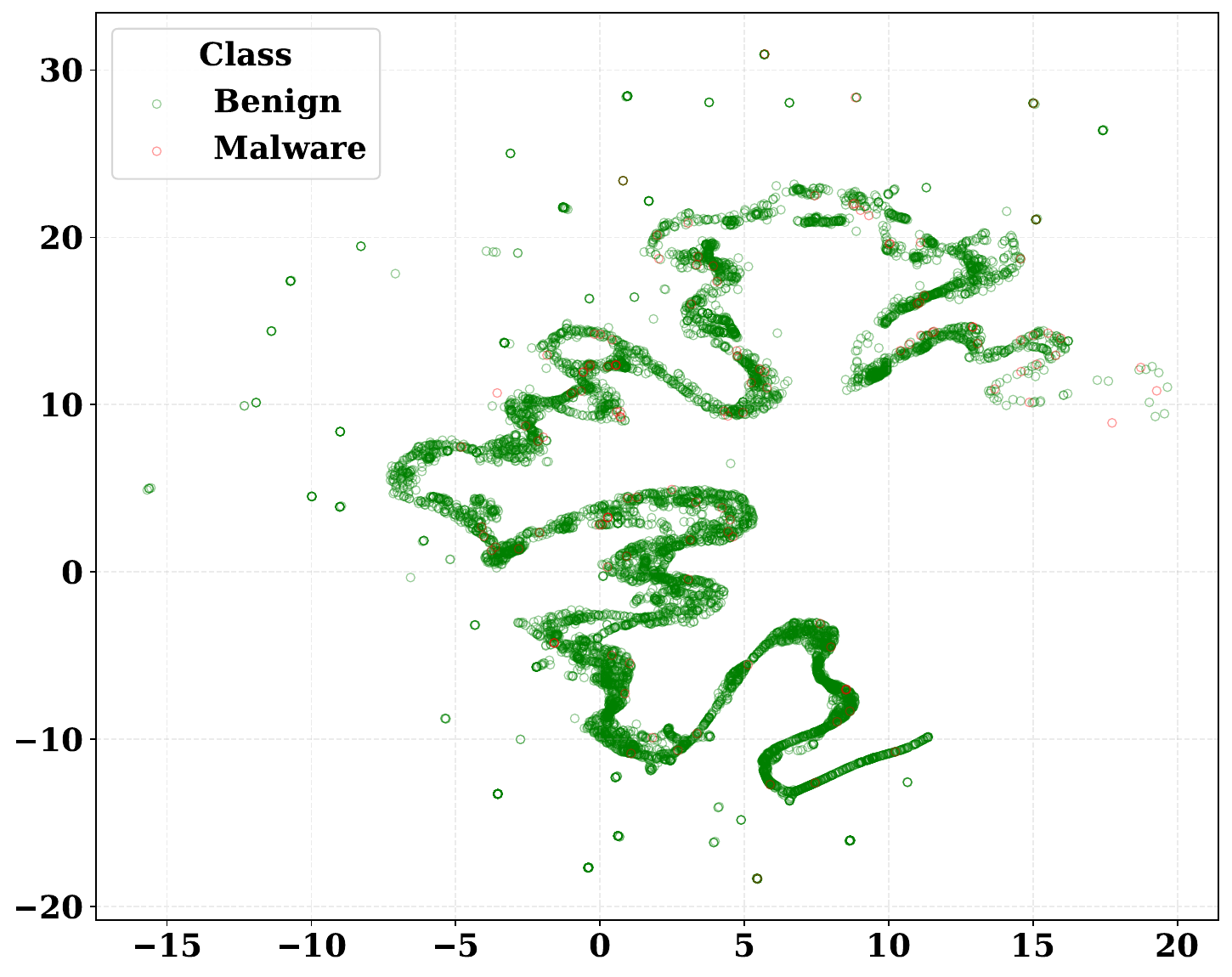}
    \caption{2025}
    \label{fig:data-year-2025}
\end{subfigure}

\caption{Year-wise visualization of the static feature space for \textcolor{red}{malware} and \textcolor{green!50!black}{benign} samples from 2013 to 2025, excluding 2015.}

\label{fig:data-year-feat-class}
\end{figure*}

\begin{figure*}[t]
\centering

\begin{subfigure}[t]{0.155\textwidth}
    \centering
    \includegraphics[width=\linewidth]{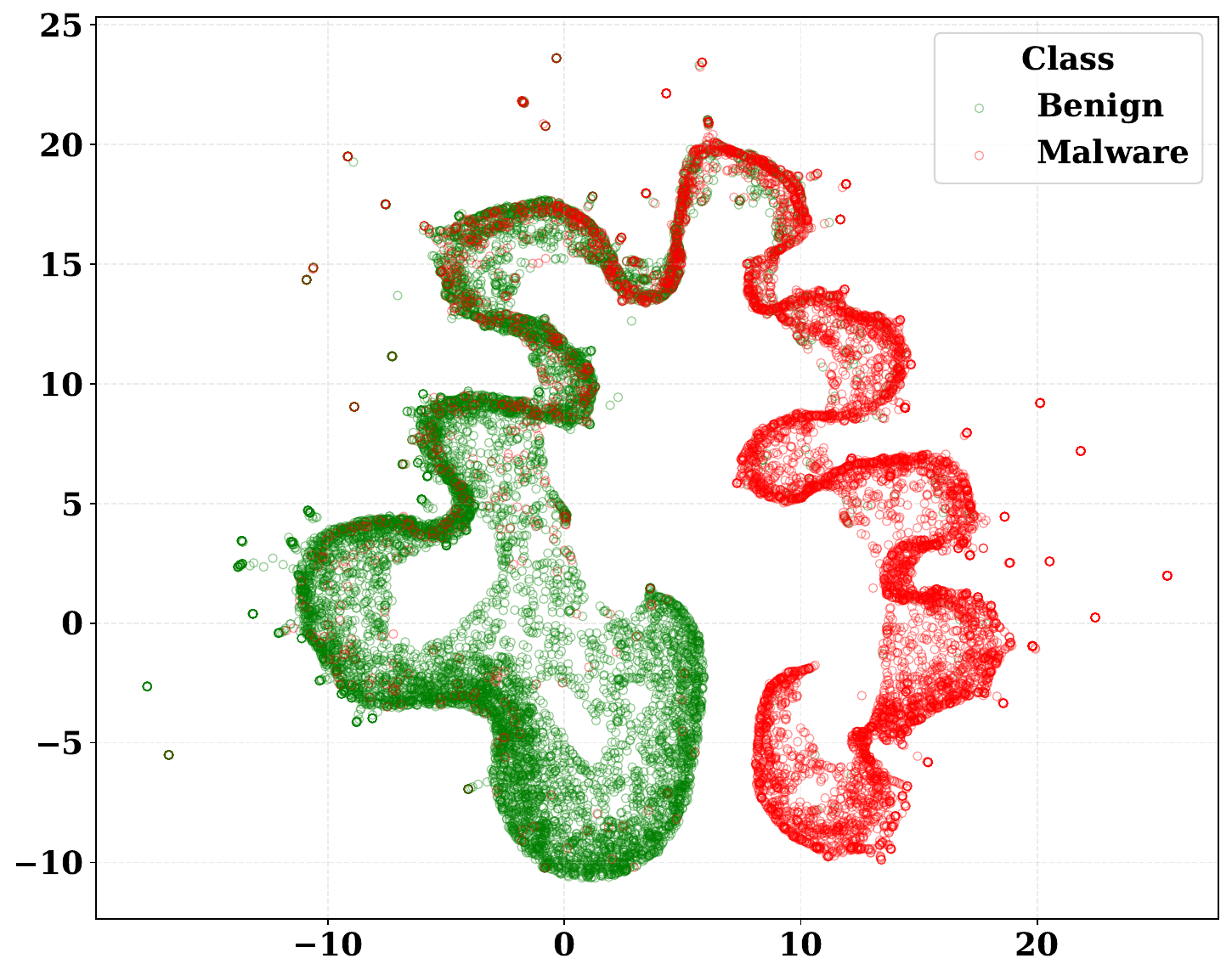}
    \caption{2013}
    \label{fig:json-year-2013}
\end{subfigure}
\begin{subfigure}[t]{0.155\textwidth}
    \centering
    \includegraphics[width=\linewidth]{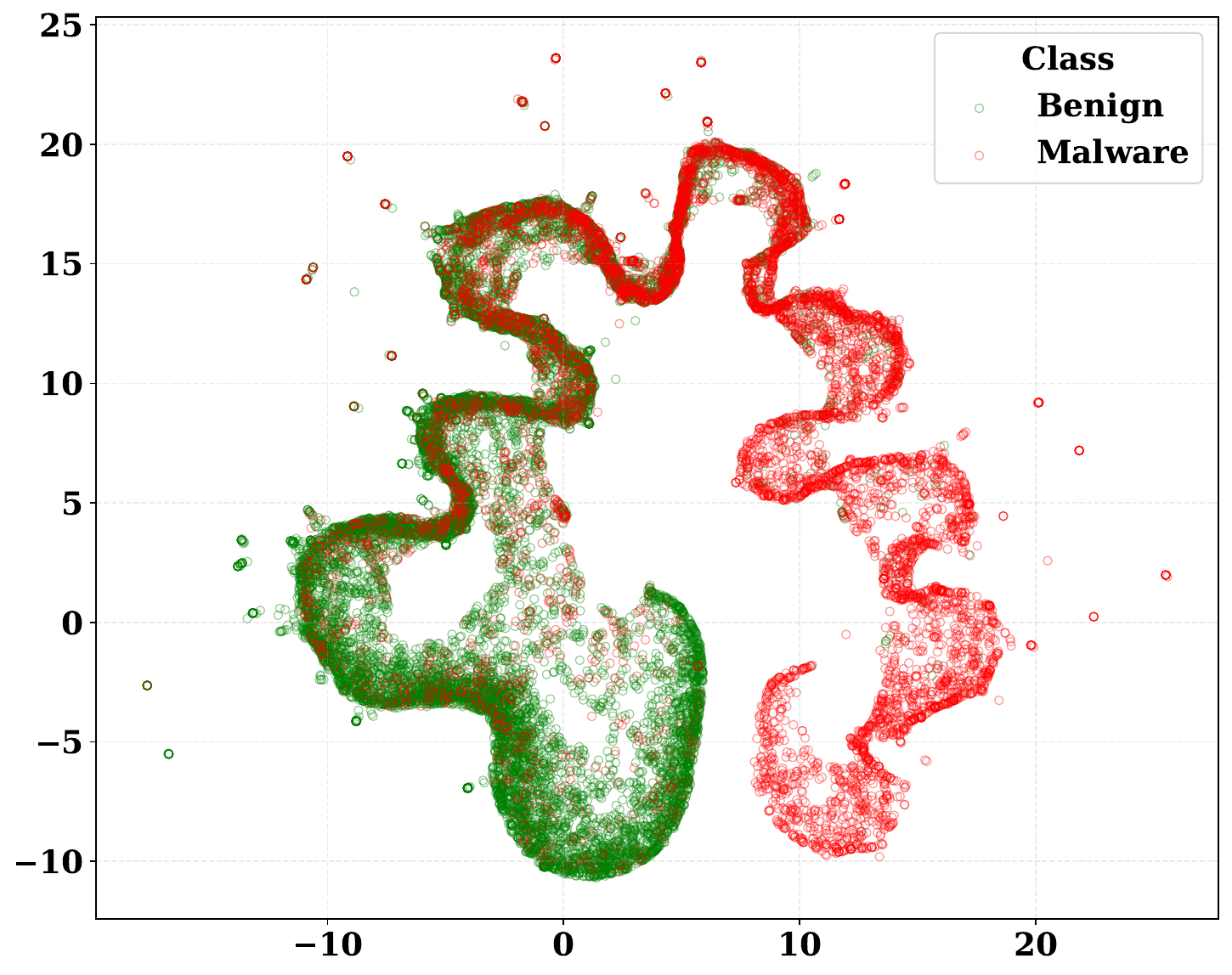}
    \caption{2014}
    \label{fig:json-year-2014}
\end{subfigure}
\begin{subfigure}[t]{0.155\textwidth}
    \centering
    \includegraphics[width=\linewidth]{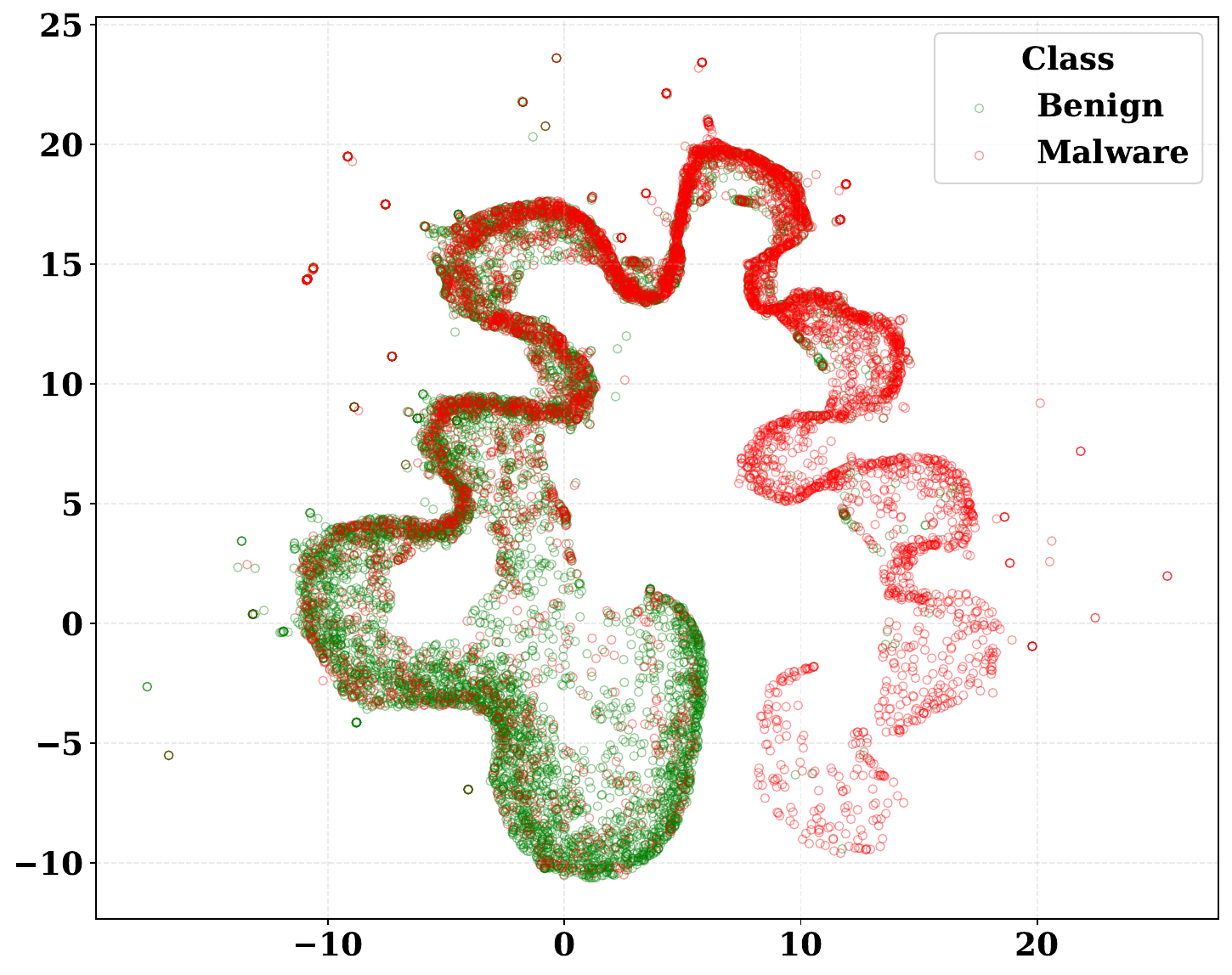}
    \caption{2016}
    \label{fig:json-year-2016}
\end{subfigure}
\begin{subfigure}[t]{0.155\textwidth}
    \centering
    \includegraphics[width=\linewidth]{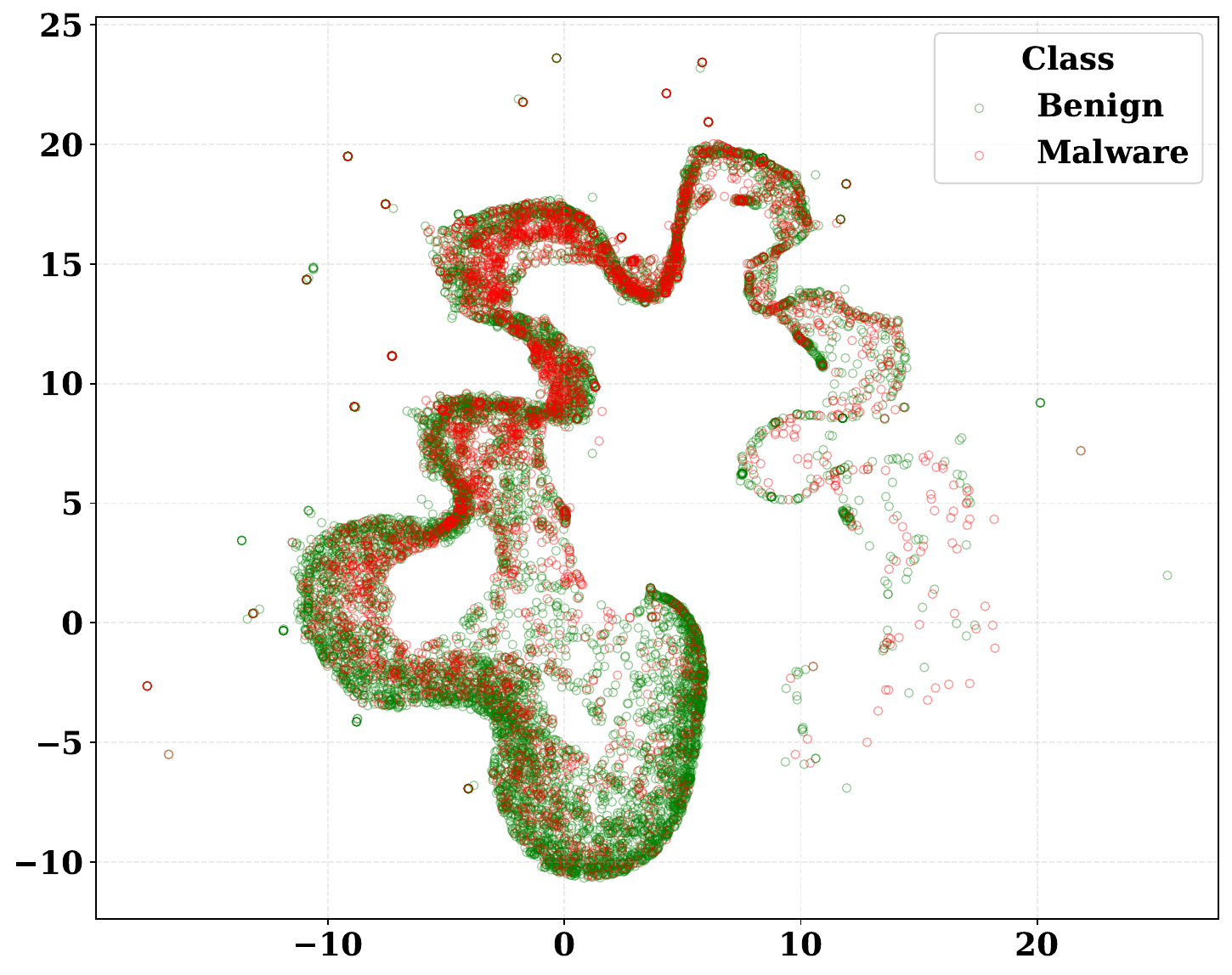}
    \caption{2017}
    \label{fig:json-year-2017}
\end{subfigure}
\begin{subfigure}[t]{0.155\textwidth}
    \centering
    \includegraphics[width=\linewidth]{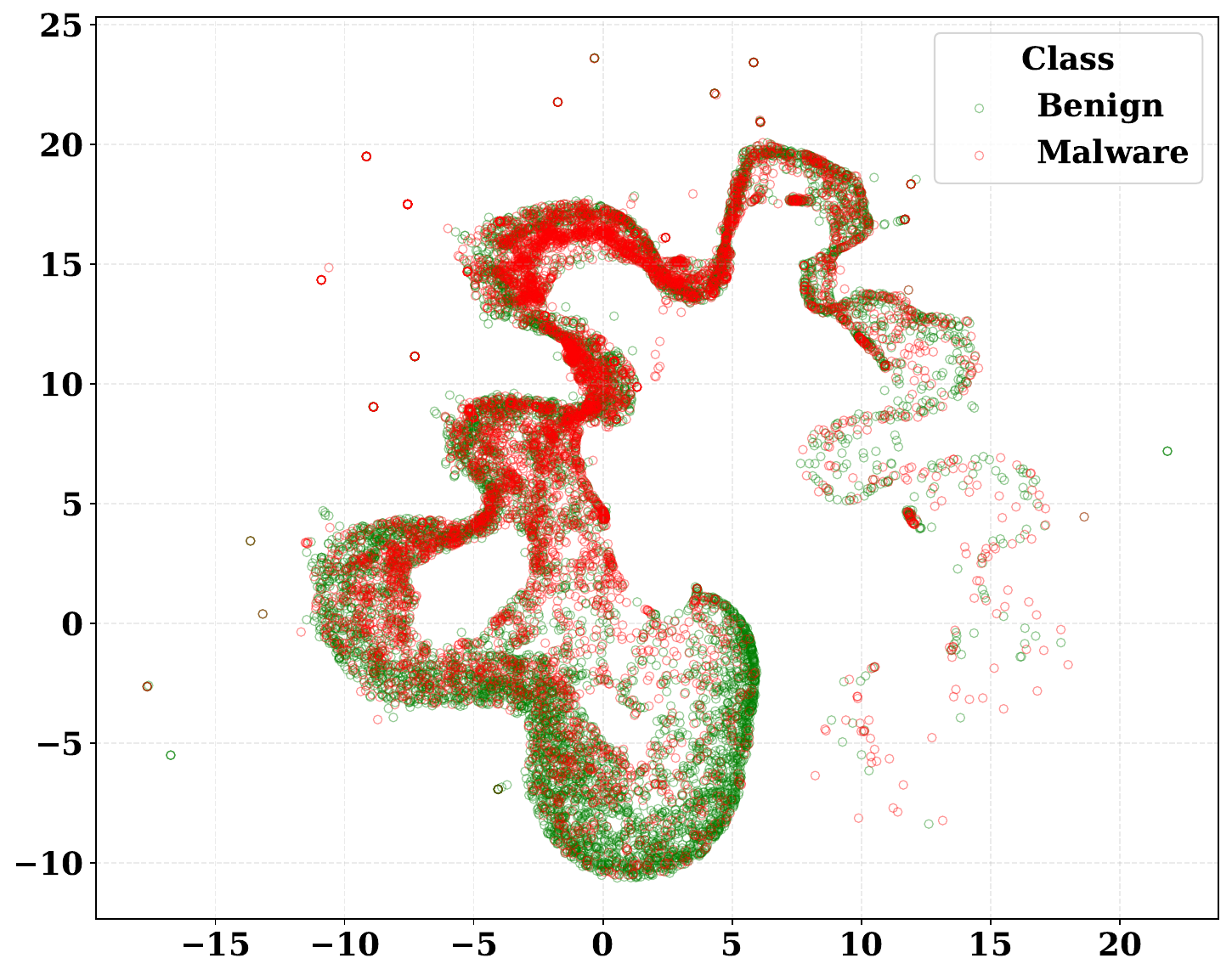}
    \caption{2018}
    \label{fig:json-year-2018}
\end{subfigure}
\begin{subfigure}[t]{0.155\textwidth}
    \centering
    \includegraphics[width=\linewidth]{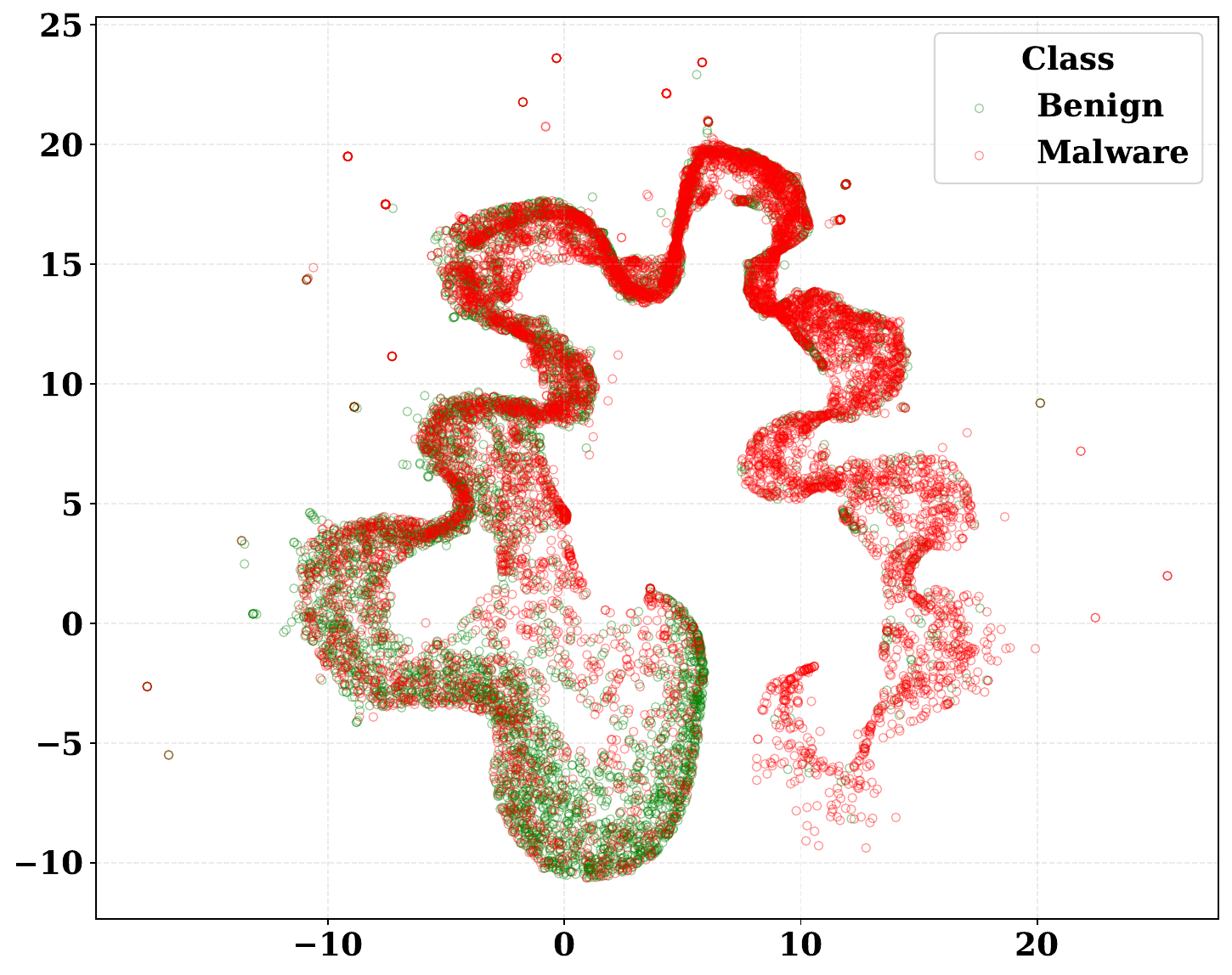}
    \caption{2019}
    \label{fig:json-year-2019}
\end{subfigure}

\medskip

\begin{subfigure}[t]{0.155\textwidth}
    \centering
    \includegraphics[width=\linewidth]{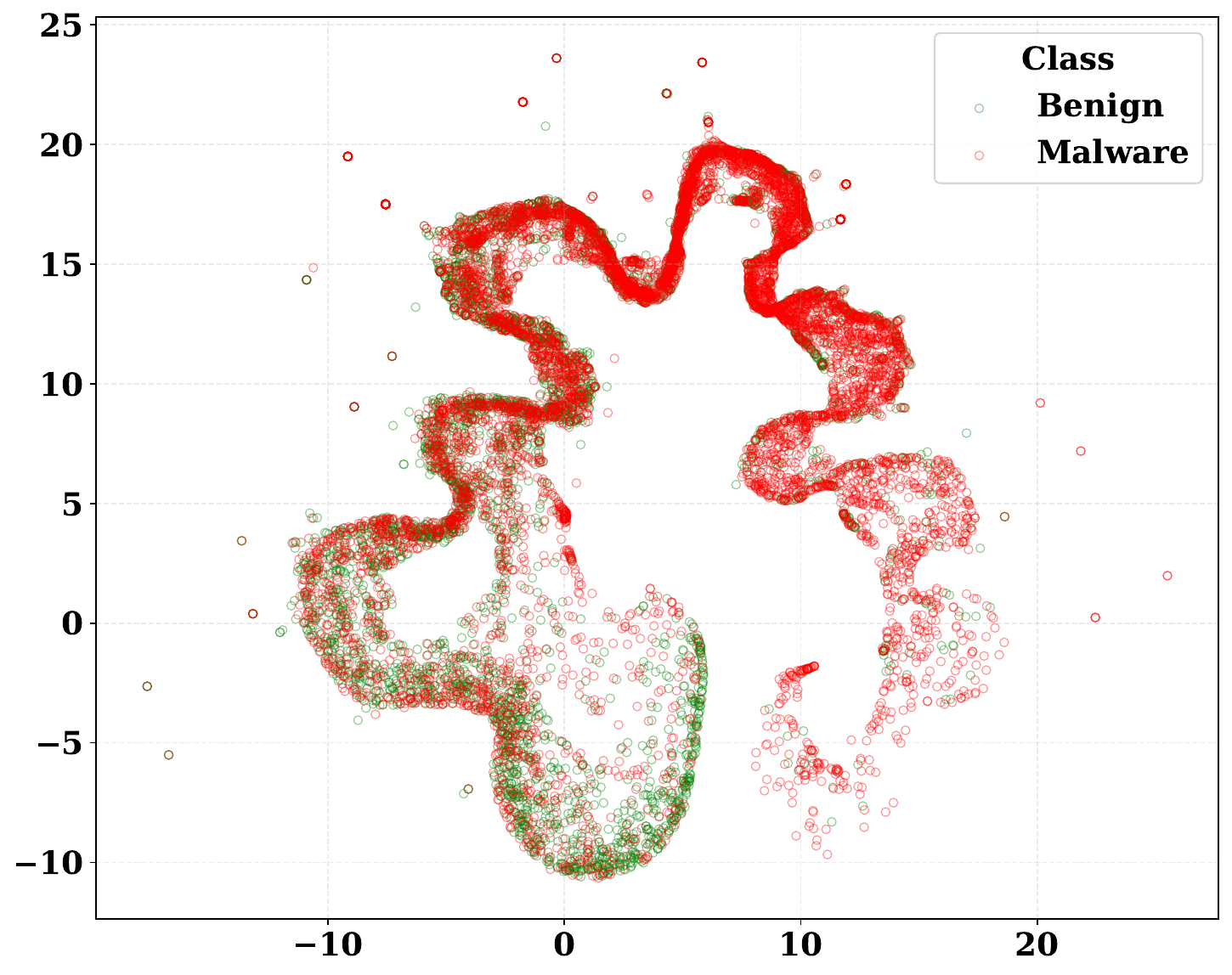}
    \caption{2020}
    \label{fig:json-year-2020}
\end{subfigure}
\begin{subfigure}[t]{0.155\textwidth}
    \centering
    \includegraphics[width=\linewidth]{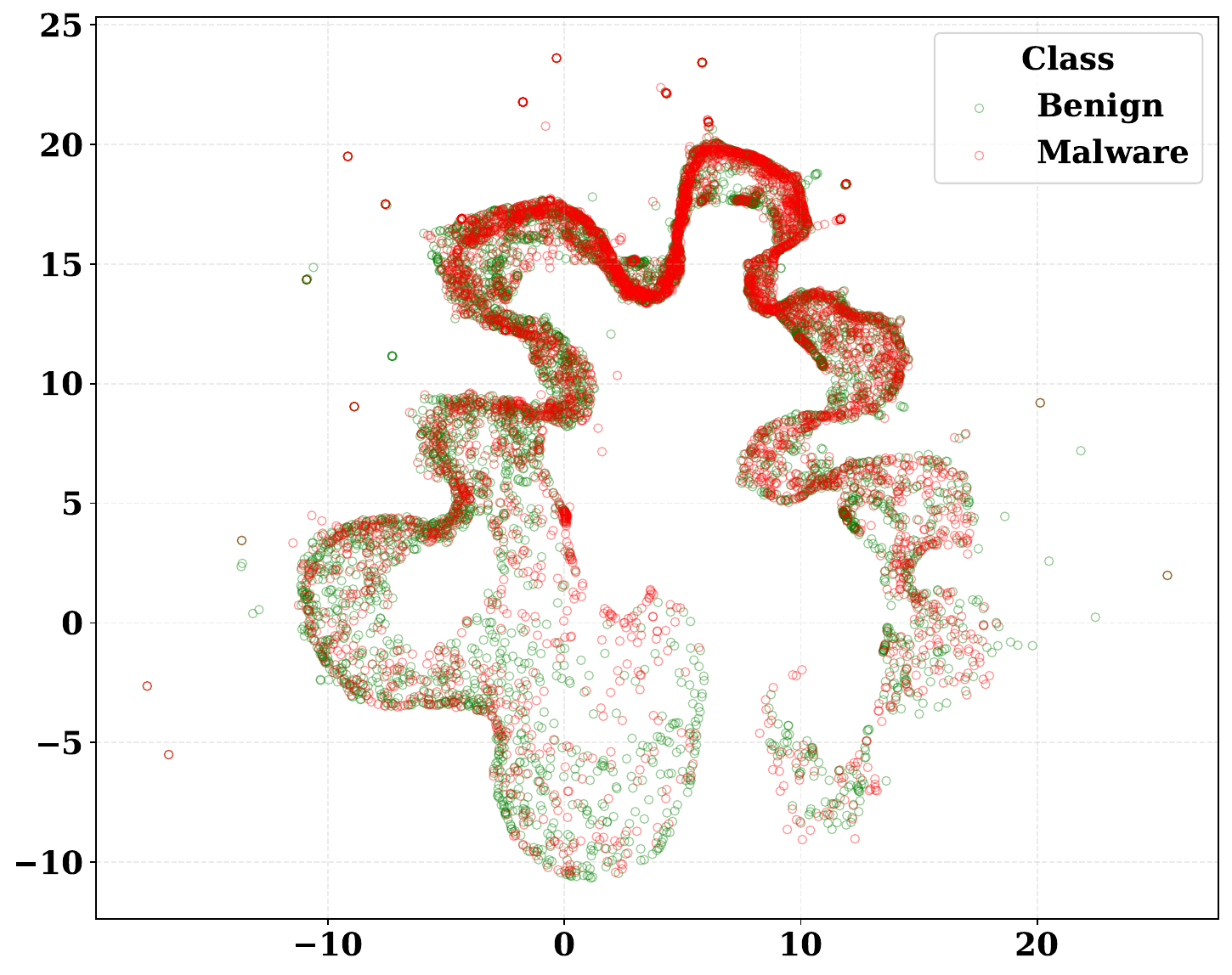}
    \caption{2021}
    \label{fig:json-year-2021}
\end{subfigure}
\begin{subfigure}[t]{0.155\textwidth}
    \centering
    \includegraphics[width=\linewidth]{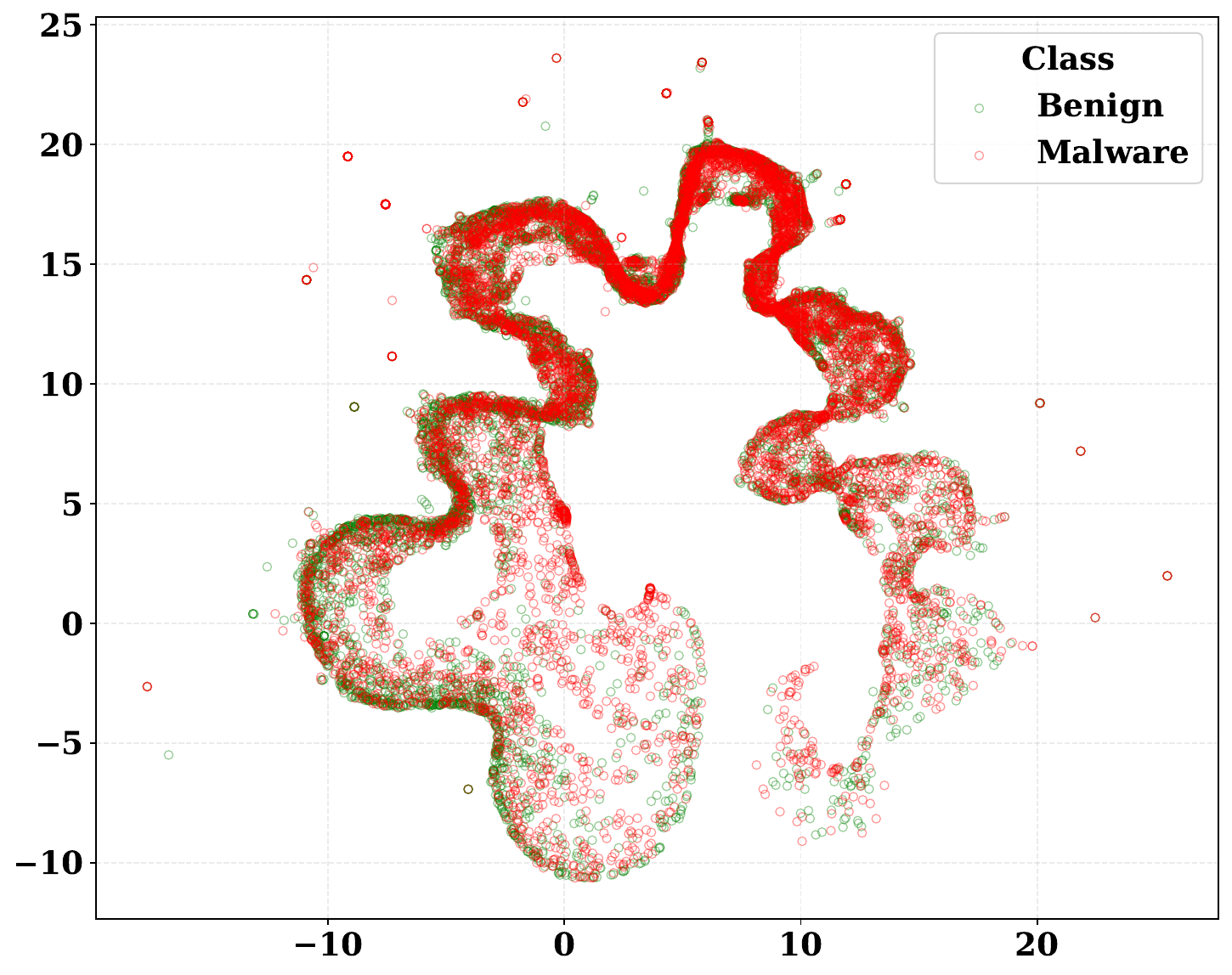}
    \caption{2022}
    \label{fig:json-year-2022}
\end{subfigure}
\begin{subfigure}[t]{0.155\textwidth}
    \centering
    \includegraphics[width=\linewidth]{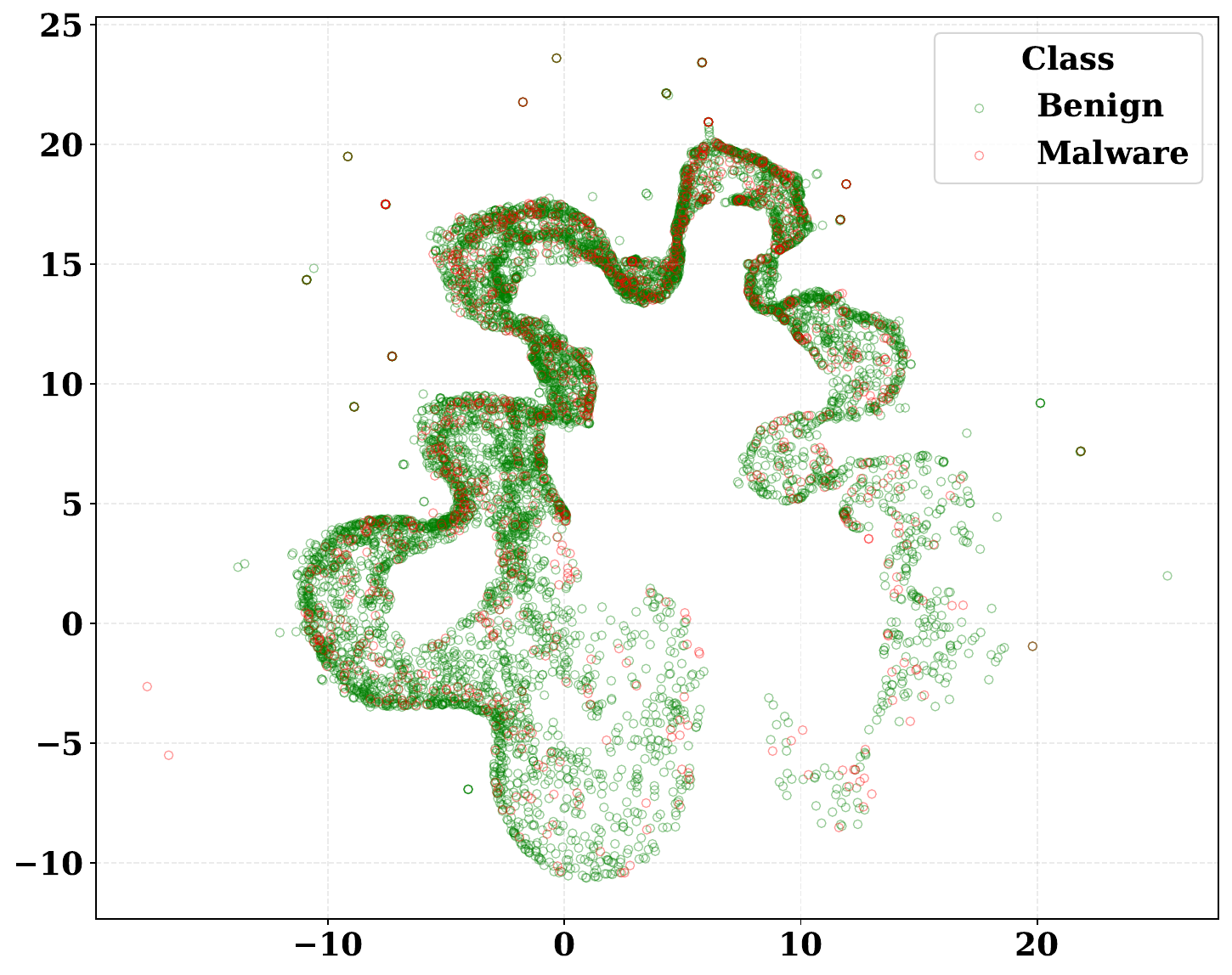}
    \caption{2023}
    \label{fig:json-year-2023}
\end{subfigure}
\begin{subfigure}[t]{0.155\textwidth}
    \centering
    \includegraphics[width=\linewidth]{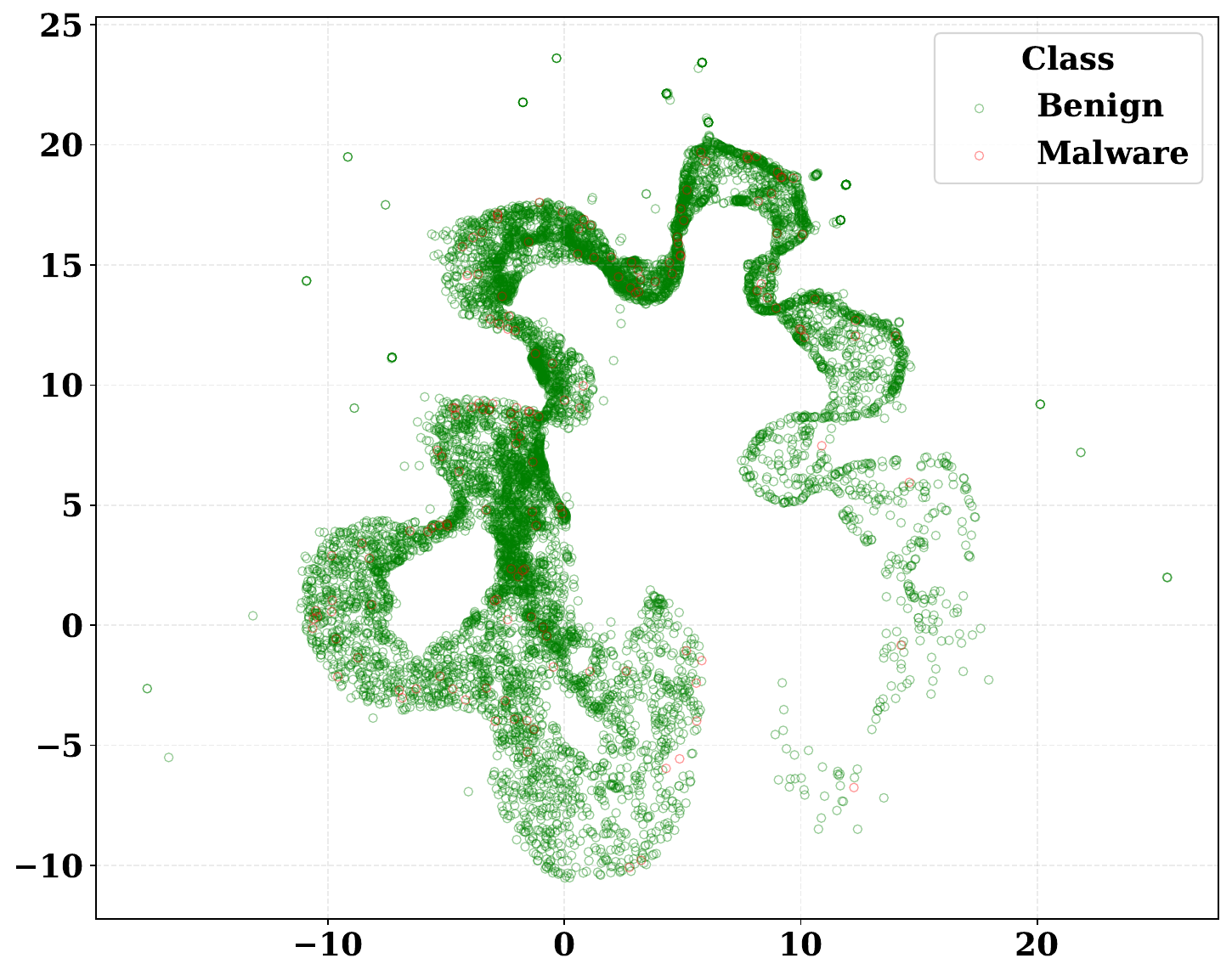}
    \caption{2024}
    \label{fig:json-year-2024}
\end{subfigure}
\begin{subfigure}[t]{0.155\textwidth}
    \centering
    \includegraphics[width=\linewidth]{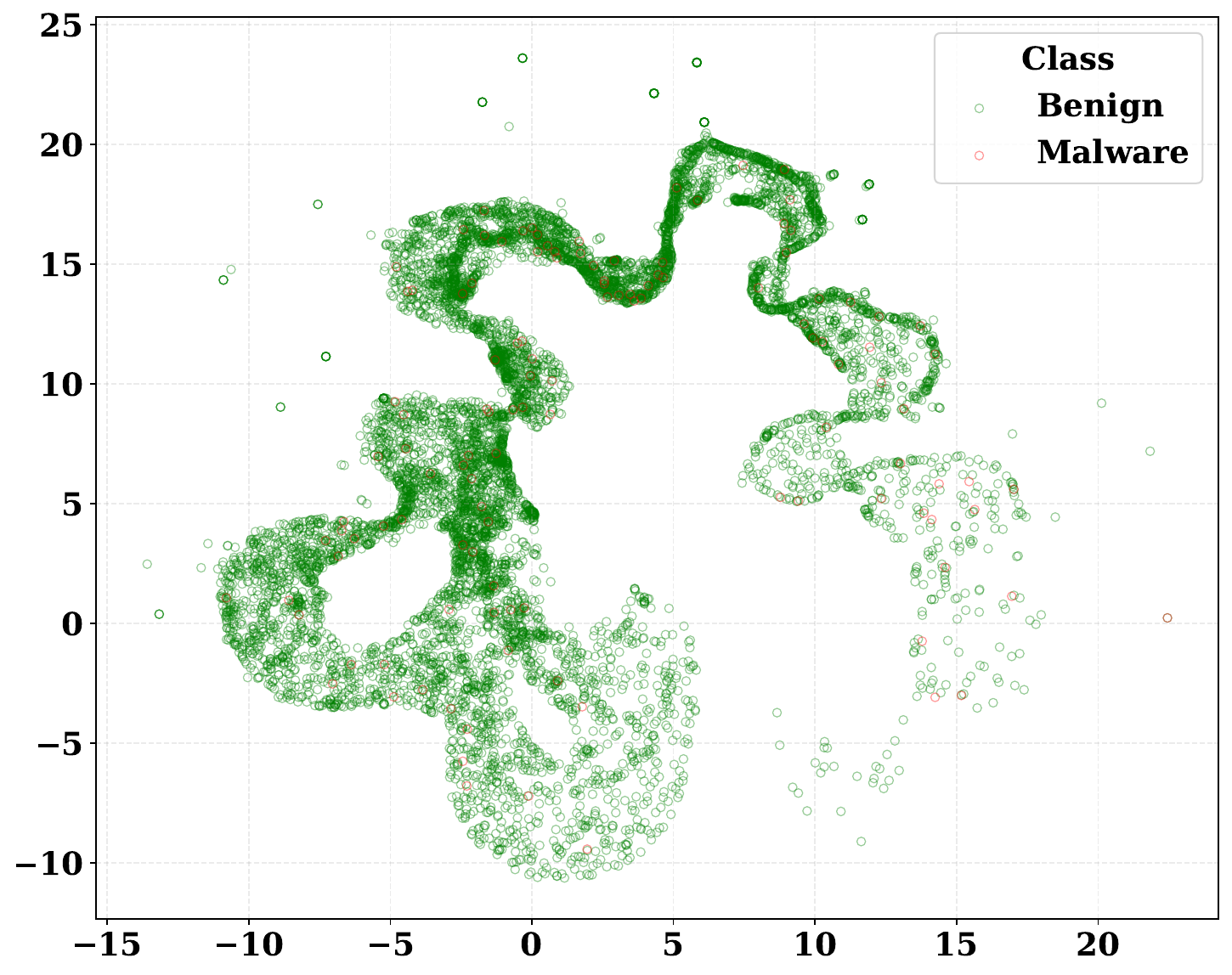}
    \caption{2025}
    \label{fig:json-year-2025}
\end{subfigure}

\caption{Year-wise dynamic feature space visualization of \textcolor{red}{malware} and \textcolor{green!50!black}{benign} samples from 2013 to 2025, excluding 2015.}
\label{fig:json-year-feat-class-json}
\end{figure*}

\begin{figure*}[t]
\centering

\begin{subfigure}[t]{0.155\textwidth}
    \centering
    \includegraphics[width=\linewidth]{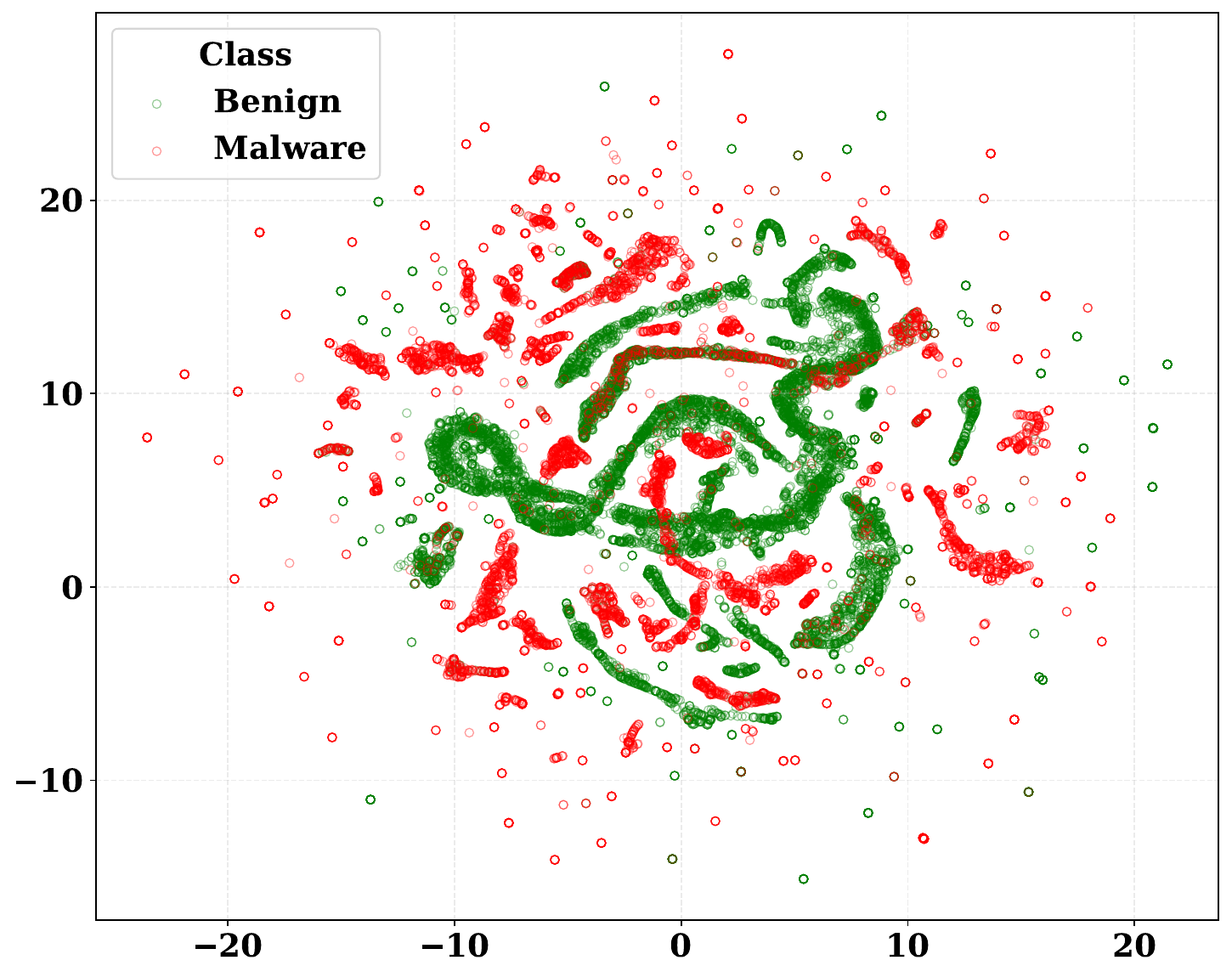}
    \caption{2013}
    \label{fig:gml-year-2013}
\end{subfigure}
\begin{subfigure}[t]{0.155\textwidth}
    \centering
    \includegraphics[width=\linewidth]{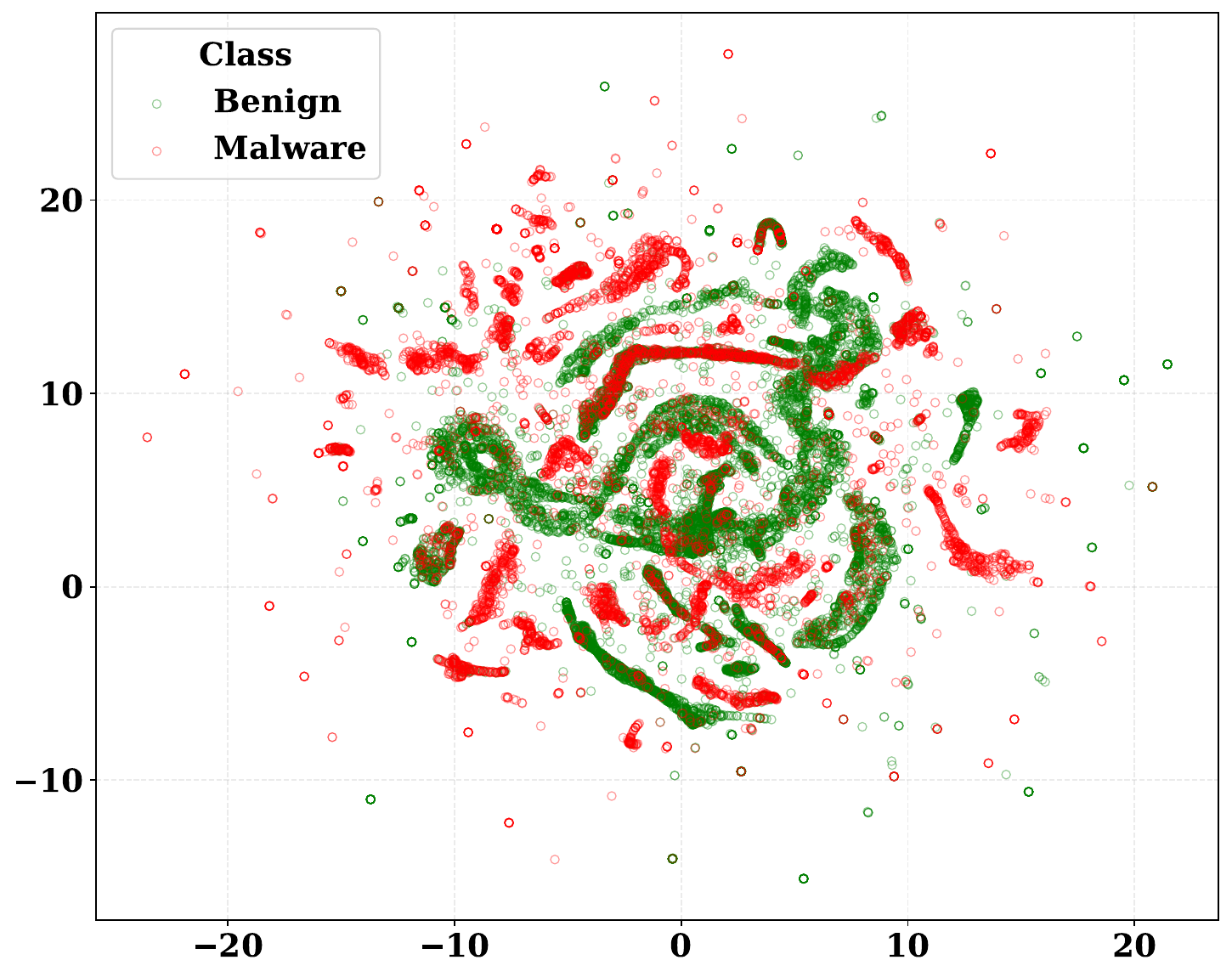}
    \caption{2014}
    \label{fig:gml-year-2014}
\end{subfigure}
\begin{subfigure}[t]{0.155\textwidth}
    \centering
    \includegraphics[width=\linewidth]{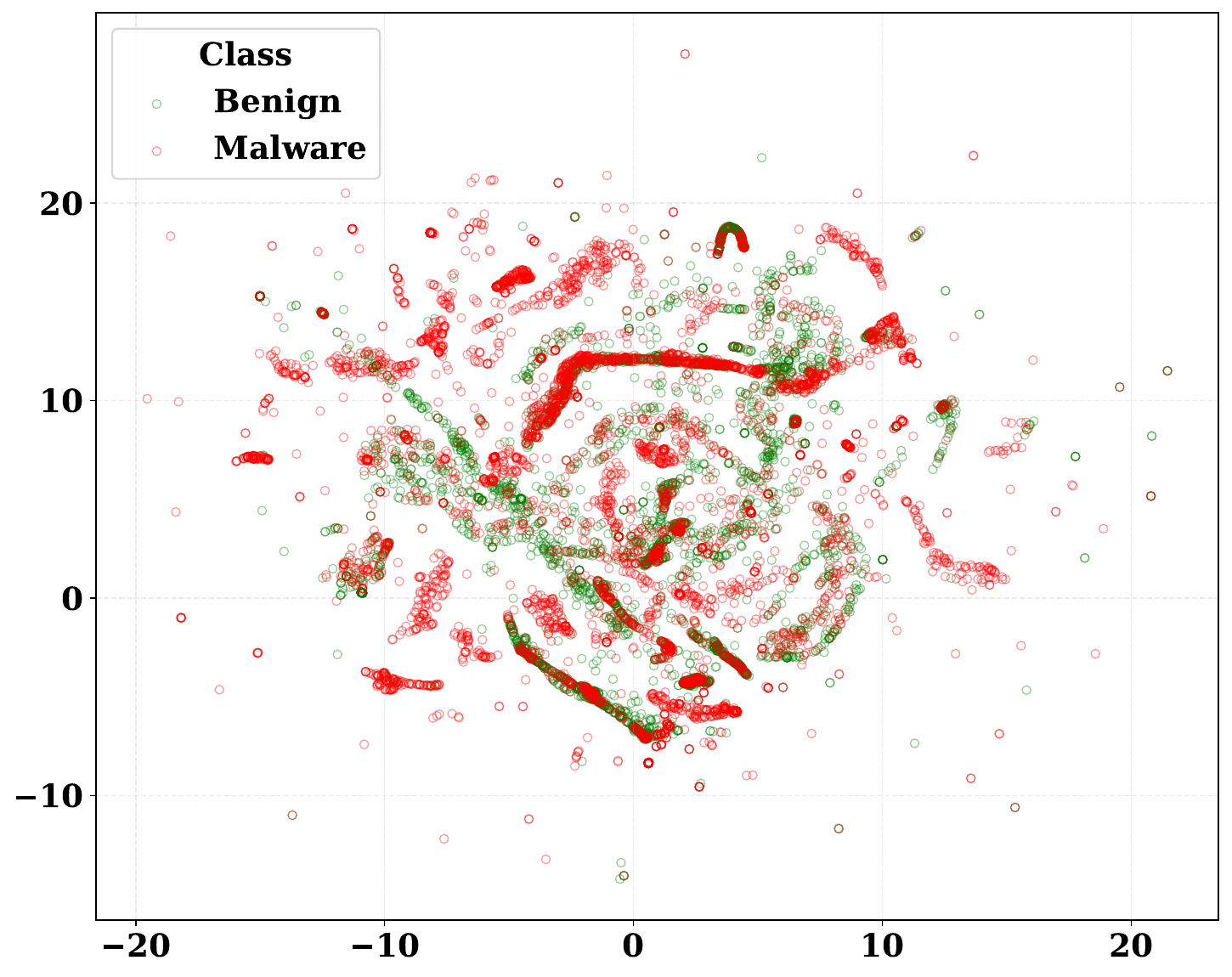}
    \caption{2016}
    \label{fig:gml-year-2016}
\end{subfigure}
\begin{subfigure}[t]{0.155\textwidth}
    \centering
    \includegraphics[width=\linewidth]{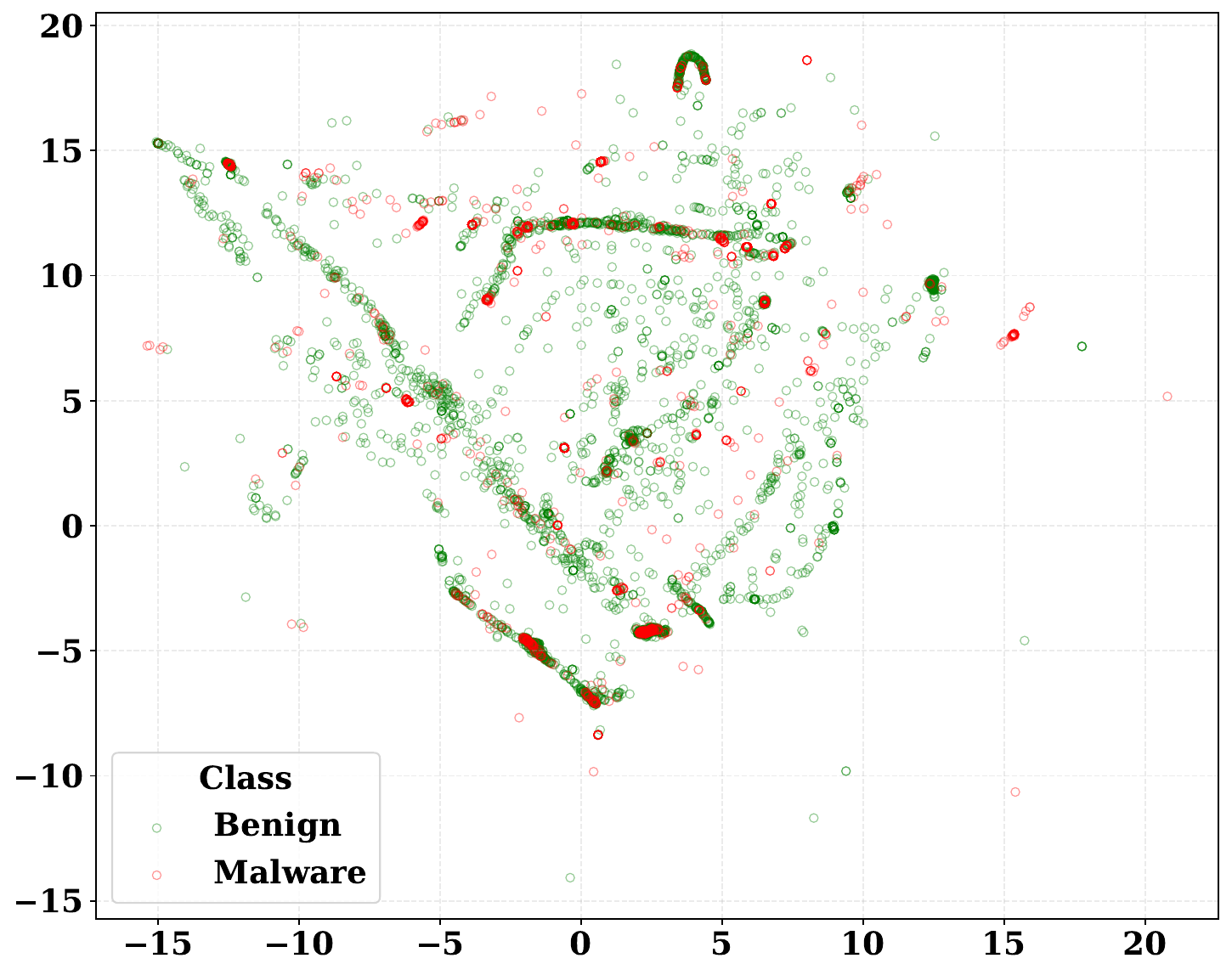}
    \caption{2017}
    \label{fig:gml-year-2017}
\end{subfigure}
\begin{subfigure}[t]{0.155\textwidth}
    \centering
    \includegraphics[width=\linewidth]{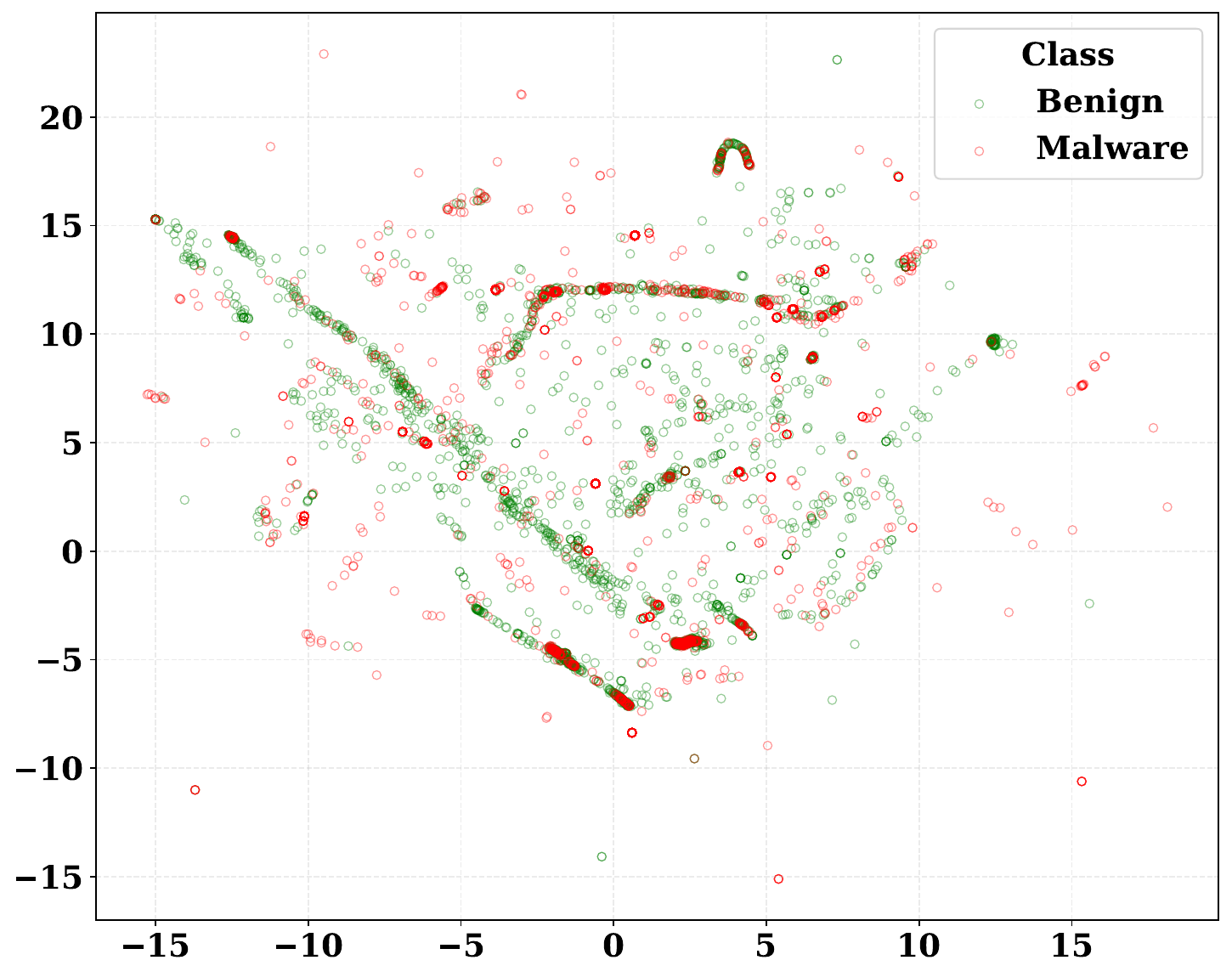}
    \caption{2018}
    \label{fig:gml-year-2018}
\end{subfigure}
\begin{subfigure}[t]{0.155\textwidth}
    \centering
    \includegraphics[width=\linewidth]{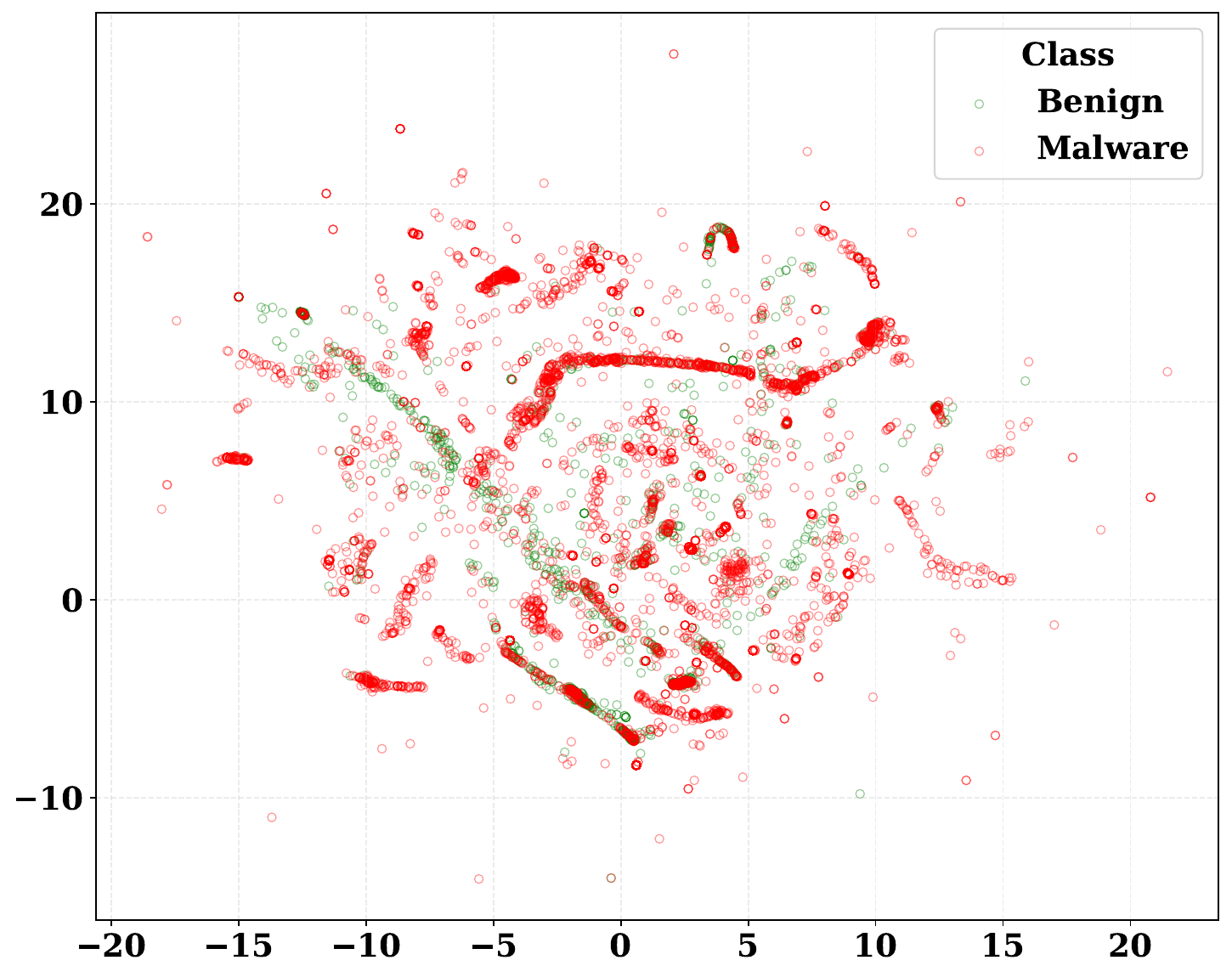}
    \caption{2019}
    \label{fig:gml-year-2019}
\end{subfigure}

\medskip

\begin{subfigure}[t]{0.155\textwidth}
    \centering
    \includegraphics[width=\linewidth]{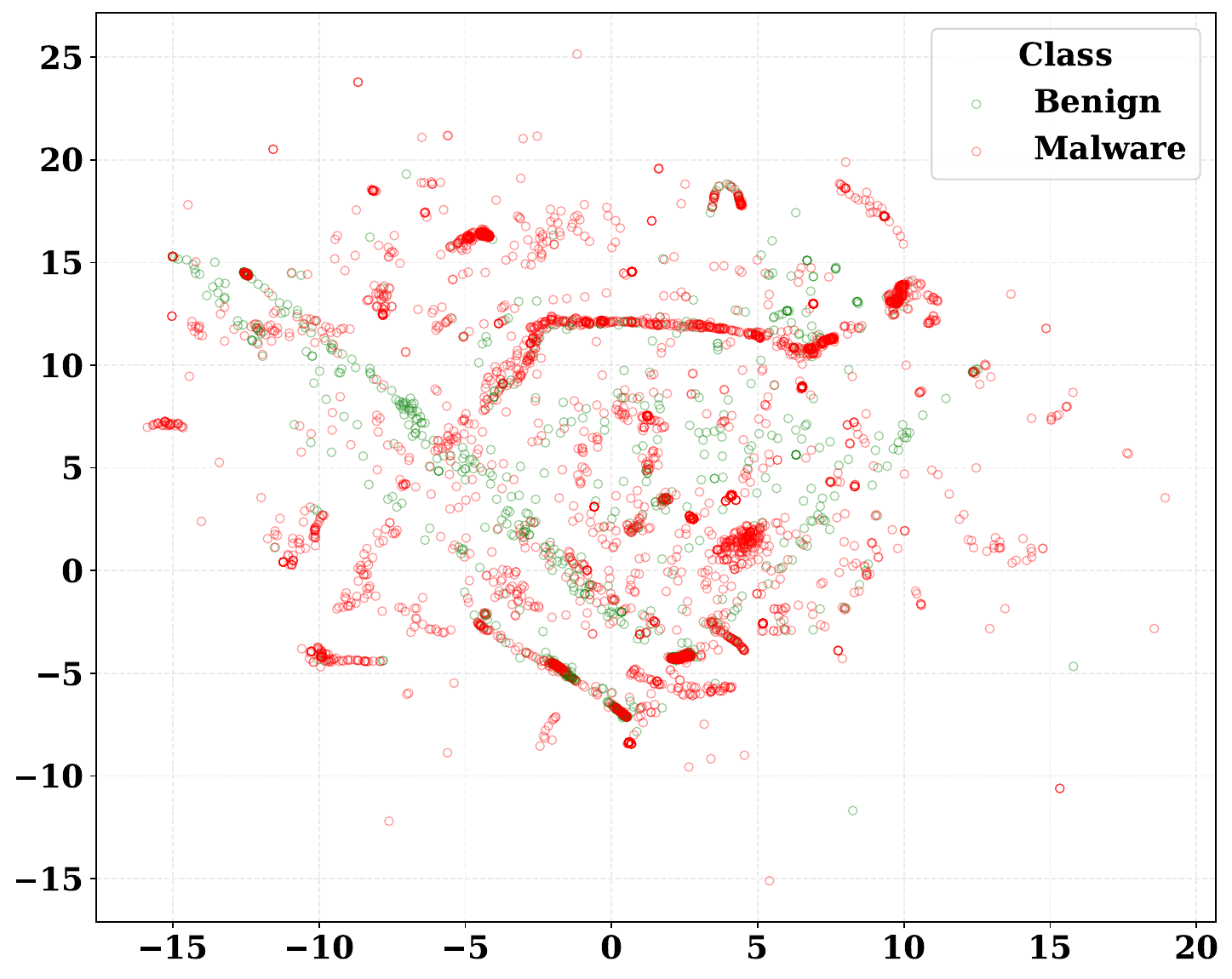}
    \caption{2020}
    \label{fig:gml-year-2020}
\end{subfigure}
\begin{subfigure}[t]{0.155\textwidth}
    \centering
    \includegraphics[width=\linewidth]{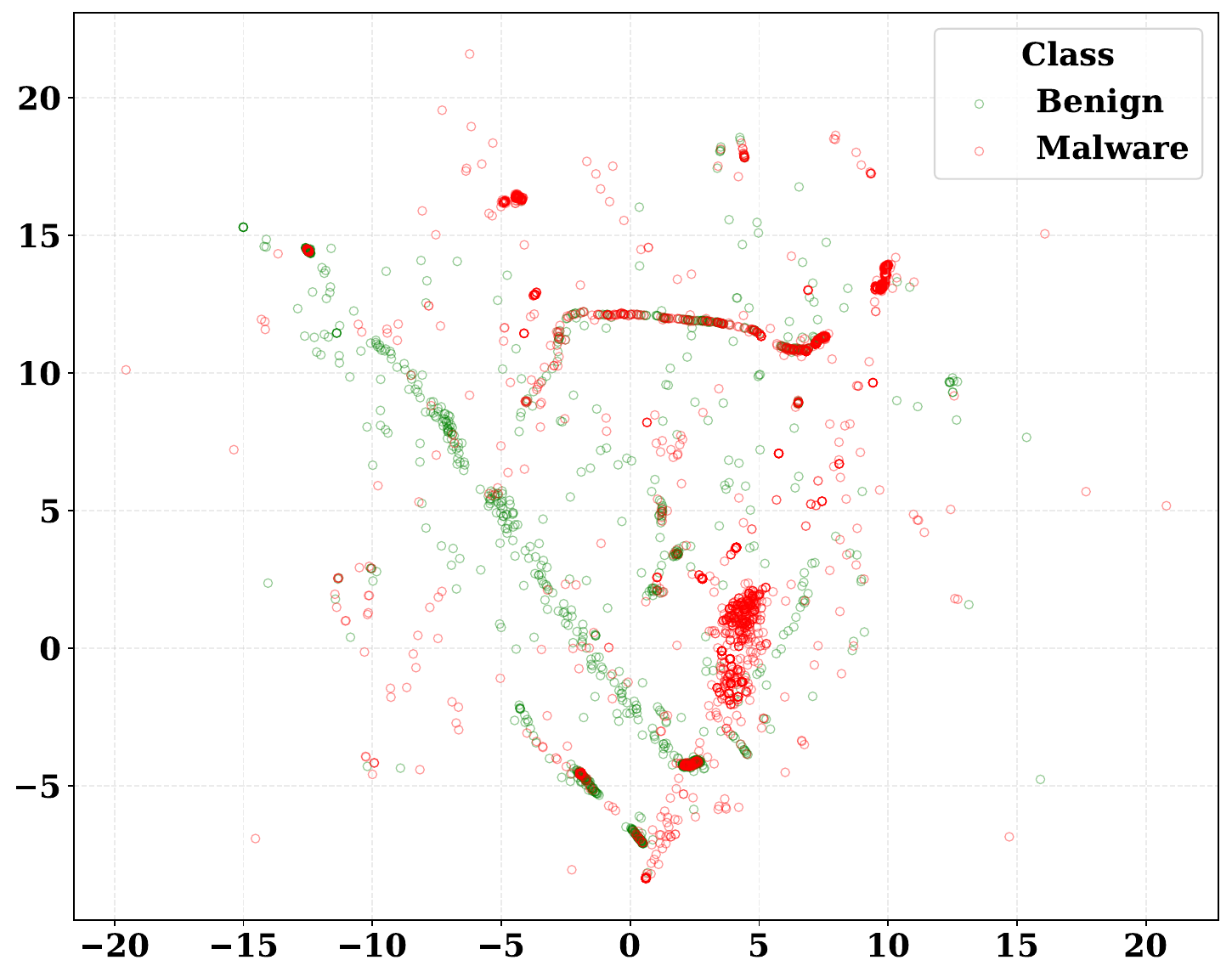}
    \caption{2021}
    \label{fig:gml-year-2021}
\end{subfigure}
\begin{subfigure}[t]{0.155\textwidth}
    \centering
    \includegraphics[width=\linewidth]{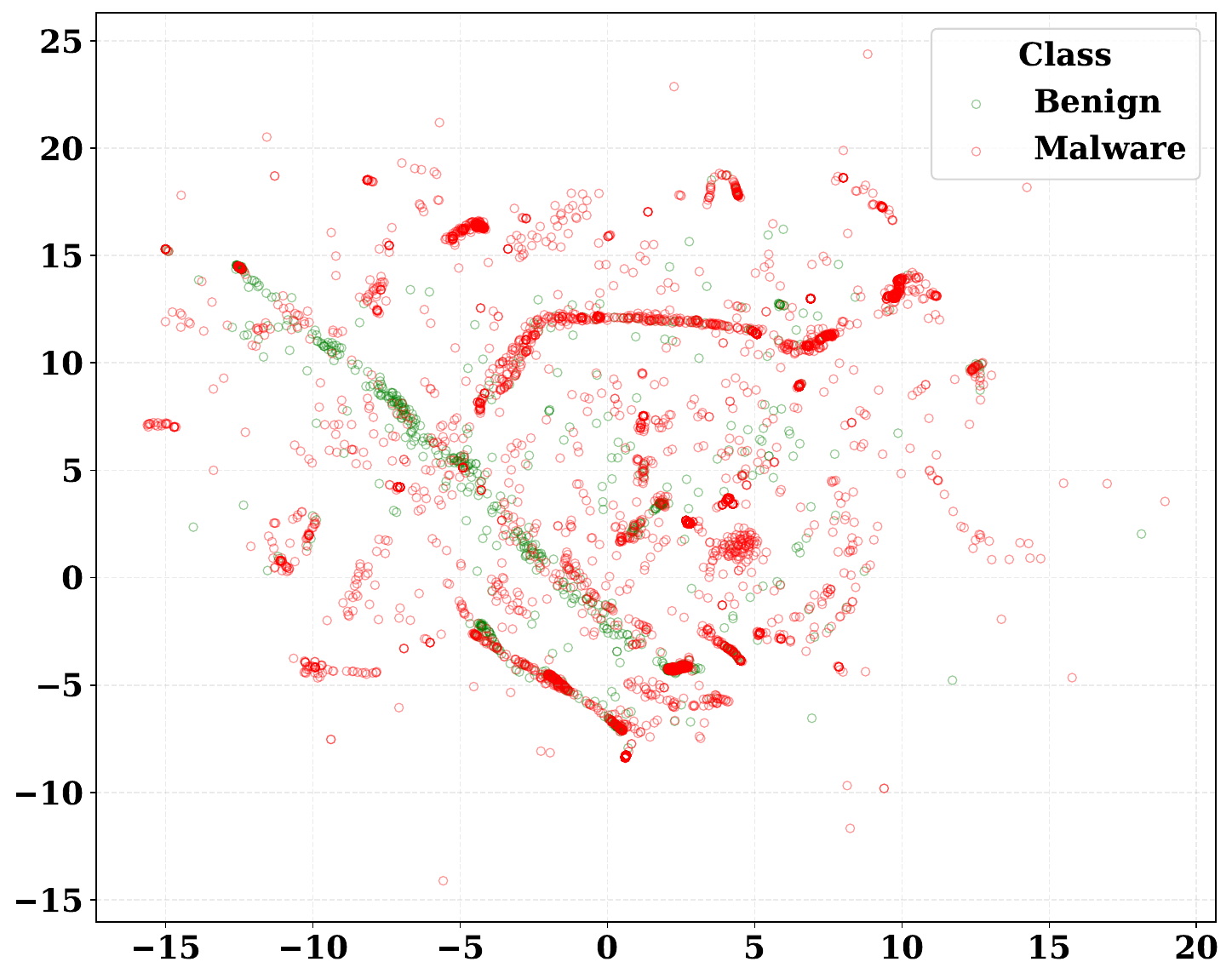}
    \caption{2022}
    \label{fig:gml-year-2022}
\end{subfigure}
\begin{subfigure}[t]{0.155\textwidth}
    \centering
    \includegraphics[width=\linewidth]{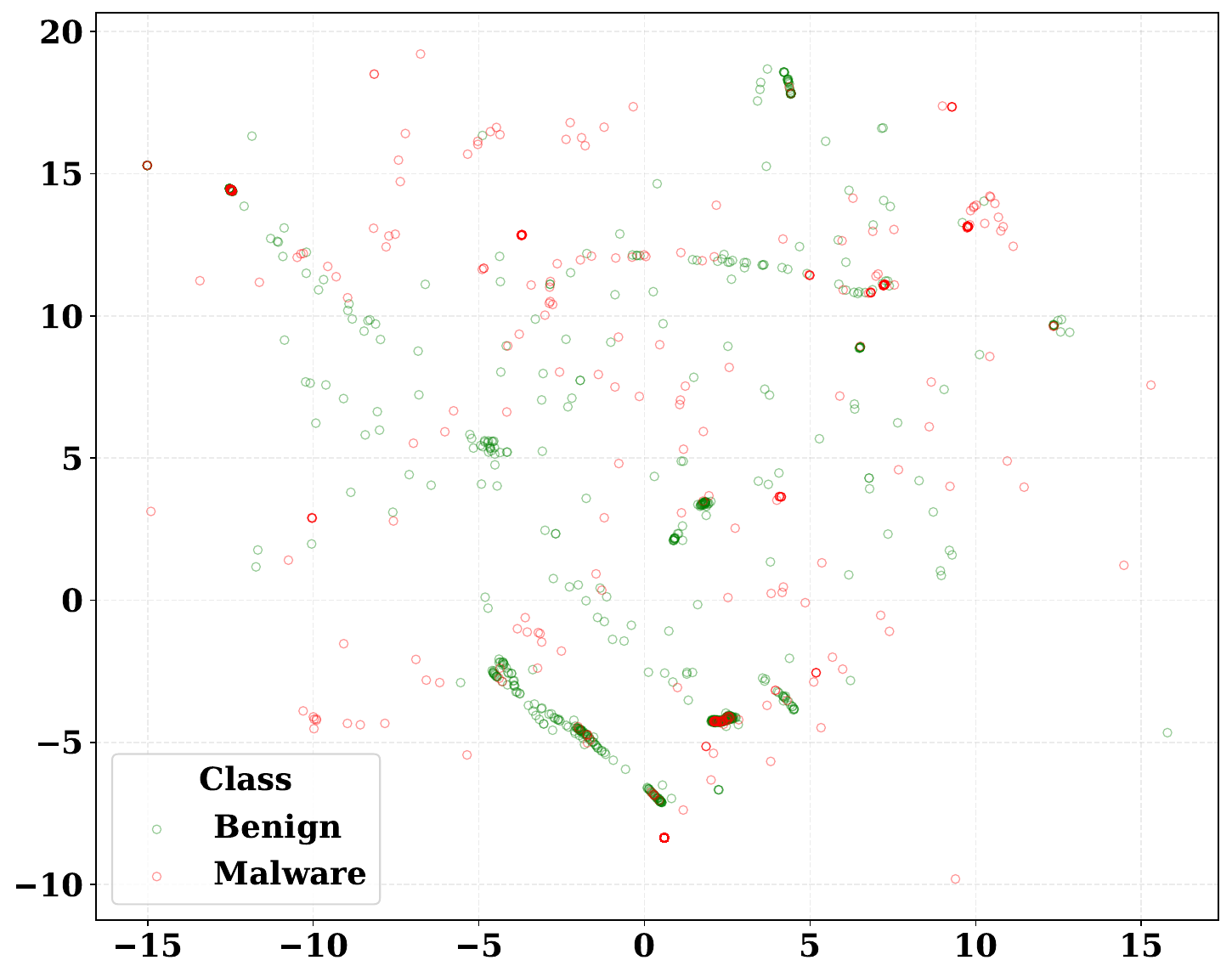}
    \caption{2023}
    \label{fig:gml-year-2023}
\end{subfigure}
\begin{subfigure}[t]{0.155\textwidth}
    \centering
    \includegraphics[width=\linewidth]{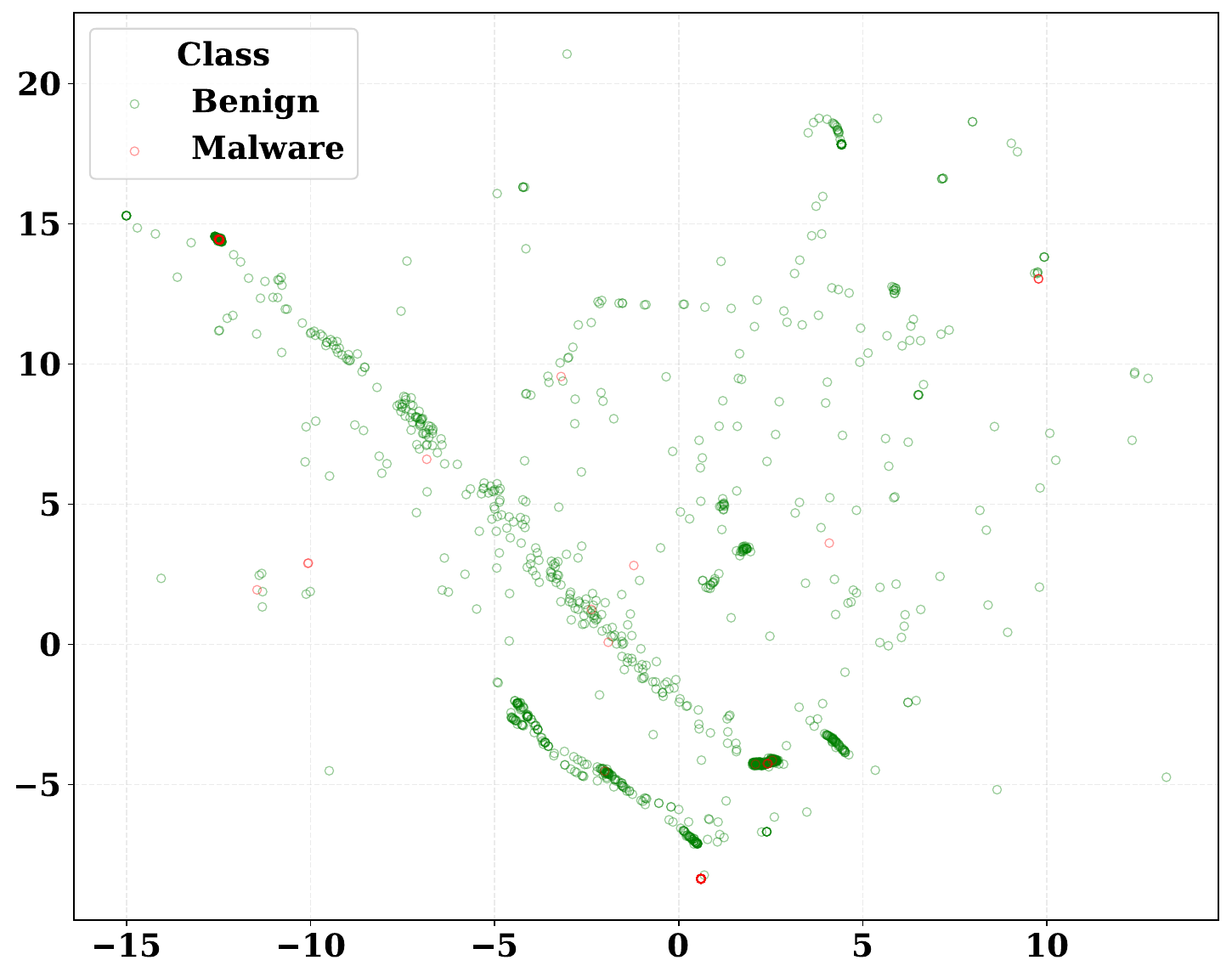}
    \caption{2024}
    \label{fig:gml-year-2024}
\end{subfigure}
\begin{subfigure}[t]{0.155\textwidth}
    \centering
    \includegraphics[width=\linewidth]{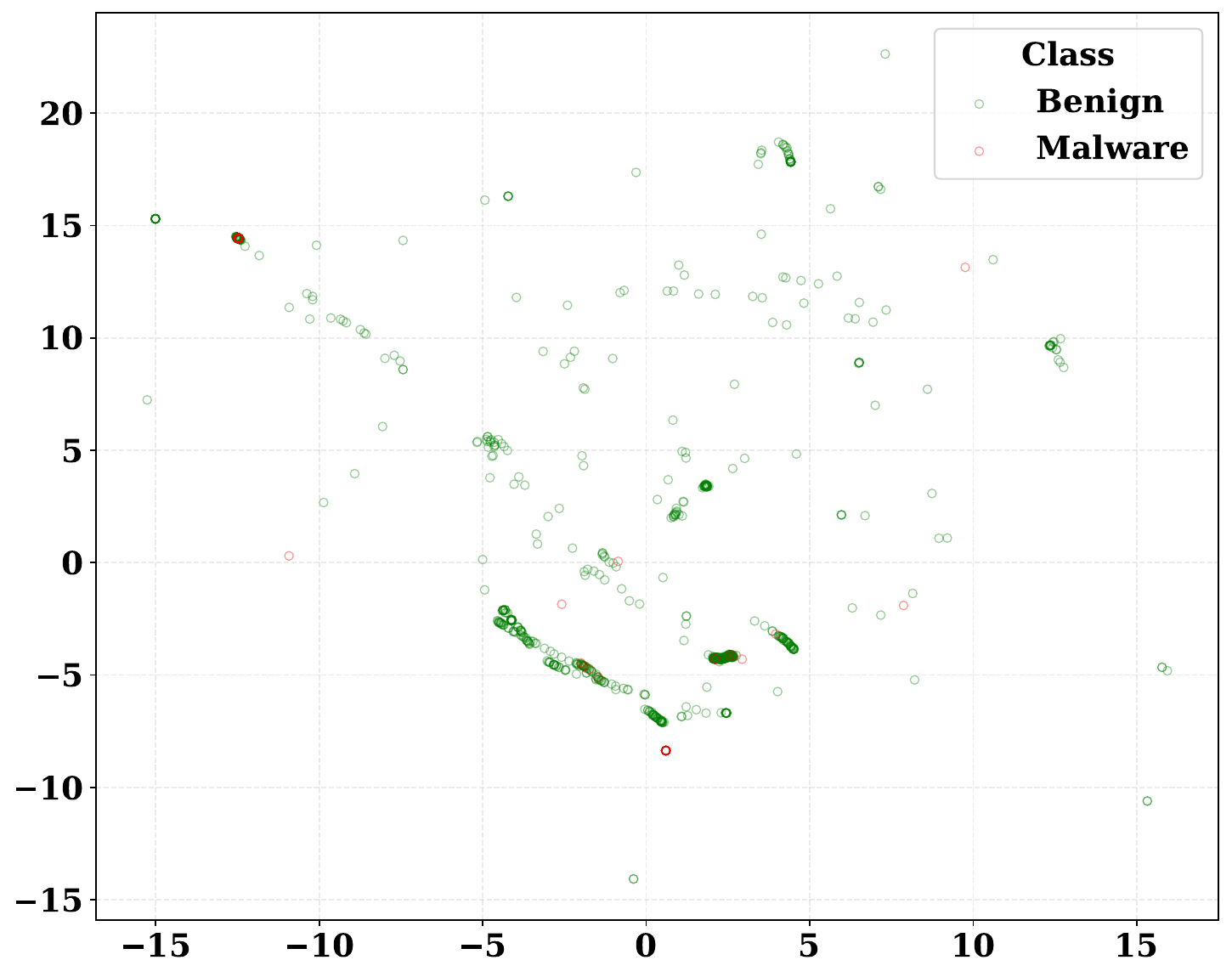}
    \caption{2025}
    \label{fig:gml-year-2025}
\end{subfigure}

\caption{Year-wise graph-based feature space visualization of \textcolor{red}{malware} and \textcolor{green!50!black}{benign} samples from 2013 to 2025, excluding 2015.}
\label{fig:gml-year-feat-class}
\end{figure*}

Figures~\ref{fig:data-year-feat-class}, \ref{fig:gml-year-feat-class}, and \ref{fig:json-year-feat-class-json} show yearly UMAP projections for static, graph-based, and dynamic representations, respectively. In 2013--2014, benign and malware samples are relatively separable across modalities, consistent with the strong IID results in Table~\ref{tab:comparison_unimodal}. From 2016 onward, the embeddings become increasingly mixed, with the strongest overlap in 2017--2018, where malware samples occupy regions previously dominated by benign samples. This provides a representation-level explanation for the sharp degradation observed in Figure~\ref{fig:fusion_f1_score}. Static features preserve the most coherent class structure among the three modalities, whereas graph-based features exhibit the largest structural instability. In particular, graph representations become less class-separable after 2017 and degrade further in later years, plausibly due to shifts in Android APIs, framework usage, third-party libraries, and malware implementation patterns that alter function-call structures~\cite{arp2014drebin,apigraph,he2023efficient,mpdroid,zhao2021structural,haque2025lamda}.

\section{Feature Importance Across Modalities}
\label{app:feature_importance}

\BfPara{Setting}To understand which feature groups drive model decisions, we conduct a grouped SHAP-based importance analysis. We analyze feature importance across three modalities: static, graph-based, and dynamic features. Model explanations are computed using SHAP~\cite{shap} values, where importance score is measured as the mean absolute SHAP contribution of each feature~\cite{haque2025lamda}. Feature-level scores are aggregated into semantic groups, and the top five groups are retained for each modality and temporal split.

\BfPara{Analysis} Table~\ref{tab:feature_group_importance} summarizes the most influential feature groups across modalities under increasing temporal shift. For static features, importance is relatively distributed, with URL domains, permissions, and restricted APIs consistently dominant, though their contributions vary across NEAR and FAR settings. Graph-based features show a clear shift in reliance: API-level signals (e.g., TelephonyManager, WebView) decrease over time, while structural properties such as graph statistics become more prominent in the FAR setting. Dynamic features remain largely driven by string and path-based signals across all splits. Overall, these trends demonstrate that different modalities capture complementary aspects of the data under distribution shift.

\begin{table}[t]
\centering
\caption{Top feature groups by normalized SHAP score, shown as \% within each modality and split.}
\footnotesize
\resizebox{\textwidth}{!}{
\setlength{\tabcolsep}{3.5pt}
\renewcommand{\arraystretch}{1.08}
\begin{tabular}{l|l|c|c|c}
\toprule
\textbf{Modality} & \textbf{Top feature groups} & \textbf{IID} & \textbf{NEAR} & \textbf{FAR} \\
\midrule
Static
& \textcolor{red}{URL domains} / \textcolor{blue}{Requested permissions} / \textcolor{brown}{Restricted APIs}

& \textcolor{red}{\meanstd{24.38}{0.16}} / \textcolor{blue}{\meanstd{19.09}{0.17}} / \textcolor{brown}{\meanstd{14.86}{0.27}}

& \textcolor{red}{\meanstd{20.90}{0.30}} / \textcolor{blue}{\meanstd{18.37}{0.15}} / \textcolor{brown}{\meanstd{25.45}{0.36}}

& \textcolor{red}{\meanstd{22.94}{0.04}} / \textcolor{blue}{\meanstd{21.87}{0.13}} / \textcolor{brown}{\meanstd{16.99}{0.07}} \\

& \textcolor{orange}{Suspicious APIs} / \textcolor{purple}{Intent filters}

& \textcolor{orange}{\meanstd{10.23}{0.27}} / \textcolor{purple}{\meanstd{12.29}{0.02}}

& \textcolor{orange}{\meanstd{12.69}{0.11}} / \textcolor{purple}{\meanstd{8.20}{0.02}}

& \textcolor{orange}{\meanstd{16.31}{0.06}} / \textcolor{purple}{\meanstd{9.76}{0.02}} \\

\midrule
Graph-based
& \textcolor{red}{TelephonyManager} / \textcolor{blue}{WebView} / \textcolor{brown}{Activity APIs}

& \textcolor{red}{\meanstd{14.30}{0.07}} / \textcolor{blue}{\meanstd{7.78}{0.14}} / \textcolor{brown}{\meanstd{6.21}{0.09}}

& \textcolor{red}{\meanstd{7.64}{0.05}} / \textcolor{blue}{\meanstd{7.41}{0.11}} / \textcolor{brown}{\meanstd{5.73}{0.10}}

& \textcolor{red}{\meanstd{4.81}{0.03}} / \textcolor{blue}{\meanstd{5.30}{0.08}} / \textcolor{brown}{\meanstd{4.93}{0.12}} \\

& \textcolor{orange}{Graph statistics} / \textcolor{purple}{View APIs}

& \textcolor{orange}{\meanstd{5.35}{0.07}} / \textcolor{purple}{\meanstd{4.47}{0.14}}

& \textcolor{orange}{\meanstd{7.87}{0.20}} / \textcolor{purple}{\meanstd{5.29}{0.20}}

& \textcolor{orange}{\meanstd{15.50}{0.16}} / \textcolor{purple}{\meanstd{7.47}{0.26}} \\

\midrule
Dynamic
& \textcolor{red}{String} / \textcolor{blue}{Path} / \textcolor{brown}{Domain}

& \textcolor{red}{\meanstd{26.73}{0.05}} / \textcolor{blue}{\meanstd{26.80}{0.10}} / \textcolor{brown}{\meanstd{16.39}{0.09}}

& \textcolor{red}{\meanstd{29.26}{0.03}} / \textcolor{blue}{\meanstd{24.10}{0.06}} / \textcolor{brown}{\meanstd{7.11}{0.10}}

& \textcolor{red}{\meanstd{27.52}{0.04}} / \textcolor{blue}{\meanstd{18.57}{0.24}} / \textcolor{brown}{\meanstd{20.81}{0.11}} \\

& \textcolor{orange}{Component} / \textcolor{purple}{Permissions}

& \textcolor{orange}{\meanstd{5.27}{0.03}} / \textcolor{purple}{\meanstd{3.34}{0.20}}

& \textcolor{orange}{\meanstd{11.65}{0.05}} / \textcolor{purple}{\meanstd{0.00}{0.00}}

& \textcolor{orange}{\meanstd{7.50}{0.07}} / \textcolor{purple}{\meanstd{0.00}{0.00}} \\

\bottomrule
\end{tabular}
}
\vspace{2pt}
\label{tab:feature_group_importance}
\end{table}

\section{Sensitivity to Missing Modalities}
\label{app:missing_modal}

\BfPara{Setting}In real-world scenarios, a detection system cannot always rely on the availability of all feature modalities due to noise, extraction failures, or incomplete data sources~\cite{liu2026unraveling}. To assess sensitivity under such conditions, we simulate missing-feature settings by selectively removing or masking one or more modalities at inference time while keeping the model trained on complete data~\cite{feng2023fedmultimodal}. This setup allows us to systematically assess how performance degrades when specific features are unavailable and to identify which modalities are most critical for reliable detection.

\BfPara{Analysis}
Figure~\ref{fig:missing_feat} evaluates the sensitivity of \system under missing-feature conditions. The results show that static features are the most indispensable modality: removing them causes F1 to collapse across nearly all test years, indicating that static representations provide the dominant signal for detection. Graph-based features are also beneficial, with their removal producing consistent but less severe degradation, while dynamic features appear less critical and in several years their removal slightly improves F1.

\begin{figure}[!t]
    \centering

    \begin{minipage}[t]{0.48\linewidth}
        \centering
        \resizebox{\linewidth}{!}{%
    \begin{tikzpicture}
    \begin{axis}[
        width=8cm,
        height=5cm,
        xtick pos=bottom,
        ytick pos=left,
        enlargelimits=0.04,
        grid=both,
        grid style={line width=0.15pt, draw=gray!25},
        major grid style={line width=0.2pt, draw=gray!35},
        legend style={
            at={(0.98,0.98)},
            anchor=north east,
            legend columns=2,
            fill=white,
            draw=black,
            line width=0.3pt,
            font=\bfseries\tiny
        },
        legend image post style={scale=0.7},
        ylabel={F1 Score},
        xlabel={Year},
        xlabel style={font=\bfseries\tiny},
        ylabel style={font=\bfseries\tiny, xshift=2pt, yshift=-5pt},
        symbolic x coords={2013,2014,2016,2017,2018,2019,2020,2021,2022,2023,2024,2025},
        xtick=data,
        x tick label style={rotate=45, anchor=east, font=\bfseries\tiny},
        y tick label style={font=\bfseries\tiny},
        ymin=0,
        ymax=1.05,
        ytick={0,0.2,0.4,0.6,0.8,1.0},
    ]

    \addplot[
        thick,
        blue!70!black,
        mark=*,
        mark size=1.4pt,
        error bars/.cd,
        y dir=both,
        y explicit,
        error bar style={draw=blue!70!black, line width=0.3pt},
        error mark options={draw=blue!70!black, line width=0.3pt, mark size=0.8pt}
    ] coordinates {
        (2013,0.9725) +- (0,0.0006)
        (2014,0.8631) +- (0,0.0017)
        (2016,0.6903) +- (0,0.0047)
        (2017,0.1031) +- (0,0.0030)
        (2018,0.0796) +- (0,0.0037)
        (2019,0.7246) +- (0,0.0073)
        (2020,0.7201) +- (0,0.0182)
        (2021,0.6324) +- (0,0.0245)
        (2022,0.6507) +- (0,0.0067)
        (2023,0.4584) +- (0,0.0351)
        (2024,0.1245) +- (0,0.0327)
        (2025,0.0234) +- (0,0.0056)
    };

    \addplot[
        thick,
        red!70!black,
        mark=square*,
        mark size=1.3pt,
        error bars/.cd,
        y dir=both,
        y explicit,
        error bar style={draw=red!70!black, line width=0.3pt},
        error mark options={draw=red!70!black, line width=0.3pt, mark size=0.8pt}
    ] coordinates {
        (2013,0.2188) +- (0,0.0209)
        (2014,0.1093) +- (0,0.0119)
        (2016,0.0434) +- (0,0.0106)
        (2017,0.0017) +- (0,0.0008)
        (2018,0.0036) +- (0,0.0007)
        (2019,0.0275) +- (0,0.0085)
        (2020,0.0238) +- (0,0.0043)
        (2021,0.0046) +- (0,0.0010)
        (2022,0.0086) +- (0,0.0037)
        (2023,0.0096) +- (0,0.0037)
        (2024,0.0039) +- (0,0.0067)
        (2025,0.0000) +- (0,0.0000)
    };

    \addplot[
        thick,
        green!50!black,
        mark=triangle*,
        mark size=1.5pt,
        error bars/.cd,
        y dir=both,
        y explicit,
        error bar style={draw=green!50!black, line width=0.3pt},
        error mark options={draw=green!50!black, line width=0.3pt, mark size=0.8pt}
    ] coordinates {
        (2013,0.8474) +- (0,0.0129)
        (2014,0.7043) +- (0,0.0094)
        (2016,0.4930) +- (0,0.0287)
        (2017,0.0582) +- (0,0.0017)
        (2018,0.0428) +- (0,0.0028)
        (2019,0.6247) +- (0,0.0138)
        (2020,0.6111) +- (0,0.0150)
        (2021,0.5386) +- (0,0.0222)
        (2022,0.5527) +- (0,0.0196)
        (2023,0.3006) +- (0,0.0559)
        (2024,0.1144) +- (0,0.0290)
        (2025,0.0452) +- (0,0.0137)
    };

    \addplot[
        thick,
        orange!80!black,
        mark=diamond*,
        mark size=1.5pt,
        error bars/.cd,
        y dir=both,
        y explicit,
        error bar style={draw=orange!80!black, line width=0.3pt},
        error mark options={draw=orange!80!black, line width=0.3pt, mark size=0.8pt}
    ] coordinates {
        (2013,0.9611) +- (0,0.0002)
        (2014,0.8774) +- (0,0.0058)
        (2016,0.7251) +- (0,0.0118)
        (2017,0.1260) +- (0,0.0144)
        (2018,0.1045) +- (0,0.0102)
        (2019,0.7681) +- (0,0.0167)
        (2020,0.7772) +- (0,0.0141)
        (2021,0.6967) +- (0,0.0127)
        (2022,0.7186) +- (0,0.0099)
        (2023,0.4894) +- (0,0.0137)
        (2024,0.1530) +- (0,0.0383)
        (2025,0.0433) +- (0,0.0054)
    };

    \legend{
        Full features,
        w/o Static,
        w/o Graph-based,
        w/o Dynamic
    }

    \end{axis}
    \end{tikzpicture}
        }
        \vspace{-0.75em}
        \caption{Year-wise F1 score comparison for full features vs missing modalities.}
        \label{fig:missing_feat}
    \end{minipage}
    \hfill
    \begin{minipage}[t]{0.48\linewidth}
        \centering
        \resizebox{\linewidth}{!}{%
    \begin{tikzpicture}
    \begin{axis}[
        width=8cm,
        height=5cm,
        xtick pos=bottom,
        ytick pos=left,
        enlargelimits=0.04,
        grid=both,
        grid style={line width=0.15pt, draw=gray!25},
        major grid style={line width=0.2pt, draw=gray!35},
        legend style={
            at={(0.98,0.98)},
            anchor=north east,
            legend columns=2,
            fill=white,
            draw=black,
            line width=0.3pt,
            font=\bfseries\tiny
        },
        legend image post style={scale=0.7},
        ylabel={Entropy},
        xlabel={Year},
        xlabel style={font=\bfseries\tiny, yshift=5pt},
        ylabel style={font=\bfseries\tiny, xshift=2pt, yshift=-5pt},
        xtick={0,1,2,3,4,5,6,7,8,9,10,11},
        xticklabels={2013,2014,2016,2017,2018,2019,2020,2021,2022,2023,2024,2025},
        x tick label style={rotate=45, anchor=east, font=\bfseries\tiny},
        y tick label style={font=\bfseries\tiny},
        ymin=0,
        ymax=0.75
    ]


    \addplot[name path=dynamicdrift, draw=none, forget plot] coordinates {
        (0,0.543663) (1,0.606716) (2,0.669459) (3,0.370465)
        (4,0.432819) (5,0.517479) (6,0.438795) (7,0.447582)
        (8,0.272572) (9,0.260411) (10,0.083880) (11,0.129776)
    };
    \addplot[name path=dynamicundrift, draw=none, forget plot] coordinates {
        (0,0.466450) (1,0.440161) (2,0.555272) (3,0.343188)
        (4,0.393359) (5,0.284293) (6,0.276128) (7,0.308282)
        (8,0.145427) (9,0.246767) (10,0.044131) (11,0.056238)
    };
    \addplot[
        blue!25,
        fill opacity=0.18,
        draw=none,
        forget plot
    ] fill between[of=dynamicdrift and dynamicundrift];

    \addplot[name path=graphdrift, draw=none, forget plot] coordinates {
        (0,0.390881) (1,0.409529) (2,0.488025) (3,0.210987)
        (4,0.220272) (5,0.392871) (6,0.260348) (7,0.328248)
        (8,0.232593) (9,0.131403) (10,0.073795) (11,0.080929)
    };
    \addplot[name path=graphundrift, draw=none, forget plot] coordinates {
        (0,0.264055) (1,0.229517) (2,0.341194) (3,0.153135)
        (4,0.238816) (5,0.150273) (6,0.134590) (7,0.175290)
        (8,0.079224) (9,0.114142) (10,0.032229) (11,0.040324)
    };
    \addplot[
        green!20,
        fill opacity=0.18,
        draw=none,
        forget plot
    ] fill between[of=graphdrift and graphundrift];

    \addplot[name path=staticdrift, draw=none, forget plot] coordinates {
        (0,0.404889) (1,0.436361) (2,0.478750) (3,0.273890)
        (4,0.337542) (5,0.407018) (6,0.280832) (7,0.361876)
        (8,0.230138) (9,0.152640) (10,0.086145) (11,0.073814)
    };
    \addplot[name path=staticundrift, draw=none, forget plot] coordinates {
        (0,0.256642) (1,0.237753) (2,0.340190) (3,0.207567)
        (4,0.308696) (5,0.166533) (6,0.147679) (7,0.165760)
        (8,0.084382) (9,0.126944) (10,0.038318) (11,0.050725)
    };
    \addplot[
        red!25,
        fill opacity=0.18,
        draw=none,
        forget plot
    ] fill between[of=staticdrift and staticundrift];

    \addlegendimage{
        thick,
        blue!70!black,
        solid,
        mark=none
    }
    \addlegendentry{Dynamic}

    \addlegendimage{
        thick,
        green!75!black,
        dashed,
        mark=none
    }
    \addlegendentry{Graph-based}

    \addlegendimage{
        thick,
        red!85!black,
        dash dot,
        mark=none
    }
    \addlegendentry{Static}

    \addplot[
        thick,
        blue!70!black,
        solid,
        mark=*,
        mark size=0.8pt,
        mark options={fill=red!80!black, draw=red!80!black}
    ] coordinates {
        (0,0.543663) (1,0.606716) (2,0.669459) (3,0.370465)
        (4,0.432819) (5,0.517479) (6,0.438795) (7,0.447582)
        (8,0.272572) (9,0.260411) (10,0.083880) (11,0.129776)
    };

    \addplot[
        thick,
        blue!70!black,
        solid,
        mark=*,
        mark size=0.8pt,
        mark options={fill=green!70!black, draw=green!70!black}
    ] coordinates {
        (0,0.466450) (1,0.440161) (2,0.555272) (3,0.343188)
        (4,0.393359) (5,0.284293) (6,0.276128) (7,0.308282)
        (8,0.145427) (9,0.246767) (10,0.044131) (11,0.056238)
    };

    \addplot[
        thick,
        green!75!black,
        dashed,
        mark=triangle*,
        mark size=0.75pt,
        mark options={fill=red!80!black, draw=red!80!black}
    ] coordinates {
        (0,0.390881) (1,0.409529) (2,0.488025) (3,0.210987)
        (4,0.220272) (5,0.392871) (6,0.260348) (7,0.328248)
        (8,0.232593) (9,0.131403) (10,0.073795) (11,0.080929)
    };

    \addplot[
        thick,
        green!75!black,
        dashed,
        mark=triangle*,
        mark size=0.75pt,
        mark options={fill=green!70!black, draw=green!70!black}
    ] coordinates {
        (0,0.264055) (1,0.229517) (2,0.341194) (3,0.153135)
        (4,0.238816) (5,0.150273) (6,0.134590) (7,0.175290)
        (8,0.079224) (9,0.114142) (10,0.032229) (11,0.040324)
    };

    \addplot[
        thick,
        red!85!black,
        dash dot,
        mark=diamond*,
        mark size=0.75pt,
        mark options={fill=red!80!black, draw=red!80!black}
    ] coordinates {
        (0,0.404889) (1,0.436361) (2,0.478750) (3,0.273890)
        (4,0.337542) (5,0.407018) (6,0.280832) (7,0.361876)
        (8,0.230138) (9,0.152640) (10,0.086145) (11,0.073814)
    };

    \addplot[
        thick,
        red!85!black,
        dash dot,
        mark=diamond*,
        mark size=0.75pt,
        mark options={fill=green!70!black, draw=green!70!black}
    ] coordinates {
        (0,0.256642) (1,0.237753) (2,0.340190) (3,0.207567)
        (4,0.308696) (5,0.166533) (6,0.147679) (7,0.165760)
        (8,0.084382) (9,0.126944) (10,0.038318) (11,0.050725)
    };

    \end{axis}
    \end{tikzpicture}

        }
        \vspace{-0.75em}
        \caption{Entropy comparison across modalities and \textcolor{red!50!black}{drifted}--\textcolor{green!50!black}{undrifted} pairs.}
        \label{fig:label_drift}
    \end{minipage}

\end{figure}

\section{Prediction Uncertainty Under Label Drift}
\label{app:label_drift}

\BfPara{Setting} To evaluate the effect of label drift on model uncertainty, we analyze samples spanning the years 2013 to 2025 whose ground-truth labels were revised after initial annotation. For each year and each of the three feature modalities (static, graph-based, and dynamic), an XGBoost classifier is trained on 70\% of the non-drifted samples and evaluated separately on the remaining 30\% of non-drifted samples and on the full set of drifted samples. We measure prediction uncertainty using binary entropy and margin distance from the decision boundary~\cite{baier2021detecting}, and compare these distributions across drifted and non-drifted groups to assess whether label-drifted samples are systematically harder for the model to classify.

\BfPara{Analysis}Figure~\ref{fig:label_drift} reports the mean prediction entropy of drifted and non-drifted samples across the three feature modalities over time. The results show that uncertainty is consistently higher for drifted samples in the early years, with the gap being most pronounced in 2016, where static, graph-based, and dynamic modalities report entropy values of 0.479, 0.488, and 0.669 respectively, compared to 0.340, 0.341, and 0.555 for non-drifted samples. However, this separation narrows in later years and nearly disappears by 2023--2025. This trend should be interpreted cautiously, as the later years contain smaller drifted-sample sizes, which can make the entropy estimates less stable and reduce the reliability of the observed uncertainty gap.

\section{Additional Concept-Drift Adaptation Results}
\label{app:extra-cda}
Tables~\ref{tab:adaptation-cd-50}, \ref{tab:adaptation-cd-100}, and \ref{tab:adaptation-cd-200} show that the trends reported in Section~\ref{sec:cda-main} persist under smaller labeling budgets. Chen et al.~\cite{chen2023continuous} consistently provides stronger adaptation than CADE~\cite{cade}, but both methods exhibit substantial degradation when moving from IID to NEAR and FAR evaluation. Comparing these results with the 400-sample setting in Table~\ref{tab:adaptation-cd} further shows that increasing the monthly labeling budget improves some IID and NEAR results, but does not resolve the FAR-setting degradation. Thus, the failure mode is not simply a consequence of insufficient labels in a single adaptation cycle; rather, existing CDA methods remain limited under accumulated temporal drift across static, dynamic, graph-based, and fused representations. 

\begin{table}[t]
\caption{Performance of adaptation methods on the McNdroid dataset across IID, NEAR, and FAR settings using three modalities and feature fusion (budget = 50).}
\label{tab:adaptation-cd-50}
\centering
\scriptsize
\setlength{\tabcolsep}{4pt}
\renewcommand{\arraystretch}{1.25}

\begin{tabular}{p{2.2cm}|c|c|c|c|c}
\hline
\textbf{Method}
& \textbf{Setting}
& \textbf{Static}
& \textbf{Dynamic}
& \textbf{Graph-based}
& \textbf{Feature Fusion} \\
\hline

\multirow{3}{*}{\centering Chen et al.~\cite{chen2023continuous}}
& IID  
& \meanstd{89.38}{0.96} 
& \meanstd{72.59}{0.28}
& \meanstd{83.63}{0.88}
& \meanstd{86.78}{0.25} \\

& NEAR 
& \meanstd{77.47}{1.05}
& \meanstd{56.21}{2.51}
& \meanstd{65.99}{1.32}
& \meanstd{70.62}{1.17} \\

& FAR  
& \meanstd{61.68}{0.50}
& \meanstd{34.21}{0.27}
& \meanstd{53.68}{0.59}
& \meanstd{62.98}{1.30} \\
\hline

\multirow{3}{*}{\centering{CADE~\cite{cade}}}
& IID
& \meanstd{87.12}{0.69}
& \meanstd{72.65}{0.99}
& \meanstd{79.72}{0.88}
& \meanstd{80.34}{1.85} \\

& NEAR
& \meanstd{47.17}{0.70}
& \meanstd{30.99}{2.90}
& \meanstd{30.40}{6.83}
& \meanstd{34.97}{8.82} \\

& FAR
& \meanstd{18.57}{0.09}
& \meanstd{11.54}{1.13}
& \meanstd{7.30}{1.23}
& \meanstd{9.94}{3.21} \\
\hline

\end{tabular}
\end{table}

\begin{table}[t]
\caption{Performance of adaptation methods on the McNdroid dataset across IID, NEAR, and FAR settings using three modalities and feature fusion (budget = 100).}
\label{tab:adaptation-cd-100}
\centering
\scriptsize
\setlength{\tabcolsep}{4pt}
\renewcommand{\arraystretch}{1.25}

\begin{tabular}{p{2.2cm}|c|c|c|c|c}
\hline
\textbf{Method}
& \textbf{Setting}
& \textbf{Static}
& \textbf{Dynamic}
& \textbf{Graph-based}
& \textbf{Feature Fusion} \\
\hline

\multirow{3}{*}{\centering Chen et al.~\cite{chen2023continuous}}
& IID  
& \meanstd{91.70}{0.54} 
& \meanstd{73.78}{0.25}
& \meanstd{87.09}{0.67}
& \meanstd{89.58}{1.23} \\

& NEAR 
& \meanstd{82.63}{0.45}
& \meanstd{62.03}{6.60}
& \meanstd{61.01}{0.73}
& \meanstd{77.65}{0.93} \\

& FAR  
& \meanstd{67.15}{1.51}
& \meanstd{39.27}{2.35}
& \meanstd{61.01}{0.73}
& \meanstd{68.23}{1.46} \\
\hline

\multirow{3}{*}{\centering{CADE~\cite{cade}}}
& IID
& \meanstd{87.69}{0.32}
& \meanstd{70.68}{1.27}
& \meanstd{80.82}{1.63}
& \meanstd{83.92}{2.06} \\

& NEAR
& \meanstd{46.44}{5.12}
& \meanstd{23.46}{5.91}
& \meanstd{29.98}{2.71}
& \meanstd{42.16}{1.16} \\

& FAR
& \meanstd{19.53}{1.43}
& \meanstd{10.37}{0.13}
& \meanstd{5.66}{0.67}
& \meanstd{12.62}{2.70} \\
\hline

\end{tabular}
\end{table}

\begin{table}[t]
\caption{Performance of adaptation methods on the McNdroid dataset across IID, NEAR, and FAR settings using three modalities and feature fusion (budget = 200).}
\label{tab:adaptation-cd-200}
\centering
\scriptsize
\setlength{\tabcolsep}{4pt}
\renewcommand{\arraystretch}{1.25}

\begin{tabular}{p{2.2cm}|c|c|c|c|c}
\hline
\textbf{Method}
& \textbf{Setting}
& \textbf{Static}
& \textbf{Dynamic}
& \textbf{Graph-based}
& \textbf{Feature Fusion} \\
\hline

\multirow{3}{*}{\centering Chen et al.~\cite{chen2023continuous}}
& IID  
& \meanstd{93.67}{0.36} 
& \meanstd{74.94}{0.28}
& \meanstd{87.94}{1.63}
& \meanstd{91.61}{0.33} \\

& NEAR 
& \meanstd{86.19}{1.52}
& \meanstd{67.10}{3.65}
& \meanstd{66.95}{1.55}
& \meanstd{79.41}{0.81} \\

& FAR  
& \meanstd{70.13}{1.20}
& \meanstd{45.27}{0.49}
& \meanstd{64.93}{0.60}
& \meanstd{70.42}{0.15} \\
\hline

\multirow{3}{*}{\centering{CADE~\cite{cade}}}
& IID
& \meanstd{90.11}{0.21}
& \meanstd{74.15}{0.29}
& \meanstd{81.97}{1.07}
& \meanstd{83.92}{2.06} \\

& NEAR
& \meanstd{52.17}{5.90}
& \meanstd{31.92}{1.39}
& \meanstd{28.66}{1.80}
& \meanstd{42.16}{1.16} \\

& FAR
& \meanstd{21.12}{1.26}
& \meanstd{11.30}{0.51}
& \meanstd{5.29}{0.38}
& \meanstd{12.62}{2.70} \\
\hline

\end{tabular}
\end{table}

\section{Unsupervised Family Structure Analysis}
\label{app:unsupervised-family-classification}

We evaluate whether \system representations preserve malware-family structure without supervision using HDBSCAN~\cite{hdbscan} clustering across years and modalities. This analysis complements supervised detection by measuring whether static, dynamic, and graph-based features naturally form clusters aligned with malware-family labels. 

\begin{figure}[t]
\centering

\begin{minipage}[b]{0.48\textwidth}
\centering
\begin{tikzpicture}[baseline=(current bounding box.south)]
\begin{axis}[
    width=0.9\linewidth,
    height=4.8cm,
    scale only axis,
    ybar,
    xtick pos=bottom,
    ytick pos=left,
    bar width=3.5pt,
    enlarge x limits=0.18,
    enlarge y limits=false,
    legend style={
        at={(0.9999,.9905)},
        anchor=north east,
        legend columns=3,
        font=\tiny,
        inner xsep=1pt,
        inner ysep=1pt,
        row sep=0pt,
        cells={anchor=west},
    },
    legend image code/.code={
        \draw[#1] (0cm,-0.08cm) rectangle (0.1cm,0.10cm);
    },
    ylabel={Mean Temporal Drift Index},
    xlabel={Family},
    xlabel style={
        at={(axis description cs:0.5,-0.22)},
        anchor=north,
        font=\bfseries\footnotesize
    },
    ylabel style={font=\bfseries\footnotesize, xshift=-6pt, yshift=-5pt},
    symbolic x coords={root,mecor,tapcore,fusob,wkload,appsgeyser,apptrack,frupi,virtualapp,ewind},
    xtick=data,
    x tick label style={rotate=45, anchor=east, font=\bfseries\tiny},
    y tick label style={font=\bfseries\tiny},
    ytick={0,0.5,1.0,1.5,2.0,2.5},
    ymin=0,
    ymax=2.7,
]

\addplot[
    fill=blue!30,
    draw=blue!30!black,
    bar shift=-4.5pt,
] coordinates {
    (root,2.5064576962283387)
    (mecor,2.2950000000000004)
    (tapcore,1.4863497455831072)
    (fusob,2.2667010309278357)
    (wkload,1.1857213052780158)
    (appsgeyser,1.1716450503287488)
    (apptrack,1.1069869074083782)
    (frupi,1.2238091622129263)
    (virtualapp,1.5607270570704588)
    (ewind,1.1370250223116511)
};

\addplot[
    fill=red!60,
    draw=red!30!black,
    bar shift=0pt,
] coordinates {
    (root,2.1067635066258923)
    (mecor,1.2530525631012372)
    (tapcore,1.291102407053612)
    (fusob,0.2563659793814433)
    (wkload,1.485233597605287)
    (appsgeyser,1.4669138282913066)
    (apptrack,1.089051046767652)
    (frupi,1.0029672923604342)
    (virtualapp,1.3350478496591116)
    (ewind,0.9854522774657433)
};

\addplot[
    fill=green!30,
    draw=green!30!black,
    bar shift=4.5pt,
] coordinates {
    (root,2.1102671637290054)
    (mecor,1.1625708123606262)
    (tapcore,1.3311575467678805)
    (fusob,1.4817564420824505)
    (wkload,0.9143795562479923)
    (appsgeyser,0.75286230455146)
    (apptrack,1.17480505099086)
    (frupi,1.0139983303430913)
    (virtualapp,0.16226044506215573)
    (ewind,0.8462856110434835)
};

\legend{Static, Graph-based, Dynamic}
\end{axis}
\end{tikzpicture}
\end{minipage}
\hfill
\begin{minipage}[b]{0.48\textwidth}
\centering
\begin{tikzpicture}[baseline=(current bounding box.south)]
\begin{axis}[
    width=0.9\linewidth,
    height=4.8cm,
    scale only axis,
    ybar,
    xtick pos=bottom,
    ytick pos=left,
    bar width=3pt,
    enlarge x limits=0.10,
    enlarge y limits=false,
    legend style={
        at={(0.9999,.9905)},
        anchor=north east,
        legend columns=3,
        font=\tiny,
        inner xsep=1pt,
        inner ysep=1pt,
        row sep=0pt,
        cells={anchor=west},
    },
    legend image code/.code={
        \draw[#1] (0cm,-0.08cm) rectangle (0.1cm,0.10cm);
    },
    ylabel={Family Identification Score},
    xlabel={Year},
    xlabel style={
        at={(axis description cs:0.5,-0.22)},
        anchor=north,
        font=\bfseries\footnotesize
    },
    ylabel style={font=\bfseries\footnotesize, xshift=-6pt, yshift=-5pt},
    symbolic x coords={2013,2014,2016,2017,2018,2019,2020,2021,2022,2023,2024,2025},
    xtick=data,
    x tick label style={rotate=45, anchor=east, font=\bfseries\tiny},
    y tick label style={font=\bfseries\tiny},
    ytick={0,0.2,0.4,0.6,0.8},
    ymin=0,
    ymax=0.8,
]

\addplot[
    fill=blue!30,
    draw=blue!30!black,
    bar shift=-4pt,
] coordinates {
    (2013,0.5471242187445535)
    (2014,0.4781225433645766)
    (2016,0.45414346297011576)
    (2017,0.7575500421813247)
    (2018,0.7233113748180925)
    (2019,0.5950385374523448)
    (2020,0.3911791358744058)
    (2021,0.4775198823795086)
    (2022,0.4439729575365351)
    (2023,0.3142642772038778)
    (2024,0.1)
    (2025,0.1)
};

\addplot[
    fill=red!60,
    draw=red!30!black,
    bar shift=0pt,
] coordinates {
    (2013,0.5081302871246351)
    (2014,0.47106659849946986)
    (2016,0.45362017853947956)
    (2017,0.7508683382747261)
    (2018,0.7510182284242624)
    (2019,0.5626227244128702)
    (2020,0.40506241786840824)
    (2021,0.46448054668414607)
    (2022,0.5142421090706906)
    (2023,0.3839201504922576)
    (2024,0.1)
    (2025,0.5957889556884766)
};

\addplot[
    fill=green!30,
    draw=green!30!black,
    bar shift=4pt,
] coordinates {
    (2013,0.3844579480815138)
    (2014,0.3813876321962847)
    (2016,0.36103853372326467)
    (2017,0.6017009927742158)
    (2018,0.6398306575880182)
    (2019,0.4439418301051796)
    (2020,0.29124717463099176)
    (2021,0.3305027784724719)
    (2022,0.3294872486756022)
    (2023,0.3161181167576048)
    (2024,0.1)
    (2025,0.1)
};

\legend{Static, Graph-based, Dynamic}
\end{axis}
\end{tikzpicture}
\end{minipage}

\caption{Comparison of modality-specific behavior across malware families and years. Left: mean temporal drift index by family. Right: family identification score by modality across years.}
\label{fig:drift-and-family-identification}
\end{figure}
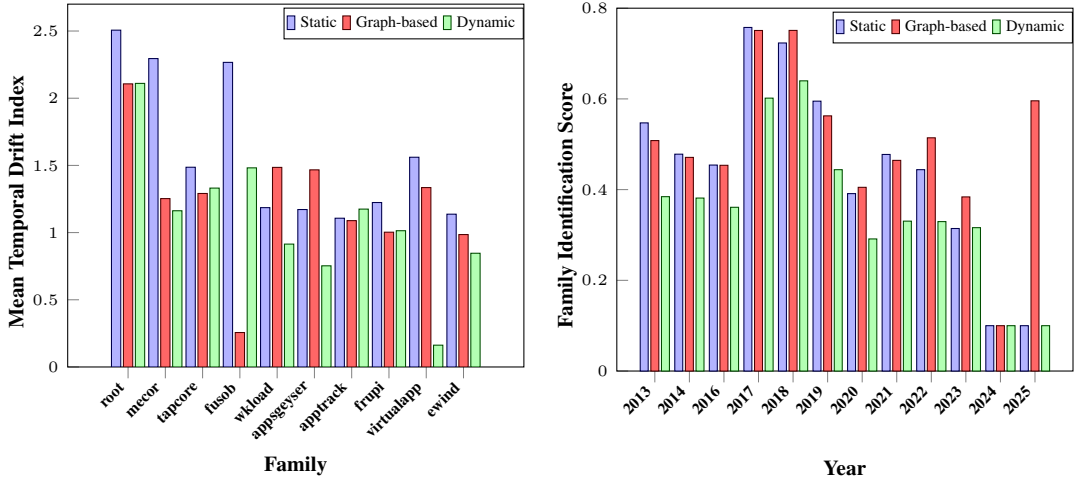

Figure~\ref{fig:drift-and-family-identification} summarizes modality-specific behavior at the family and year levels. The left panel shows that temporal drift varies substantially across malware families and modalities. Families such as \texttt{root}, \texttt{mecor}, and \texttt{fusob} exhibit high drift in static features, indicating large changes in manifest- and code-level artifacts over time. In contrast, dynamic and graph-based drift are lower for some families but higher for others, suggesting that malware evolution is not uniform across representations. This heterogeneity shows that family-level drift cannot be captured reliably by a single modality.

The right panel reports the family identification score across years. Static and graph-based representations generally provide stronger family-level structure than dynamic features, while dynamic features remain consistently lower, likely due to the sparsity and variability of sandbox-observed behavior. All modalities peak around 2017--2018, indicating stronger family separability in these years, but decline in later periods, especially after 2019. This trend suggests that family structure becomes less stable under long-term temporal shift. Together, these results show that McNdroid supports fine-grained analysis of both family-specific drift and modality-dependent family identification, complementing the supervised detection results.

\section{Continual Learning on \system}
\label{app:continual-learning}

Continual learning (CL) studies how a model learns from a sequence of tasks while retaining knowledge from previous tasks~\cite{van2024continual, continual-learning-malware}. This is important for Android malware analysis because malware data is non-stationary: benign apps, malware behaviors, and malware families evolve over time~\cite{continual-learning-malware, malcl}. Following standard CL definitions, we consider Domain Incremental Learning (Domain-IL) and Class Incremental Learning (Class-IL), which are two widely used CL scenarios in prior work.

We evaluate CL on our multimodal Android malware dataset using four feature settings: static, dynamic, graph-based, and fusion. In the fusion setting, samples are aligned by hash and features from all modalities are concatenated. We compare three CL baselines: None/Naive, Joint, and Experience Replay (ER). The None baseline trains sequentially on the current task without using past data. Joint retrains from scratch using all cumulative training data observed so far. ER trains on the current task together with samples stored from previous tasks, following the replay-based strategy used to reduce catastrophic forgetting~\cite{er}.

\BfPara{Domain-IL} In Domain-IL, each task corresponds to one calendar year from 2013 to 2025, excluding 2015. Each task contains both benign and malware samples, and the prediction target remains fixed as binary malware detection. Thus, the model must learn to distinguish benign from malicious apps while adapting to temporal distribution shift. 

For None, the model is trained sequentially on each year without access to previous years. For Joint, the model is retrained using all training data from the first year up to the current year. For ER, the model trains on the current year together with replay samples from earlier years. We report task-wise performance for each modality and for the fused multimodal representation.

\newcommand{\std}[1]{{\scriptscriptstyle\pm#1}}

\begin{table*}[ht]
\centering
\caption{Average F1 score (\%) comparison across modalities (static, graph-based, dynamic, and fusion) for domain incremental continual learning settings. Results are reported as mean $\pm$ std over 5 runs.}
\label{tab:domail-il-f1}
\resizebox{\textwidth}{!}{%
\begin{tabular}{l|ccc|ccc|ccc|ccc}
\toprule
\textbf{Year} &
\multicolumn{3}{c|}{\textbf{Static}} &
\multicolumn{3}{c|}{\textbf{Graph-based}} &
\multicolumn{3}{c|}{\textbf{Dynamic}} &
\multicolumn{3}{c}{\textbf{Feature Fusion}} \\
\cmidrule(lr){2-4}
\cmidrule(lr){5-7}
\cmidrule(lr){8-10}
\cmidrule(lr){11-13}
& \textbf{None} & \textbf{Joint} & \textbf{ER}
& \textbf{None} & \textbf{Joint} & \textbf{ER}
& \textbf{None} & \textbf{Joint} & \textbf{ER}
& \textbf{None} & \textbf{Joint} & \textbf{ER} \\
\midrule
2013 & \meanstd{41.8}{0.2} & \meanstd{42.0}{0.1} & \meanstd{40.1}{0.3} & \meanstd{27.0}{0.2} & \meanstd{27.6}{0.1} & \meanstd{26.7}{0.3} & \meanstd{37.9}{0.2} & \meanstd{39.5}{0.1} & \meanstd{37.8}{0.3} & \meanstd{44.8}{0.2} & \meanstd{43.2}{0.1} & \meanstd{44.7}{0.3} \\
2014 & \meanstd{47.4}{0.2} & \meanstd{48.6}{0.1} & \meanstd{48.3}{0.3} & \meanstd{31.5}{0.2} & \meanstd{32.9}{0.1} & \meanstd{33.3}{0.3} & \meanstd{s44.0}{0.2} & \meanstd{39.9}{0.2} & \meanstd{42.3}{0.3} & \meanstd{42.4}{0.2} & \meanstd{46.5}{0.1} & \meanstd{42.0}{0.3} \\
2016 & \meanstd{67.3}{0.1} & \meanstd{69.5}{0.1} & \meanstd{69.8}{0.2} & \meanstd{59.8}{0.2} & \meanstd{68.2}{0.1} & \meanstd{58.2}{0.3} & \meanstd{59.5}{0.2} & \meanstd{58.1}{0.2} & \meanstd{57.2}{0.3} & \meanstd{65.4}{0.1} & \meanstd{65.8}{0.1} & \meanstd{65.5}{0.2} \\
2017 & \meanstd{66.6}{0.2} & \meanstd{78.6}{0.1} & \meanstd{74.2}{0.3} & \meanstd{62.7}{0.2} & \meanstd{80.2}{0.1} & \meanstd{67.3}{0.3} & \meanstd{47.2}{0.3} & \meanstd{62.1}{0.2} & \meanstd{56.3}{0.3} & \meanstd{74.0}{0.2} & \meanstd{81.1}{0.1} & \meanstd{74.2}{0.3} \\
2018 & \meanstd{69.6}{0.1} & \meanstd{80.5}{0.1} & \meanstd{73.7}{0.3} & \meanstd{62.3}{0.2} & \meanstd{80.9}{0.1} & \meanstd{73.7}{0.3} & \meanstd{44.8}{0.3} & \meanstd{54.3}{0.2} & \meanstd{55.1}{0.3} & \meanstd{68.9}{0.2} & \meanstd{79.4}{0.1} & \meanstd{72.8}{0.3} \\
2019 & \textbf{\meanstd{76.1}{0.3}} & \meanstd{83.8}{0.2} & \textbf{\meanstd{78.4}{0.4}} & \meanstd{73.4}{0.3} & \meanstd{82.5}{0.2} & \meanstd{75.1}{0.4} & \textbf{\meanstd{62.7}{0.3}} & \meanstd{69.5}{0.2} & \meanstd{63.9}{0.4} & \textbf{\meanstd{75.7}{0.3}} & \meanstd{83.2}{0.2} & \meanstd{78.4}{0.4} \\
2020 & \meanstd{75.9}{0.1} & \meanstd{84.8}{0.1} & \meanstd{78.2}{0.3} & \meanstd{69.9}{0.2} & \meanstd{83.5}{0.1} & \meanstd{72.7}{0.3} & \meanstd{61.5}{0.2} & \meanstd{70.3}{0.1} & \meanstd{62.2}{0.3} & \meanstd{74.0}{0.1} & \meanstd{84.1}{0.1} & \meanstd{76.7}{0.3} \\
2021 & \meanstd{75.4}{0.2} & \meanstd{86.1}{0.1} & \meanstd{77.8}{0.3} & \meanstd{67.7}{0.2} & \meanstd{83.5}{0.1} & \meanstd{70.2}{0.3} & \meanstd{58.3}{0.2} & \meanstd{71.5}{0.1} & \meanstd{62.1}{0.3} & \meanstd{70.6}{0.2} & \meanstd{85.4}{0.1} & \meanstd{72.4}{0.3} \\
2022 & \meanstd{72.0}{0.2} & \meanstd{86.8}{0.1} & \meanstd{74.3}{0.3} & \meanstd{67.8}{0.2} & \meanstd{84.9}{0.1} & \meanstd{70.1}{0.3} & \meanstd{59.4}{0.2} & \meanstd{72.1}{0.1} & \meanstd{63.5}{0.3} & \meanstd{69.4}{0.2} & \meanstd{85.9}{0.1} & \meanstd{73.2}{0.3} \\
2023 & \meanstd{72.7}{0.1} & \meanstd{88.3}{0.1} & \textbf{\meanstd{78.3}{0.3}} & \textbf{\meanstd{73.5}{0.2}} & \meanstd{86.2}{0.1} & \textbf{\meanstd{75.6}{0.3}} & \meanstd{61.8}{0.2} & \meanstd{74.8}{0.1} & \textbf{\meanstd{65.7}{0.3}} & \meanstd{74.6}{0.1} & \meanstd{86.2}{0.1} & \meanstd{76.5}{0.3} \\
2024 & \meanstd{50.9}{0.3} & \meanstd{89.3}{0.1} & \meanstd{76.3}{0.4} & \meanstd{12.0}{0.4} & \meanstd{87.8}{0.1} & \meanstd{73.5}{0.4} & \meanstd{42.2}{0.4} & \meanstd{74.9}{0.1} & \meanstd{64.6}{0.4} & \meanstd{54.4}{0.3} & \meanstd{87.5}{0.1} & \meanstd{77.6}{0.4} \\
2025 & \meanstd{31.8}{0.4} & \textbf{\meanstd{91.4}{0.1}} & \meanstd{74.6}{0.4} & \meanstd{21.0}{0.4} & \textbf{\meanstd{89.3}{0.1}} & \meanstd{72.9}{0.4} & \meanstd{27.3}{0.4} & \textbf{\meanstd{75.6}{0.1}} & \meanstd{63.3}{0.4} & \meanstd{49.3}{0.3} & \textbf{\meanstd{89.2}{0.1}} & \textbf{\meanstd{79.3}{0.4}} \\
\bottomrule
\end{tabular}%
}
\end{table*}

\BfPara{Class-IL} In Class-IL, we focus on malware family lassification. We select the top 100 malware families with the largest number of samples following prior work~\cite{malcl}. We then create a stratified 80/20 train/test split by malware family. The task sequence is constructed as follows: Task 1 contains the first 50 malware families, and each following task introduces 5 new malware families until all 100 families are included.

At each task, the model is trained according to the corresponding CL strategy and evaluated on the cumulative test set containing all malware families observed so far. None trains only on the current task, Joint trains on all observed tasks, and ER trains on the current task plus replay samples from previous tasks. This setting evaluates whether the model can learn newly introduced malware families while preserving performance on previously seen families. As in Domain-IL, we run this setup separately for static, Graph-based, Dynamic, and features fusion.

\begin{table*}[ht]
\centering
\caption{Macro-F1 scores (\%) across 11 tasks for Static, Graph-based, Dynamic, and Fusion modalities on class incremental continual learning settings. Results are reported as mean $\pm$ std over 5 runs.}
\label{tab:class-il-f1}
\resizebox{\textwidth}{!}{%
\begin{tabular}{c|ccc|ccc|ccc|ccc}
\toprule
\textbf{Task}
& \multicolumn{3}{c|}{\textbf{Static}}
& \multicolumn{3}{c|}{\textbf{Graph-based}}
& \multicolumn{3}{c|}{\textbf{Dynamic}}
& \multicolumn{3}{c}{\textbf{Feature Fusion}} \\
\cmidrule(lr){2-4}
\cmidrule(lr){5-7}
\cmidrule(lr){8-10}
\cmidrule(lr){11-13}
& \textbf{None} & \textbf{Joint} & \textbf{ER}
& \textbf{None} & \textbf{Joint} & \textbf{ER}
& \textbf{None} & \textbf{Joint} & \textbf{ER}
& \textbf{None} & \textbf{Joint} & \textbf{ER} \\
\midrule
1  & \textbf{\meanstd{67.1}{0.6}} & \meanstd{67.1}{0.5} & \meanstd{67.1}{0.8} & \textbf{\meanstd{17.3}{0.7}} & \meanstd{17.3}{0.6} & \meanstd{17.3}{0.9} & \meanstd{51.7}{0.6} & \meanstd{51.7}{0.5} & \meanstd{51.7}{0.8} & \meanstd{33.8}{0.7} & \meanstd{35.1}{0.5} & \meanstd{32.7}{0.9} \\
2  & \meanstd{31.3}{0.8} & \meanstd{68.7}{0.5} & \meanstd{59.3}{1.0} & \meanstd{0.2}{0.2}  & \meanstd{15.1}{0.7} & \meanstd{15.4}{1.0} & \meanstd{8.3}{0.5}  & \meanstd{52.7}{0.6} & \meanstd{45.2}{0.9} & \meanstd{2.1}{0.3}  & \meanstd{29.5}{0.7} & \meanstd{30.5}{1.1} \\
3  & \meanstd{9.0}{0.5}  & \meanstd{66.4}{0.6} & \meanstd{53.6}{1.0} & \meanstd{0.0}{0.1}  & \meanstd{16.6}{0.7} & \meanstd{12.9}{1.1} & \meanstd{1.0}{0.2}  & \meanstd{49.0}{0.6} & \meanstd{41.7}{1.0} & \meanstd{0.1}{0.1}  & \meanstd{29.4}{0.7} & \meanstd{27.1}{1.1} \\
4  & \meanstd{1.3}{0.3}  & \meanstd{64.5}{0.6} & \meanstd{49.7}{1.1} & \meanstd{0.0}{0.1}  & \meanstd{16.1}{0.8} & \meanstd{10.0}{1.0} & \meanstd{2.1}{0.3}  & \meanstd{48.1}{0.7} & \meanstd{39.5}{1.0} & \meanstd{0.0}{0.1}  & \meanstd{29.2}{0.7} & \meanstd{22.7}{1.2} \\
5  & \meanstd{0.0}{0.1}  & \meanstd{63.5}{0.6} & \meanstd{45.9}{1.1} & \meanstd{0.0}{0.1}  & \meanstd{14.0}{0.8} & \meanstd{10.2}{1.0} & \meanstd{0.9}{0.2}  & \meanstd{48.1}{0.7} & \meanstd{36.4}{1.1} & \meanstd{0.0}{0.1}  & \meanstd{28.3}{0.8} & \meanstd{19.7}{1.2} \\
6  & \meanstd{0.0}{0.1}  & \meanstd{62.5}{0.7} & \meanstd{41.6}{1.2} & \meanstd{0.0}{0.1}  & \meanstd{12.3}{0.8} & \meanstd{6.6}{0.9}  & \meanstd{0.7}{0.2}  & \meanstd{44.5}{0.7} & \meanstd{34.6}{1.1} & \meanstd{0.0}{0.1}  & \meanstd{22.5}{0.8} & \meanstd{13.7}{1.1} \\
7  & \meanstd{0.0}{0.1}  & \meanstd{61.7}{0.7} & \meanstd{39.7}{1.2} & \meanstd{0.0}{0.1}  & \meanstd{11.2}{0.8} & \meanstd{5.9}{0.9}  & \meanstd{1.0}{0.2}  & \meanstd{45.3}{0.8} & \meanstd{31.7}{1.1} & \meanstd{0.0}{0.1}  & \meanstd{22.1}{0.8} & \meanstd{10.6}{1.0} \\
8  & \meanstd{0.0}{0.1}  & \meanstd{60.4}{0.7} & \meanstd{36.3}{1.2} & \meanstd{0.0}{0.1}  & \meanstd{10.9}{0.8} & \meanstd{4.8}{0.8}  & \meanstd{0.4}{0.2}  & \meanstd{44.7}{0.8} & \meanstd{29.0}{1.1} & \meanstd{0.0}{0.1}  & \meanstd{20.0}{0.8} & \meanstd{13.7}{1.1} \\
9  & \meanstd{0.0}{0.1}  & \meanstd{59.0}{0.8} & \meanstd{34.0}{1.2} & \meanstd{0.0}{0.1}  & \meanstd{9.3}{0.7}  & \meanstd{3.6}{0.8}  & \meanstd{0.1}{0.1}  & \meanstd{43.0}{0.8} & \meanstd{26.9}{1.1} & \meanstd{0.0}{0.1}  & \meanstd{20.6}{0.8} & \meanstd{8.9}{0.9} \\
10 & \meanstd{0.0}{0.1}  & \meanstd{59.8}{0.8} & \meanstd{32.0}{1.3} & \meanstd{0.0}{0.1}  & \meanstd{9.8}{0.7}  & \meanstd{2.0}{0.7}  & \meanstd{0.0}{0.1}  & \meanstd{43.2}{0.8} & \meanstd{25.7}{1.2} & \meanstd{0.0}{0.1}  & \meanstd{18.6}{0.8} & \meanstd{9.3}{0.9} \\
11 & \meanstd{0.0}{0.1}  & \meanstd{59.3}{0.8} & \meanstd{31.2}{1.3} & \meanstd{0.0}{0.1}  & \meanstd{11.0}{0.8} & \meanstd{3.6}{0.8}  & \meanstd{0.0}{0.1}  & \meanstd{43.3}{0.8} & \meanstd{25.2}{1.2} & \meanstd{0.0}{0.1}  & \meanstd{16.4}{0.8} & \meanstd{8.8}{0.9} \\
\bottomrule
\end{tabular}%
}
\end{table*}
\BfPara{Results} Table~\ref{tab:domail-il-f1} reports the F1 scores under the Domain-IL setting. Joint training gives the best F1 scores for almost all modalities, while ER gives moderate improvement over None. Static and Feature Fusion perform strongest overall, reaching around 91.4\% and 89.2\% in 2025. Dynamic features are lower than others, but still improve with Joint training. This shows that domain changes are easier to handle when old and new data are available together.

Table~\ref{tab:class-il-f1} reports the F1 scores under the Class-IL setting. This setting is substantially more challenging than Domain-IL: without a CL mechanism, performance rapidly collapses to near 0\%, indicating severe catastrophic forgetting. Joint training achieves the best overall performance, particularly for Static features, whereas ER provides partial mitigation but still degrades as tasks are introduced. Overall, class incremental learning is more difficult than domain incremental learning and needs better forgetting control.


\section{Malware Family Temporal Stability}
\label{app:family-temporal-stability}

\BfPara{Setting} We analyze the temporal stability of malware families by quantifying how their feature distributions evolve across years withing each modality. For each modality, we load yearly feature representations and associated metadata, and restrict the analysis to samples labeled as malware with valid family annotations. 


To ensure statistical reliability, we retain only families that appear consistently across multiple years with sufficient representation. For each feature dimension, empirical distributions are estimated using normalized histograms, and their discrepancy is measured using the Jeffrey divergence ~\cite{Jeffreysdivergence}. The divergence is computed independently for each feature and then averaged across all features to obtain a single drift score per family and year pair.

\begin{figure}[h]
    \centering

    \vspace{-0.1cm}
    \resizebox{\linewidth}{!}{\input{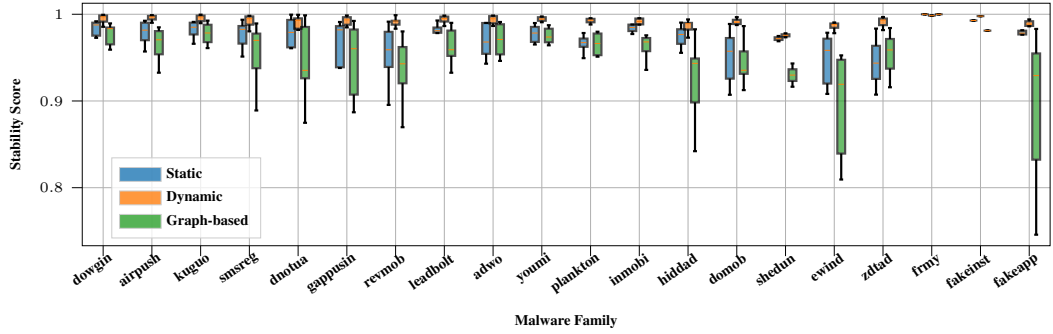}}
    \caption{
        Per-family stability score distributions across three feature modalities for the
        top-20 Android malware families. Each box represents the interquartile
        range over all samples within a family.}    
    \label{fig:stability}
\end{figure}

\BfPara{Analysis} Figure~\ref{fig:stability} presents a comparative analysis of stability scores across twenty distinct malware families, segmented by feature modality. Dynamic features consistently exhibit the highest stability, with median scores approximating 1.00 across the top 20 malware families. Static features, maintain high stability with narrow inter-quartile ranges. In contrast, graph based exhibit substantial volatility, they consistently yield the lowest stability scores and demonstrate the widest distribution of values across all analyzed families. Overall, these results confirm, that all while all three modalities produce stable representations for the majority of the malware families, graph-based are significantly more sensitive to intra-family structural diversity.



\section{Computational Resources}
\label{app:computation}

All dataset processing and experiments for \system were conducted on a high-performance compute server with the following configuration:

\begin{itemize}
    \item \textbf{CPU}: Dual-socket \texttt{Intel Xeon Gold 6430} with a total of 128 logical cores (64 physical cores, 2 threads per core).
    \item \textbf{Memory}: 512 GB RAM.
    \item \textbf{GPU}: 2$\times$ NVIDIA H100 NVL GPUs with 95.8 GB memory per GPU. Experiments were conducted under \texttt{CUDA 12.8} and driver version \texttt{595.71.05}. 
\end{itemize}

This infrastructure enabled us to efficiently process near \~900k million APKs, large-scale temporal benchmarking over 12 years of Android malware data.

\section{Broader Impacts}
\label{app:broaderImpacts}

This work contributes a large-scale longitudinal multimodal benchmark for Android malware detection under temporal distribution shift. As discussed in Section~\ref{sec:discuss}, {\system} highlights the benefits of multimodal data for improving malware detection and enabling research on concept drift, modality-specific degradation, cross-modal robustness, malware-family evolution, and adaptive detection. By aligning static, dynamic, and graph-based representations over time, the benchmark can support more reliable malware detectors for mobile security, endpoint protection, app-store vetting, and adaptive threat monitoring.

The main positive societal impact of {\system} is better evaluation of malware detection systems under realistic non-stationary conditions. The benchmark exposes model aging, reduces reliance on random-split evaluation, and supports reproducible comparison of methods under temporal drift. Releasing pre-extracted features, benchmark splits, and code also lowers the barrier for researchers without large-scale malware-processing or sandboxing infrastructure.

To the best of our knowledge, this dataset does not introduce direct negative societal impacts. {\system} is intended for defensive malware analysis and releases processed feature representations rather than tools for creating, deploying, or evading malware. As with any malware benchmark, insights about detector weaknesses could be misused, but the primary value is defensive: enabling rigorous evaluation of malware detectors as software ecosystems and attacker behaviors evolve. We encourage responsible use for security research and compliance with applicable data-use restrictions.

\section{Dataset Documentation}
\label{app:datasetdocument}

\subsection{Hosted URLs}

\paragraph{Dataset on Hugging Face.} \url{https://huggingface.co/datasets/IQSeC-Lab/McNdroid}.

\paragraph{Croissant.} \url{https://github.com/IQSeC-Lab/McNdroid/mcn_metadata.json} 

\paragraph{Code Access.} \url{https://github.com/IQSeC-Lab/McNdroid/}

\subsection{Accessibility and Reproducibility}

The dataset has been made publicly available on Hugging face at \url{https://huggingface.co/datasets/IQSeC-Lab/McNdroid}. Furthermore, a dedicated GitHub project page has been created at \url{https://github.com/IQSeC-Lab/McNdroid}, which includes detailed instructions and code to reproduce the reported results.



\end{document}